\documentclass[%
 reprint,
superscriptaddress,
 amsmath,amssymb,
 aps,
longbibliography
]{revtex4-1}

\usepackage{aas_macros}
\usepackage{graphicx}
\usepackage{dcolumn}
\usepackage{bm}
\usepackage{hyperref}
\usepackage{color}
\usepackage[mathlines]{lineno}
\usepackage{tabularx}
\usepackage{subfigure}
\usepackage[normalem]{ulem}
\usepackage[export]{adjustbox}
\usepackage{url}
\usepackage{lineno}
\usepackage[flushleft]{threeparttable}
\usepackage{makecell}
\usepackage{multirow}

\usepackage{soul}
\usepackage{booktabs} 
\usepackage{array}
\usepackage{amsmath}  
\usepackage{makecell}

\hypersetup{
    pdfnewwindow=true,    
    colorlinks=true,      
    linkcolor=blue,       
    citecolor=blue,       
    filecolor=blue,       
    urlcolor=blue         
}

\def\orcid#1{\href{https://orcid.org/#1}{\includegraphics[keepaspectratio,width=1.1em]{figures/orcid.png}}}
\usepackage{natbib}

\begin{document}
\title{Precise Measurement of the Cosmic Ray Helium Spectrum above 0.1 PeV}
 
\author{Zhen Cao}
\affiliation{State Key Laboratory of Particle Astrophysics \& Experimental Physics Division \& Computing Center, Institute of High Energy Physics, Chinese Academy of Sciences, 100049 Beijing, China}
\affiliation{University of Chinese Academy of Sciences, 100049 Beijing, China}
\affiliation{TIANFU Cosmic Ray Research Center, 610000 Chengdu, Sichuan,  China}
 
\author{F. Aharonian}
\affiliation{TIANFU Cosmic Ray Research Center, 610000 Chengdu, Sichuan,  China}
\affiliation{University of Science and Technology of China, 230026 Hefei, Anhui, China}
\affiliation{Yerevan State University, 1 Alek Manukyan Street, Yerevan 0025, Armeni a}
\affiliation{Max-Planck-Institut for Nuclear Physics, P.O. Box 103980, 69029  Heidelberg, Germany}
 
\author{Y.X. Bai}
\affiliation{State Key Laboratory of Particle Astrophysics \& Experimental Physics Division \& Computing Center, Institute of High Energy Physics, Chinese Academy of Sciences, 100049 Beijing, China}
\affiliation{TIANFU Cosmic Ray Research Center, 610000 Chengdu, Sichuan,  China}
 
\author{Y.W. Bao}
\affiliation{Tsung-Dao Lee Institute \& School of Physics and Astronomy, Shanghai Jiao Tong University, 200240 Shanghai, China}
 
\author{D. Bastieri}
\affiliation{Center for Astrophysics, Guangzhou University, 510006 Guangzhou, Guangdong, China}
 
\author{X.J. Bi}
\affiliation{State Key Laboratory of Particle Astrophysics \& Experimental Physics Division \& Computing Center, Institute of High Energy Physics, Chinese Academy of Sciences, 100049 Beijing, China}
\affiliation{University of Chinese Academy of Sciences, 100049 Beijing, China}
\affiliation{TIANFU Cosmic Ray Research Center, 610000 Chengdu, Sichuan,  China}
 
\author{Y.J. Bi}
\affiliation{State Key Laboratory of Particle Astrophysics \& Experimental Physics Division \& Computing Center, Institute of High Energy Physics, Chinese Academy of Sciences, 100049 Beijing, China}
\affiliation{TIANFU Cosmic Ray Research Center, 610000 Chengdu, Sichuan,  China}
 
\author{W. Bian}
\affiliation{Tsung-Dao Lee Institute \& School of Physics and Astronomy, Shanghai Jiao Tong University, 200240 Shanghai, China}
 
\author{J. Blunier}
\affiliation{APC, Universit\'e Paris Cit\'e, CNRS/IN2P3, CEA/IRFU, Observatoire de Paris, 119 75205 Paris, France}
 
\author{A.V. Bukevich}
\affiliation{Institute for Nuclear Research of Russian Academy of Sciences, 117312 Moscow, Russia}
 
% ...
\collaboration{The LHAASO Collaboration}

\email{wanglp@ihep.ac.cn\\
llma@ihep.ac.cn\\
zhangss@ihep.ac.cn\\
chensh@ihep.ac.cn}

\date{\today}

\begin{abstract}
We report a measurement of the cosmic ray helium energy spectrum in the energy interval 0.16 -- 13~PeV, derived by subtracting the proton spectrum from the light component~(proton and helium) spectrum obtained with observations made by the Large High Altitude Air Shower Observatory~(LHAASO) under a consistent energy scale. 
The helium spectrum shows a significant hardening centered at $E \simeq$ 1.1~PeV, followed by a softening at $\sim$ 7 PeV, indicating the appearance of a helium `knee'.
Comparing the proton and helium spectra in the LHAASO energy range reveals some remarkable facts. In the lower part of this range, in contrast to the behavior at lower energies, the helium spectrum is significantly softer than the proton spectrum. This results in protons overtaking helium nuclei and becoming the largest cosmic ray component at $E \simeq$ 0.7 PeV. A second crossing of the two spectra is observed at $E \simeq$ 5 PeV, above the proton knee, when helium nuclei overtake protons to become the largest cosmic ray component again. These results have important implications for our understanding of the Galactic cosmic ray sources.

\end{abstract}

\maketitle

\textbf{Introduction---}
Cosmic rays (CRs) are high-energy particles originating from outer space, with energies spanning $10^9$~eV to $10^{20}$~eV~\cite{DEttorrePiazzoli:2022fzw,Greisen:1960wc}.~Their flux $F$ decreases sharply with increasing energy $E$, roughly following a power-law $F \propto E^{\gamma}$~\cite{Lipari:2008ak}, where $\gamma$ is the spectral index.~Notably, several features were found deviating from the single power-law.~The most prominent one is the `knee' at $\sim$ 3.7 Peta-electron-Volts~(PeV; 1 PeV=10$^{15}$ electron Volts), where the all-particle spectrum steepens abruptly, with $\gamma$ changing from $-$2.7 to $-$3.1~\cite{LHAASO:2024knt}.~Although observed for over 60 years, the origin of the knee remains uncertain. It could correspond to the maximum energy limit of Galactic accelerators, or to the rigidity (kinetic energy per charge) at which the properties of cosmic ray propagation undergo a sudden change.
It is evident that measuring the energy spectra of individual CR species, particularly finding their possible knees, is crucial for understanding the origins of cosmic rays~\cite{ParticleDataGroup:2024cfk,Sciascio:2022vkb}.

In this direction, there has been significant progress in the energy range below 0.1~PeV by sending detectors to space and directly detecting the CR particles, for instance, DAMPE~\cite{DAMPE:2019gys,Alemanno:2021gpb,DAMPE:2023pjt}, CALET~\cite{CALET:2022vro,CALET:2023nif}, and AMS~\cite{AMS:2015tnn,AMS:2015azc,AMS:2017seo}.~In those experiments, individual cosmic ray elements can be separated from each other and their energy spectra are measured.~Significantly more complex structures than those inferred from the all-particle spectrum have been revealed.~Many attempts have been made to interpret complex phenomena in association with detailed mechanisms of particle acceleration and propagation; however, challenges still remain~\cite{Blasi:2013rva,Yue:2019sxt,Lagutin:2023pag}.
Extending the measurements of energy spectra in the knee region has long been challenging.

Space-borne detectors suffer from large statistical uncertainties due to their limited acceptance, while ground-based experiments, which measure the CR-induced extensive air showers, face large systematic uncertainties in reconstructing the primary energy and species of CRs. The inherent correlation between the two parameters is the major issue in the shower measurements.
Statistical unfolding techniques are developed by groups such as KASCADE~\cite{KASCADE:2005ynk} covering the range from 1 to 100 PeV, and IceTop~\cite{IceCube:2019hmk} for $E \gtrsim$ 3 PeV.
Those results give some indications for the existence of a helium knee, but with very large systematic uncertainties, e.g. $\sim$100\% in Ref.~\cite{KASCADE:2005ynk}.

The Large High Altitude Air Shower Observatory (LHAASO) has the capability to measure multiple air shower components, enabling more precise energy determination and better mass identification than previous experiments.~Consequently, proton events are isolated from all measured showers with an average purity of 89\% above 0.16 PeV, leading to a high‑precision measurement of the proton energy spectrum~\cite{LHAASO:2025byy}.~This opens the possibility of returning to the conventional way of measuring spectra of individual species of CRs, as done in direct measurements, but in a totally different energy range in which knee features may exist. 

The high altitude of LHAASO ($\sim$4410~m above sea level) enables the measurement of air showers with minimal fluctuations since showers in the knee region are typically developing to around their maxima~\cite{LHAASO:2019qwt}. The simultaneous measurements of multiple shower parameters using up to three different detector arrays enable precise determination of key variables for particle identification. The large acceptance of the detector arrays allows a strict event selection, ensuring high detection quality and thereby providing good control over systematic uncertainties. As a result, the proton spectrum measurement carries a systematic uncertainty of $\sim$17\%, which mainly originates from hadronic interaction models. Notably, this result reveals an unexpected spectral hardening below the long-expected softening known as the knee at 3.3~PeV~\cite{LHAASO:2025byy,LHAASO:2024knt}.

The composition of CRs as a function of energy encodes crucial information regarding their origins.
In the energy range 3 orders of magnitude lower than those of the LHAASO working energies, unexpected hardening of the CR spectra is found in direct measurements~\cite{DAMPE:2019gys,Alemanno:2021gpb,CALET:2022vro,CALET:2023nif}.~Namely, the helium spectrum is harder than that of protons, with a spectral index difference of order 0.1~\cite{AMS:2015azc}. As a consequence, the helium flux exceeds protons at $\sim$0.01~PeV and helium becomes the most abundant species. 
The proton spectrum shows a clear softening at an energy slightly higher than the crossing energy.~Many models, therefore, predict the helium dominance around the knee following a simple rigidity dependence assumption~\cite{Gaisser:2013bla,Hoerandel:2002yg}.~However, the LHAASO measurement of the proton spectrum around the knee seems to challenge the scenario dramatically because the strong hardening of the spectrum is observed.
The proton knee occurs at the same energy as the knee of the all-particle spectrum, thereby indicating the proton dominance at the knee. Measuring the helium spectrum in the same energy range is crucial for clarifying these spectral features, thus improving our knowledge about the composition variation with energy and corresponding interpretations of particle acceleration and propagation through the Galaxy. 

However, the method developed in Ref.~\cite{LHAASO:2025byy} for the proton spectrum measurement, i.e., collecting a pure sample and directly measuring its spectrum, is almost inapplicable to helium.~The key difficulty lies in the need to both suppress heavier nuclei contamination and separate helium events from protons.~The distribution of the composition sensitive parameter, e.g., muon content, of protons is wider than that of helium showers in the selected sample with the contamination of heavier nuclei being minimized.~The shapes of the distributions of the composition sensitive parameter for protons and helium particles are such that it is not possible to find an adequate interval in which helium is dominant with only a small proton contamination. Therefore the method developed in Ref.~\cite{LHAASO:2025byy} for obtaining the spectrum from a high purity selection is not applicable for the helium component.~An alternative approach is to collect a so-called `light CR component' sample, i.e., a mixture of protons and helium nuclei, and measure the spectrum of this light component.
The helium energy spectrum is then derived by subtracting the proton spectrum from it. This method is valid as long as two key conditions are satisfied, namely a) consistent energy scales for proton and helium showers, and b) an unchanged proton-to-helium abundance ratio (H/He) during event selection.
In effect, differences in shower development between light and heavy components are more pronounced than those between protons and others, and the light showers dominate over others in the knee energy range. This enables the collection of purer samples of the light component with higher efficiency than that of the proton sample in Ref.~\cite{LHAASO:2025byy}. This facilitates precise measurement of the light energy spectrum.

Previous attempts to extract the light samples from all cosmic ray species and measure their spectra in the knee region have been made by ground-based experiments~\cite{HAWC:2022zma,ARGO-YBJ:2015isx,Glasmacher:1999xn}.~However, neither of the two key conditions noted above was fully satisfied. Limited statistical power constrained the measurement quality, leading to low-significance hints of the knee in the light spectrum, such as the indication of softening at sub-PeV energy~\cite{ARGO-YBJ:2015isx}.

The identical data set used for the LHAASO proton energy spectrum measurement is adopted in the light spectrum measurement reported here. To ensure a consistent energy scale for the measurements of both protons and helium nuclei, the same energy reconstruction procedure is adopted~\cite{LHAASO:2025byy}.~The shower energy estimator is defined as $N_{c\mu}$~\cite{Wang:2023qeg}, combining the number of Cherenkov photons in the shower image and the shower muon content. For the proton and light showers, the energy has been reconstructed with a systematic bias of approximately 1\% and 1.7\%, respectively.~The energy resolution obeys nearly perfectly symmetrical Gaussian functions with the $\sigma$ parameters of 15\% at $\sim$0.1 PeV, gradually decreasing to 10\% around 1 PeV and then remaining constant at higher energies. 
More details are available in the Supplemental Material~\cite{Helium} (hereafter referred to as SM).

Sufficient showers ($\sim$90 million) were generated in the full simulation including shower development and detector responses to optimize the event selection for the light sample.~Three high energy interaction models, EPOS-LHC~\cite{Pierog:2013ria}, QGSJET-II-04~\cite{Ostapchenko:2013pia} and SIBYLL 2.3d~\cite{Riehn:2019jet}, are used for comparison and estimating corresponding systematic uncertainties (see Sec.~\ref{sec:MC} in the SM). 

\begin{figure}[htpb]
  \centering\includegraphics[width=0.8\linewidth]{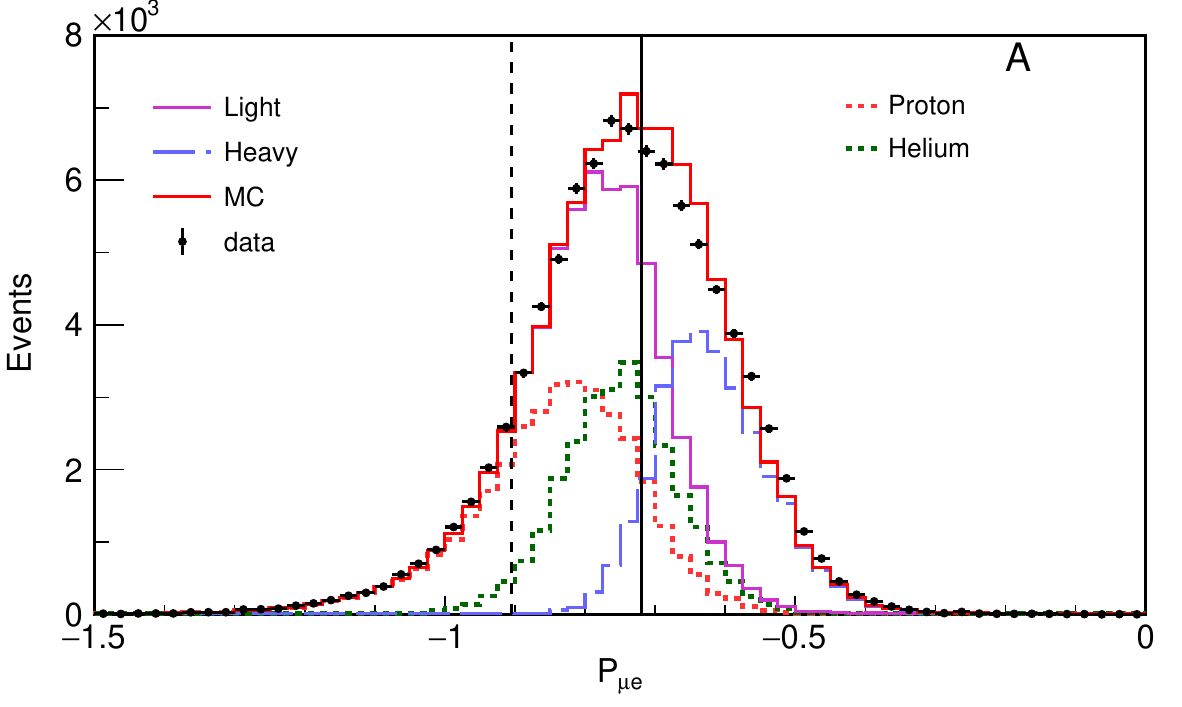}
  \centering\includegraphics[width=0.8\linewidth]{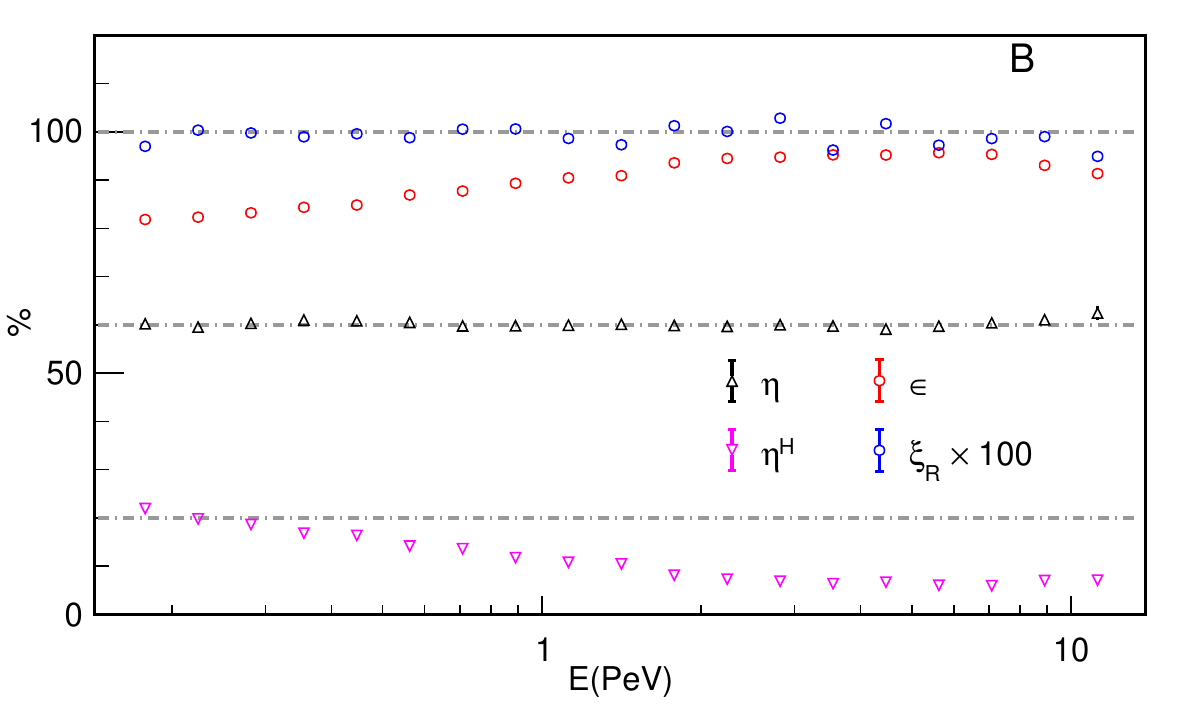}
    \caption{(A) LHAASO data (black dots) are compared with the simulated (histograms) using the GSF-LHAASO composition assumption. Namely: red dashes for protons, green dashes for helium nuclei, and their sum (light showers) in solid purple. Heavier showers in blue dashes, and the sum of all components in solid red. Good agreement with data is evident. 
    Showers distributed between the two vertical solid and dashed lines are selected as the light showers. EPOS‑LHC~\cite{Pierog:2013ria} is assumed for hadronic interaction. (B) The light component selection. The selection efficiency $\eta$ (black triangles) and purity $\epsilon$ (red circles) are plotted together with the survival probability of heavier nuclei $\eta^\text{H}$ (pink inverted triangles). The relative variation of ratio, $R$ = $\Phi_{\rm H}/\Phi_{\rm He}$, $\xi_{R} = \frac{R_{\rm after~cut}}{R_{\rm before~cut}}\times 100$ (blue circles) is also plotted. Three gray dashed lines indicate y‑axis reference levels at 20, 60, and 100.
  }
  \label{fig:divide}
\end{figure}

\textbf{ Light event selection and spectral measurement---}For a given energy, the numbers of both electromagnetic particles $N_e$ and muons $N_{\mu}$ in a shower are sensitive to the primary particle species. 
The heavier nuclei produce more muons and fewer electromagnetic particles, and Monte Carlo studies have shown that the combination $P_{\mu e} = \log_{10} (N_\mu/N_e^{0.82})$ (with both $N_{e}$ and $N_{\mu}$ counted in the annular region 40~–~200 m from the shower core)~\cite{LHAASO:2025byy} is an optimum choice for a parameter sensitive to the mass of the primary particle that is also in good approximation energy independent.
The $P_{\mu e}$-distribution of all measured showers (black dots) and simulated samples with different cosmic ray species (colored histograms) is plotted in Fig.~\ref{fig:divide}A. It is clear that the solid vertical line effectively separates the light showers from other heavier ones.
The dominance of light showers is obvious in this distribution, and their separation from heavier showers is greater than the separation of protons from other species~\cite{LHAASO:2025byy}. Indeed, the use of $P_{\mu e}$ alone suffices to reduce contamination from heavier species to under 20\% in the same energy range as that used for the proton spectrum measurement.

Note that the energy dependence of CRs must be assumed in this analysis to estimate contamination and determine the selection criteria. 
Initially, the GSF model~\cite{Dembinski:2017zsh} was adopted. Once the proton and helium spectra are measured, the GSF model is modified by replacing its spectra with the measured ones. The heavier component proportions are then re‑estimated, leading to the construction of the new GSF‑LHAASO model. The whole analysis is subsequently repeated with the new model. This iterative procedure reduces the systematic uncertainty due to the heavy component contamination. Convergence is achieved after just one iteration.~Further iteration varies the light component fluxes less than 1\% (see Sec.~\ref{sec:new_model} in the SM).

Measurements of the light spectrum often rely on the single‑threshold criterion, to suppress the background of heavier nuclei~\cite{HAWC:2022zma,ARGO-YBJ:2015isx}, as shown by the solid vertical line in Fig.~\ref{fig:divide}A. This approach, however, yields higher selection efficiency for protons than for helium nuclei, altering the original H/He ratio. Given that protons are concentrated at the lowest $P_{\mu e}$ values, an additional criterion can be applied by setting an appropriate lower limit on $P_{\mu e}$ (dashed vertical line in Fig.~\ref{fig:divide}A), such that the combined effects of the two cuts select
protons and helium nuclei with equal efficiencies. This keeps the H/He ratio unchanged during selection.
The selection can be optimized by balancing the selection efficiency $\eta$ against the purity $\epsilon$.~The slightly energy dependent criteria are finally optimized to minimize contamination from the heavy component to be $\sim$20\% at~0.16~PeV, gradually reducing below 10\% above 1.0~PeV.
The contamination is mainly from the C-N-O mass group, which is $\sim$15\% at 0.16~PeV, and reduces to below 10\% above 1.0~PeV. 
The contamination from the Mg-Al-Si mass group and iron is less than 1\% above 0.80~PeV. 
The selection efficiency for light showers is nearly constant of 60\%. All specifications of the selection are summarized in Fig.~\ref{fig:divide}B, including the variation of the H/He ratio $\xi_{R}<5\%$.

\begin{figure}[htpb]
  \centering\includegraphics[width=1\linewidth]{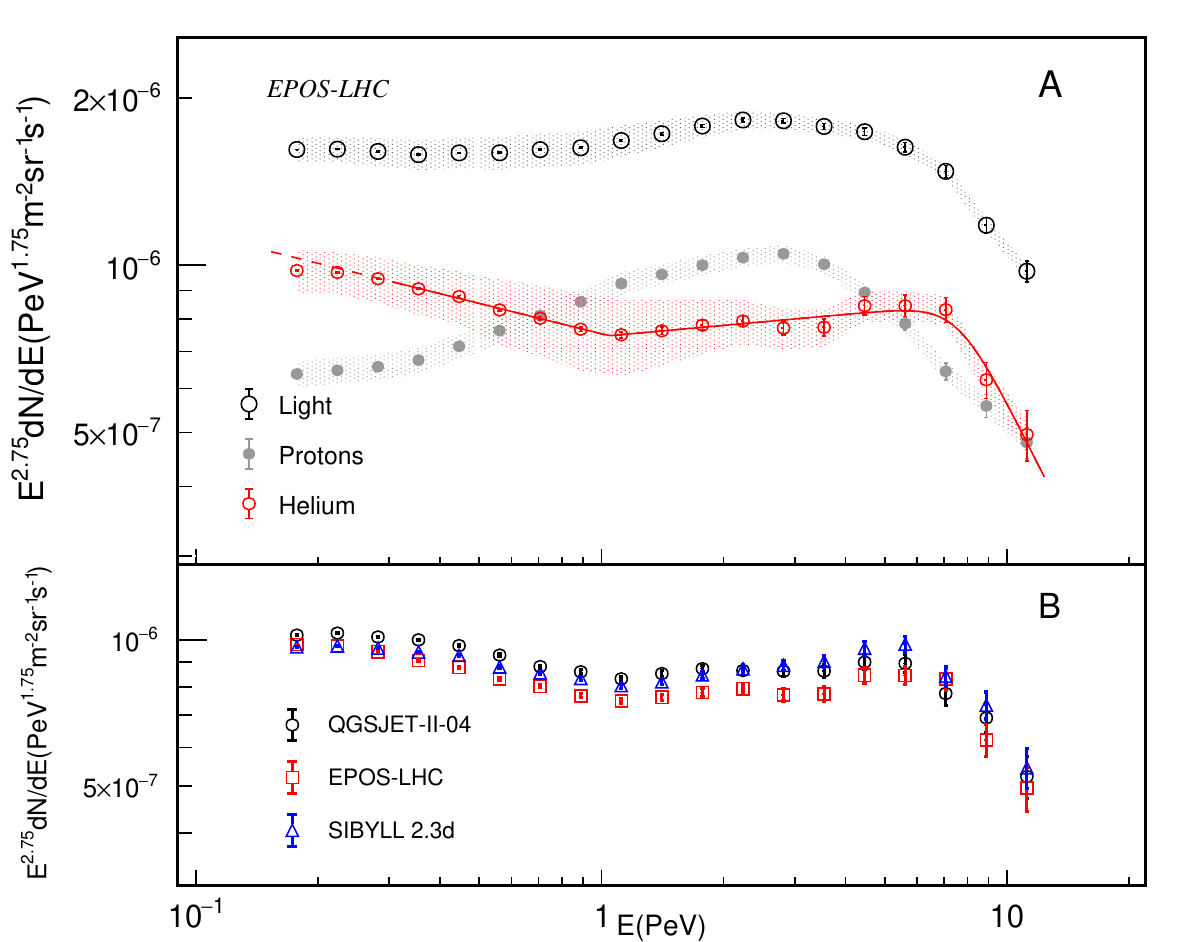}
 \caption{(A) Energy spectra. The light component flux ($\Phi_{\rm L}$) and the updated proton flux ($\Phi_{\rm H}$) are measured with the EPOS-LHC~\cite{Pierog:2013ria} hadronic interaction model assumption. The helium fluxes are derived as $\Phi_{\rm He}=\Phi_{\rm L}-\Phi_{\rm H}$ in red. All spectra are multiplied by $E^{2.75}$. Error bars represent statistical uncertainties only, while the systematic uncertainties are shown as shaded bands. The red solid line is the fit to the helium spectrum in the range 0.3~–~13~PeV. The extension in the dashed part helps to indicate how much the first three data points deviate from the single-index power-law, i.e. by 31.7$\sigma$, 8.7$\sigma$, and 0.7$\sigma$, respectively. This implies a possible structure at energies below 0.3 PeV. (B) Helium spectra derived using three hadronic interaction models, multiplied by $E^{2.75}$.}
\label{fig:Helium_spec}
\end{figure}

The light component energy spectrum from 0.16 to 13 PeV is presented in Fig.~\ref{fig:Helium_spec}A, with the flux multiplied by $E^{2.75}$.~The corresponding event counts $\Delta N_{S}(E)$, fluxes $\Phi_{\rm L}(E)$, and uncertainties are given in Table~\ref{tab:light} in the SM.
The slight hardening of the spectrum at $\sim$ 1~PeV is clearly observed.~The broad knee-like feature is also rather clearly manifested.~Furthermore, statistically speaking, the whole spectrum deviates from a single-index power law, with noticeably small systematic uncertainty as well.

As indicated by the shaded band in Fig.~\ref{fig:Helium_spec}A, the systematic error is estimated accounting for the following sources: a) Lack of knowledge about the abundance of heavy nuclei. Composition models are used in the light spectrum measurement, and the largest difference in the fluxes is $<$6\%. b) The light event selection efficiency is estimated using the simulated MC samples. 
The difference in fluxes obtained by varying the selection efficiency by 10\% from 60\%, that is, between 50\% and 70\%, is found to be within $\pm$1\%.
The systematic uncertainties due to environmental effect corrections include: c) the atmospheric pressure correction, within $\pm$2\%, d) the absolute humidity correction, within $\pm$1\%, and e) the background light correction, within $\pm$2\% (see Sec.~\ref{sec:light_sys} in the SM). 

The largest systematic uncertainty is associated with the high energy hadronic interaction model used in the simulation of shower development in the atmosphere.
As mentioned above, three interaction models are used in the independent analyses and the corresponding spectra are shown in Fig.~\ref{fig:hadronic}. The largest difference in fluxes is found to be within 9\%. 

\textbf{Helium spectrum---}
The helium spectrum is derived by subtracting the proton spectrum from the light spectrum provided that the two necessary conditions discussed above are satisfied.~Subsequently, the proton spectrum is remeasured using the improved estimate of helium contamination. This significantly reduces the systematic uncertainty associated with the composition assumption as shown in Fig.~\ref{fig:proton} in the SM, e.g. the uncertainty reduces from 7\% to 2\% around 3 PeV.
This updated proton spectrum is then used to recalculate the helium spectrum. The iterative process converges rapidly, with flux variations staying below 1\% after four cycles (see Sec.~\ref{sec:proton-iteration} in the SM).

The helium spectrum is shown in Fig.~\ref{fig:Helium_spec}A together with the updated proton spectrum.~The uncertainty is from the combination of statistical errors of proton and light component energy spectra.
The helium flux $\Phi_{\rm He}(E)$ and associated errors are tabulated in Table~\ref{tab:helium} of the SM. 
The systematic uncertainty is shown by the shaded area in Fig.~\ref{fig:Helium_spec}A.
The spectra based on SIBYLL 2.3d~\cite{Riehn:2019jet} and QGSJET-II-04~\cite{Ostapchenko:2013pia} are systematically higher than that obtained using EPOS-LHC~\cite{Pierog:2013ria} by $\sim$10\% and $\sim$13\% on average, respectively, as shown in Fig.~\ref{fig:Helium_spec}B (see Sec.~\ref{sec:helium_sys} in the SM). 

The data can be fitted with a three-component power-law functional form (Eq.~\ref{for:fit3} in Sec.~\ref{sec:Helium} in the SM) over the energy range of 0.3~–~13 PeV. Key spectral features are precisely captured, namely the sharp transients of spectral hardening and the spectral knee. The overall $\chi^2$ of 10.7 for 10 degrees of freedom indicates the goodness of the fitting, as shown in Fig.~\ref{fig:Helium_spec}A. Precise parameter values for the three hadronic interaction model assumptions are listed in Table~\ref{tab:error} in the SM.

\begin{figure}[htpb]
    \centering
    \includegraphics[width=1.0\linewidth]{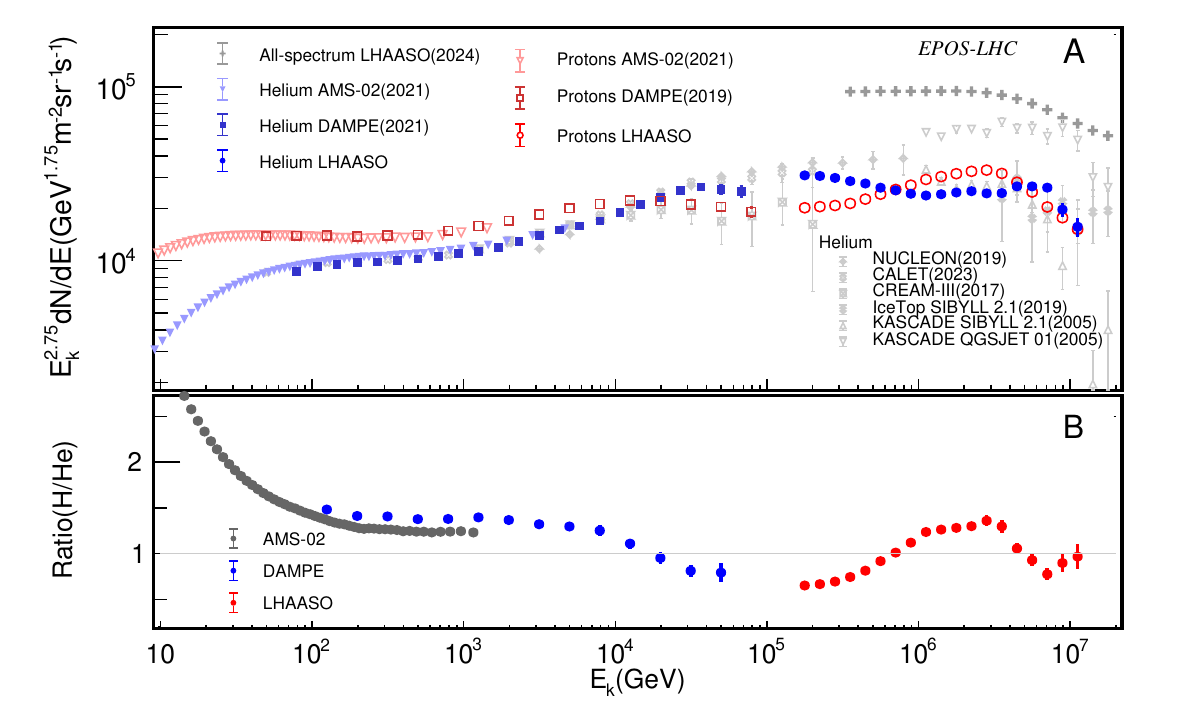}
     \caption{(A) Energy spectra of protons (red open markers) and helium (blue solid markers) scaled by $E^{2.75}$. Only statistical uncertainties are plotted. The helium data are from AMS-02~\cite{AMS:2021nhj}, NUCLEON~\cite{Gorbunov:2018stf}, DAMPE~\cite{Alemanno:2021gpb}, CREAM-III~\cite{Yoon:2017qjx}, CALET~\cite{CALET:2023nif}, IceTop~\cite{IceCube:2019hmk}, KASCADE~\cite{KASCADE:2005ynk} and LHAASO. 
     The proton data are from AMS-02~\cite{AMS:2021nhj}, DAMPE~\cite{DAMPE:2019gys} and LHAASO. 
     The all-particle spectrum from LHAASO~\cite{LHAASO:2024knt} is plotted for reference.~(B) Proton to helium flux ratio $\Phi_{\rm H}(E)/\Phi_{\rm He}(E)$. AMS-02 data in gray (proton data interpolated from Ref.~\cite{AMS:2021nhj}), DAMPE data in blue (helium data interpolated from Ref.~\cite{Alemanno:2021gpb}) and LHAASO data in red are plotted together. Only statistical uncertainties are included. The unity is indicated by the gray line for reference.}
    \label{fig:other_spec}
\end{figure}

The proton and helium spectra are summarized over the energy range from 10 GeV to 20 PeV in Fig.~\ref{fig:other_spec}, alongside data from AMS-02~\cite{AMS:2021nhj}, DAMPE~\cite{DAMPE:2019gys,Alemanno:2021gpb}, CALET~\cite{CALET:2023nif}, NUCLEON~\cite{Gorbunov:2018stf}, CREAM-III~\cite{Yoon:2017qjx}, IceTop~\cite{IceCube:2019hmk}, KASCADE~\cite{KASCADE:2005ynk}, and the LHAASO all-particle energy spectrum. The helium spectrum in the energy range above 0.1 PeV shows some interesting and unexpected features.
In the energy range of 0.3~–~1.0~PeV, the spectrum is well described by a power-law with a single spectral index of $-$2.92 $\pm$ 0.01, approximately equal to what has been measured by the DAMPE and CALET detectors at lower energies in the 0.03~–~0.1~PeV interval. The three lowest-energy points measured by LHAASO indicate that the spectrum is likely harder than the single-index power-law. However, the three points at the boundary of the observation range are not sufficient to define a new spectral component as a firm conclusion. It is clearly very important to cover the gap in energy between the satellite and air shower measurements. Further efforts involving the two detection methods to determine the spectral shape in this critical energy range will help in understanding systematic effects. It obviously is an important goal for future studies.

The helium spectrum becomes harder at the energy $E_h \simeq 1.1 \pm 0.1$ PeV, with the spectral index changing by $\Delta\gamma \simeq 0.25 \pm 0.02$.~This is a very clear feature of the spectrum with high significance of more than 12$\sigma$. The hardening feature, the energy $E_{h}$, and the spectral index at energies below 1 PeV are almost model independent, whereas the spectral index (hardness) above 3~PeV shows some uncertainty, with differences of 1~–~3$\sigma$ between models.
It is also interesting to note that a similar hardening feature has been observed in the proton spectrum with a very similar change of spectral index of $\Delta\gamma \sim$ 0.2 at energies below 0.3 PeV. 
The feature seems not to manifest a strict rigidity dependence—which would imply a factor-of-two energy shift—between proton and helium spectra, although the energy at which the proton spectrum becomes harder is not yet well determined in experiments~\cite{LHAASO:2025byy}.~It is also notable that the hardness of the helium spectrum is not as high as that of the proton spectrum. This results in a significantly lower helium flux compared to protons at the `proton knee' ($\sim$ 3.3 PeV).
This directly demonstrates the dominance of protons at the all-particle knee. 

At the energy $E_k\sim$ 7~PeV, there is a broad softening in the spectrum, referred to as the `helium knee', with the spectral index decreasing by about 1.0. The spectral index and $E_k$ are slightly dependent on the hadronic interaction model assumption within 2$\sigma$. The significance of the knee-like structure is greater than 6$\sigma$. Nevertheless, due to the limited coverage of high energies above 7~PeV in the current analysis, this knee feature indicates a rigidity dependence within the error when compared with the proton spectrum~\cite{LHAASO:2025byy}. 

\textbf{H/He ratio---} 
Above the knee, the spectral indices of proton and helium spectra are very similar within 1$\sigma$ and the spectra are softer than the all-particle spectrum~\cite{LHAASO:2024knt}, as shown in Fig.~\ref{fig:other_spec}.~This implies that there is room to accommodate the rigidity-dependent knees of heavier components.~If the hardening of the proton spectrum at $\sim$0.1~PeV was a surprising feature found by LHAASO, the softening of the helium spectrum in the same energy range is somewhat `expected' since the all-particle spectrum has a significant simple power-law feature with the single index of $\sim-$2.74~\cite{LHAASO:2024knt}.~These features result in a very complex energy dependence of the H/He ratio, which transits unity twice in the range below the helium knee, as shown in Fig.~\ref{fig:other_spec}B. Note that all kinetic energy-rigidity conversions in this work are based on the assumption of a pure $^4$He sample.

This is a remarkable feature of the LHAASO measurements in contrast to the behavior at lower energies, i.e., the helium spectrum is softer than the proton spectrum in a wide energy range below the proton knee. This results in the H/He ratio increasing with energy from a minimum value of $\sim$0.7 at $E \sim 0.2$~PeV to a maximum value of $\sim$1.4. 
The crossing point at $E \sim 0.7$~PeV marks protons restoring as the dominant cosmic ray component, a surprise following the strong helium spectrum hardening observed in space-borne data.~Another crossing at $E \sim 5$~PeV is then observed, the helium population overtaking protons again. The reversal is interpreted as driven by the proton knee occurring at a lower energy than the helium knee.

\begin{figure}[htpb]
    \centering
    \includegraphics[width=1.0\linewidth]{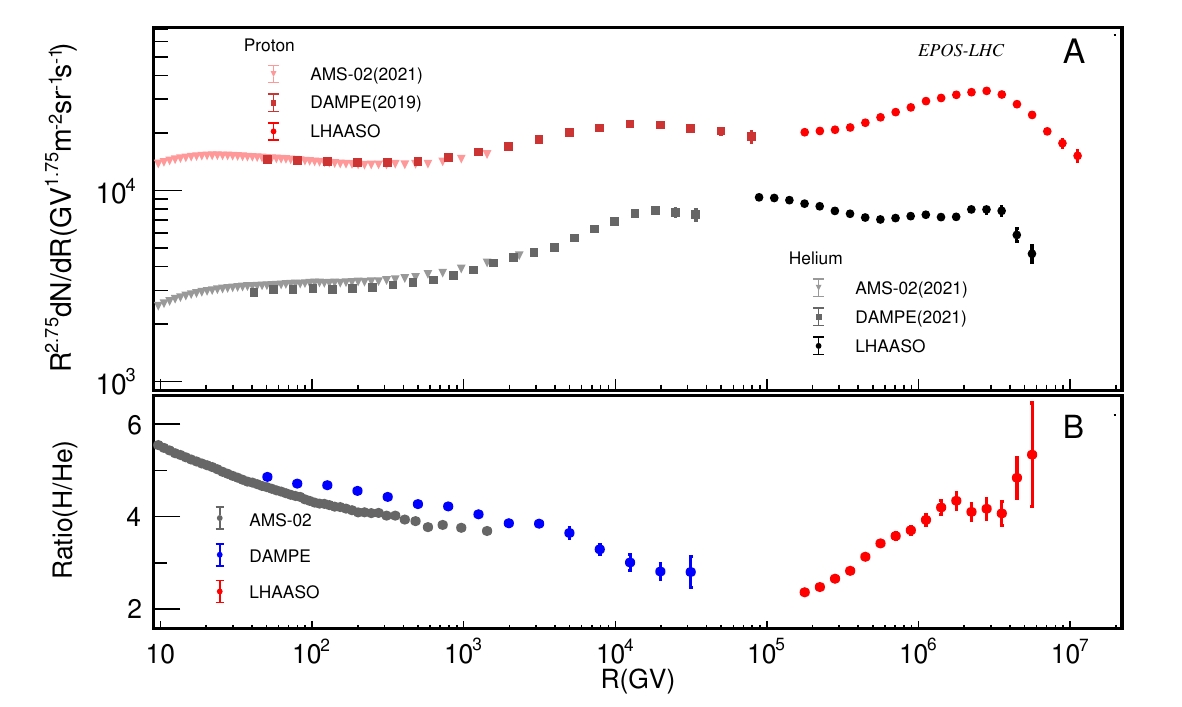}
     \caption{(A) 
     Rigidity spectra. Data are from LHAASO for helium (black circles) and protons (red circles), AMS-02~\cite{AMS:2021nhj} and DAMPE~\cite{Alemanno:2021gpb, DAMPE:2019gys}. The fluxes are scaled by $R^{2.75}$. Only statistical uncertainties are included.~(B) Proton to helium flux ratio $\Phi_{\rm H}(R)/\Phi_{\rm He}(R)$. Only statistical uncertainties are included. Data are from AMS-02~\cite{AMS:2021nhj} in gray, from DAMPE (helium data interpolated from Ref.~\cite{Alemanno:2021gpb}) in blue and from LHAASO in red.}
    \label{fig:other_spec_R}
\end{figure}

\textbf{Implications---} 
LHAASO measurements reveal distinct spectral shapes in proton and helium spectra, with at least three energy-ordered transitions in cosmic ray component dominance.~Such a complex energy dependence pattern variation is not expected based on the all-particle spectrum measurement alone, which manifests nearly a perfect smooth transition from one single-index power-law ($\gamma$$\sim-$2.74) to another ($-$3.13) over the same energy range. These findings challenge current theoretical frameworks about cosmic ray origins. Some relevant features are better discerned from the rigidity spectra of protons and helium nuclei, as shown in Fig.~\ref{fig:other_spec_R}.~For simplicity, only AMS-02, DAMPE and LHAASO data, which have the statistical error $<10\%$ for all points, are plotted. It is interesting to observe that the H/He ratio decreases monotonically over a wide range from 10 GV to an energy of approximately 100 TV, then increases monotonically up to a few PV. 

The most remarkable implication is that Galactic cosmic rays do not originate from a single source population. Other than the small hump at the lowest rigidity range $<$ (20$\sim$30) GV, in which particles are not sufficiently energetic and are modulated by solar activity, one observes two major humps in the rigidity ranges of (0.1, 100) TV and (0.1, 10) PV, respectively. They may represent different source populations.~Notably, the lower energy component is characterized by a clearly stronger enhancement in helium flux than that in proton flux over nearly the same range of rigidity, although both spectra are undergoing hardening. This may be interpreted in terms of sources with helium enriched with respect to protons. In contrast, the higher energy component has a much larger rate of increase for the proton content than that for helium over a similar rigidity range, thus indicating a group of sources dominated by proton enriched accelerators. This may be indicated by the two opposite slopes of the H/He ratio as a function of rigidity.

Most cosmic ray acceleration models predict that the proton and helium spectra generated by the same sources exhibit identical rigidity-dependent shapes~\cite{Gabici:2019jvz,Zatsepin:2006ci}. Additionally, Galactic propagation models also anticipate rigidity-dependent distortions of the source spectra. This indicates that the differing energy and rigidity dependence of the proton and helium spectra revealed by the LHAASO observations are linked to combinations of source classes that contribute dominantly to cosmic ray acceleration across different energy ranges. The different abundances of protons and helium nuclei in different classes of CR sources might result in the complex rigidity dependence of the spectral hardening features. On the other hand, softening of the spectra might indicate the maximal acceleration energy of the sources which have similar rigidity features. However, the interpretation of the spectra in terms of an astrophysical model that includes different classes of sources with different emission spectra and different injection rates for different nuclei remains a very open problem.
The extension of the measurements of the proton and helium spectra to both lower energy (to cover the gap with the space borne measurements) and to higher energy (to fully determine the knee features), and the study of the spectra of heavier nuclei will soon add very important information to develop an understanding of those issues.

\section*{Acknowledgments} \label{sec:acknowledgements} 
We would like to thank all staff members who work at the LHAASO site above 4400 meters above sea level year round to maintain the detector and keep the water recycling system, electricity power supply and other components of the experiment operating smoothly. We are grateful to the Chengdu Management Committee of Tianfu New Area for the constant financial support for research with LHAASO data. We appreciate the computing and data service support provided by the National High Energy Physics Data Center for the data analysis in this paper.
This research work is
supported by the following grants: The National Key R\&D program of China under the grant 2024YFA1611401, 2024YFA1611402, 2024YFA1611403, 2024YFA1611404, the National Natural Science Foundation of China No.12393851, No.12393852, No.12393853, No.12393854, NSFC No.12205314, No.12105301, No.12305120, No.12261160362, No.12105294, No.U1931201, No.12375107, NSFC No.12275280, No.12105293, No.11905240, No.12375106, No.12261141691, Innovation Project of IHEP No.E25451U2, the grant of Sichuan Science and Technology Department (No. 2024JDHJ0001), the Youth Innovation Promotion Association of the Chinese Academy of Sciences (CAS YIPA) (Grant No. 2023019), Institute of High Energy Physics with the grants of E45464U2. We are grateful to the Institute of Plateau Meteorology, CMA Chengdu to maintain meteorological data, and Thailand's National Science and Technology Development Agency (NSTDA) and National Research Council of Thailand (NRCT) under the High-Potential Research Team Grant Program (N42A650868). 
\\
 
\noindent {\bf \large Author Contributions}\\
L.P. Wang, L.L. Ma, and S.S. Zhang drafted the manuscript, with Z. Cao, spokesperson of LHAASO, leading the interpretation and revisions. S.S. Zhang led the data analysis team, including L.P. Wang and L.L. Ma, who analyzed the light component and helium energy spectra. S.H. Chen and Z.Y. You performed cross-checking. L.Q. Yin handled offline matching and reconstruction of LHAASO hybrid detector data. P. Lipari contributed theoretical insights. Other authors contributed to event reconstruction, simulation, detector calibration, and the operation and maintenance of scintillator counters, muon detectors, and Cherenkov telescopes, as well as the construction and deployment of the detectors.

\noindent {\bf \large Data availability}\\
The data that support the findings of this article are not publicly available upon publication because it is not technically feasible and/or the cost of preparing, depositing, and hosting the data would be prohibitive within the terms of this research project. The data are available from the authors upon reasonable request.

\bibliography{bib}

\clearpage

\onecolumngrid

\begin{center}
{\large\textbf{Supplemental Material for precise measurement of the cosmic ray helium spectrum above 0.1 PeV}}
\end{center}
\setcounter{figure}{0}
\renewcommand{\thefigure}{S\arabic{figure}}
\renewcommand{\thetable}{S\arabic{table}}
\setcounter{page}{1}
\pagenumbering{arabic}
\renewcommand{\thepage}{S\arabic{page}}
\renewcommand{\thefootnote}{\fnsymbol{footnote}}


\section{DATA}
\subsection{A Hybrid Observation of KM2A and WFCTA System}
The KM2A consists of two detector arrays: the electromagnetic particle detector (ED) array and the muon detector (MD) array. The ED array spans a total area of 1.3 km$^{2}$ and comprises 5,216 EDs, each with an active area of 1 m$^{2}$. The MD array includes 1,188 MDs, each with a detection area of 36 m$^{2}$.
This configuration yields a total muon-sensitive area of 42,768 m$^{2}$. EDs are deployed on a 15 m grid spacing, while MDs are spaced at 30 m intervals. This configuration allows KM2A to achieve precise measurements of the electromagnetic particle and muon content in air showers, with calibration uncertainties of less than 2\% \cite{LHAASO:2022lxa} and 0.5\% \cite{LHAASO:2015pei}, respectively.

The WFCTA comprises 18 telescopes operating in two modes: Cherenkov light detection and fluorescence detection. Each telescope features a 5~m$^{2}$ spherical mirror that focuses Cherenkov photons onto its focal plane at 2.87 m. The imaging camera employs a 32$ \times $32 array of silicon photomultipliers (SiPMs), totaling 1,024 pixels, each providing a 0.5$^\circ \times$ 0.5$^\circ$ field of view. All telescopes are tilted at 45$^{\circ}$ zenith angle with 0.02$^{\circ}$ pointing accuracy~\cite{LHAASO:2023plo}, and are azimuthally oriented to provide full 360$^{\circ}$ coverage. This configuration enables measurement of air showers from all azimuthal directions. The photon response of the SiPM camera is calibrated using LEDs of five different wavelengths, mounted at the center of the mirror, to a precision better than 2.6\%~\cite{LHAASO:2022tiv}, ensuring precise shower energy measurement.

Cosmic rays entering the atmosphere induce extensive air showers, producing secondary particles such as electrons, positrons, gamma photons, and muons. These secondary particles are detected by the ED array and MD array of KM2A. Additionally, the charged secondary particles in the shower emit Cherenkov light as they travel faster than the speed of light in air. This Cherenkov light is detected by the telescopes of WFCTA. 
KM2A and WFCTA simultaneously measure the secondary particles and Cherenkov photons generated in the shower.

Both WFCTA and KM2A use a White Rabbit timing system to synchronize the arrival times of cosmic ray events. By matching the trigger times of events from both arrays, the cosmic ray shower information measured by the two arrays is merged offline.

\subsection{Experimental data}
The dataset used in this paper for measuring the light component energy spectrum is the same as that used for the proton energy spectrum~\cite{LHAASO:2025byy}. It was collected by LHAASO from October 2021 to April 2022, with an effective exposure time of approximately $T \sim 900$ hours. This exposure time is consistent with the value adopted in the proton spectrum analysis.
All data meet the following criteria: 
(1) Data recorded by abnormally behaving detectors (i.e., unphysical rates) or data that result in abnormal data files (i.e., unphysical event reconstruction) are filtered out to ensure the reliability of the KM2A data~\cite{LHAASO:2024kbg}. (2) To reduce the influence of background light on the detection shower energy threshold of the telescope, the angle between the moon and the main axis of the telescope is greater than 20$^{\circ}$. Additionally, the ADC count of the night sky background measured by the WFCTA telescope should be less than 200. (3) The infrared brightness temperature ($T_{b}$) recorded by an infrared cloud instrument mounted on LHAASO site is less than $-$65$^{\circ}$C, indicating clear weather.

\subsection{MC simulation\label{sec:MC}} 
Monte Carlo (MC) simulations are generated using the CORSIKA package version 7.7420~\cite{Heck:1998vt}. To simulate the combined observation of KM2A and WFCTA, the Cherenkov option is turned on, and both the Cherenkov information and secondary particle information of each air shower are recorded. The high-energy hadronic interaction model EPOS-LHC~\cite{Pierog:2013ria} is chosen as the baseline event generator, whereas QGSJET-II-04~\cite{Ostapchenko:2013pia} and SIBYLL 2.3d~\cite{Riehn:2019jet} are chosen for model validation.
For secondary particles with energy lower than 80~GeV, the low-energy interaction model FLUKA~\cite{Battistoni:2015epi} is used.

\begin{figure}[h]
    \centering    \includegraphics[width=0.6\linewidth]{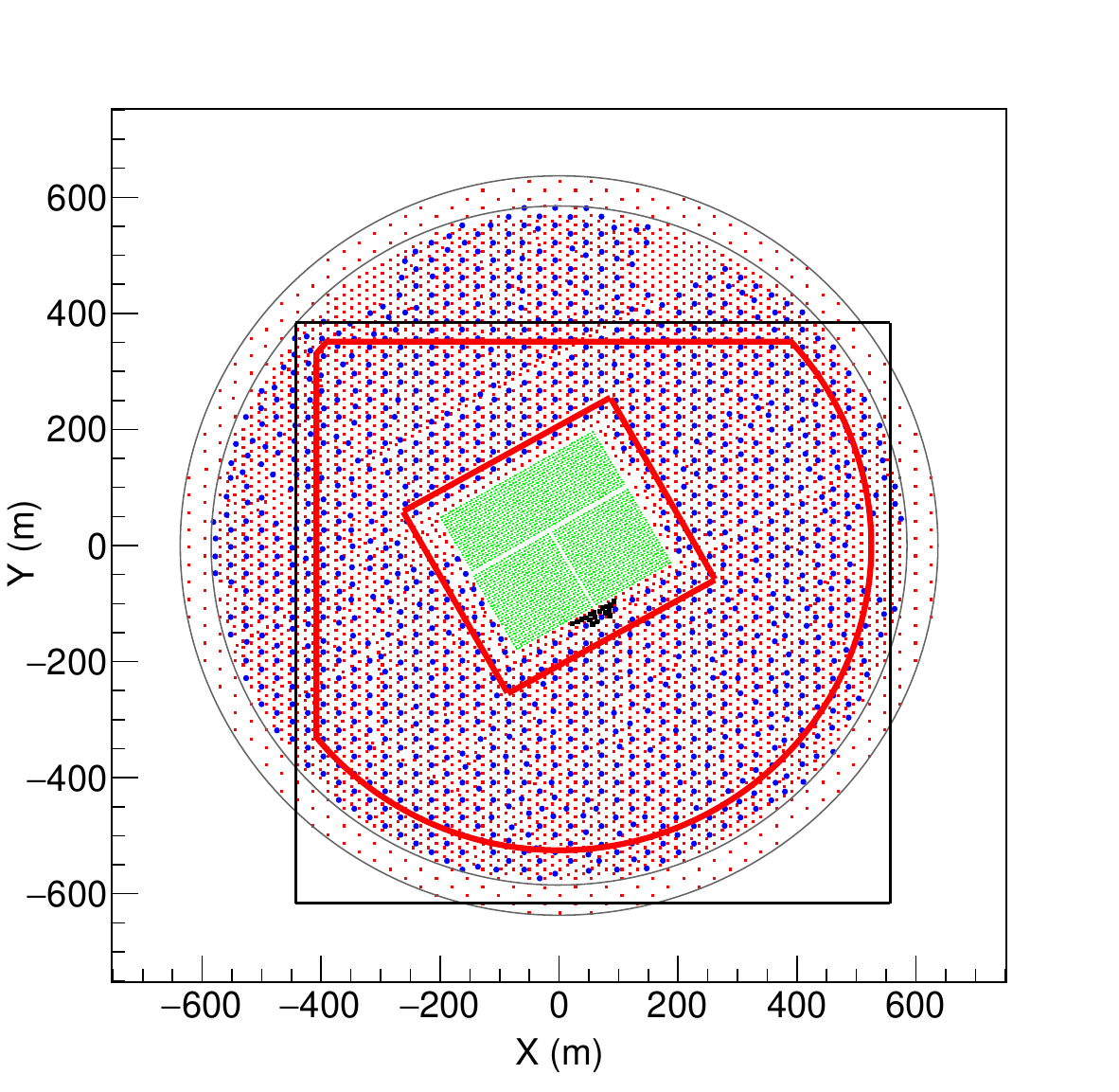}
    \caption{The red squares and blue dots indicate the operational electromagnetic detectors~(EDs) and muon detectors~(MDs) of KM2A, respectively. The black dots represent the telescopes of WFCTA. The central green squares delineate the region of the WCDA. The red solid lines delineate the area for selecting the reconstructed shower core. 
    The area enclosed by the black lines represents the throwing area for the shower core in the simulation. This picture is from~Ref.~\cite{LHAASO:2025byy}.  }
    \label{fig:lhaaso}
\end{figure}

Five mass compositions of proton, helium, nitrogen~(CNO group), aluminum~(MgAlSi group), and iron are generated in four energy ranges 0.01~–~40~PeV, following a power-law function with a spectral index of $-$1 to increase the statistics of high-energy events. Proton and helium represent the light component of cosmic rays, while the CNO group, MgAlSi group, and iron represent the heavy component. 
Each EAS is reused 20 times with a random core location in a square area of side length 1000~m, centered at WFCTA, as indicated by the area enclosed by the black line in Fig.~\ref{fig:lhaaso}.
Since the WFCTA telescopes are oriented at 45$^{\circ}$ in the zenith direction and have a field of view of 16$^{\circ} \times$ 16$^{\circ}$, the zenith angle is sampled in the range of 35$^{\circ}$$-$55$^{\circ}$ and the azimuth angle is sampled in the range of 0$^{\circ}$$-$360$^{\circ}$ to cover the field of view of the 18 telescopes.
The total numbers of simulated events for the QGSJET-II-04, EPOS-LHC, and SIBYLL 2.3d interaction models are 3.2 $\times $ 10$^7$, 3.2$ \times $10$^7$, and 2.9$ \times $10$^7$, respectively. Combined simulation for KM2A and WFCTA is carried out.
In WFCTA simulation, the attenuation of Cherenkov light caused by Rayleigh scattering and ozone absorption is considered based on the American standard atmosphere model. 
Given the atmospheric pressure at the LHAASO site and its seasonal variations, we evaluated that the Rayleigh scattering attenuation of Cherenkov photons differs by less than 0.1\% compared to the American standard atmosphere model. 
During data analysis, the absorption of Cherenkov photons by aerosols (mainly due to water vapor) was calibrated accordingly\cite{LHAASO:2025byy}. 
The KM2A detector response is simulated using G4KM2A~\cite{Chen:2020ddg,LHAASO:2024rtc}, which is based on GEANT4~\cite{GEANT4:2002zbu}.

\subsection{Criteria for Selecting High-Quality Measurement Events}
To ensure high-quality shower reconstruction, the same selection criteria as those used in the proton spectrum analysis~\cite{LHAASO:2025byy} are applied to both simulations and data in this work. These criteria are summarized below: 

1) To avoid edge effects, air showers are selected based on reconstructed cores located within the area bounded by the red lines, as illustrated in Fig.~\ref{fig:lhaaso}. The red lines delineate a region where reconstructed cores are more than 50~m from the edge of the WCDA and KM2A, and over 30~m from the boundary of the simulated events.\\
2) To minimize the effects of SiPM saturation and to ensure the full detection efficiency of the telescope, only events with a perpendicular distance from the telescope to the shower axis ($R_{p}$) within 180~–~310~m are included in the analysis. \\
3) To ensure the integrity of Cherenkov images, the center of gravity ($X_I$, $Y_I$) must satisfy the conditions $|X_I| < 5^{\circ} $ and $|Y_I| < 5^{\circ} $.\\
4) To prevent leakage of the Cherenkov image, the number of photons counted from the outermost row of SiPMs must be less than 3\% of the total number of photons counted.\\
5) To ensure high-quality reconstruction of the shower core and arrival directions, the number of fired EDs must exceed 20.\\
6) To ensure the high-quality measurement of the Cherenkov image, the number of fired pixels in the cleaned Cherenkov image must be greater than 5.\\

\subsection{Effective Aperture}
\begin{figure}[h]
    \centering
    \includegraphics[width=0.56\linewidth]{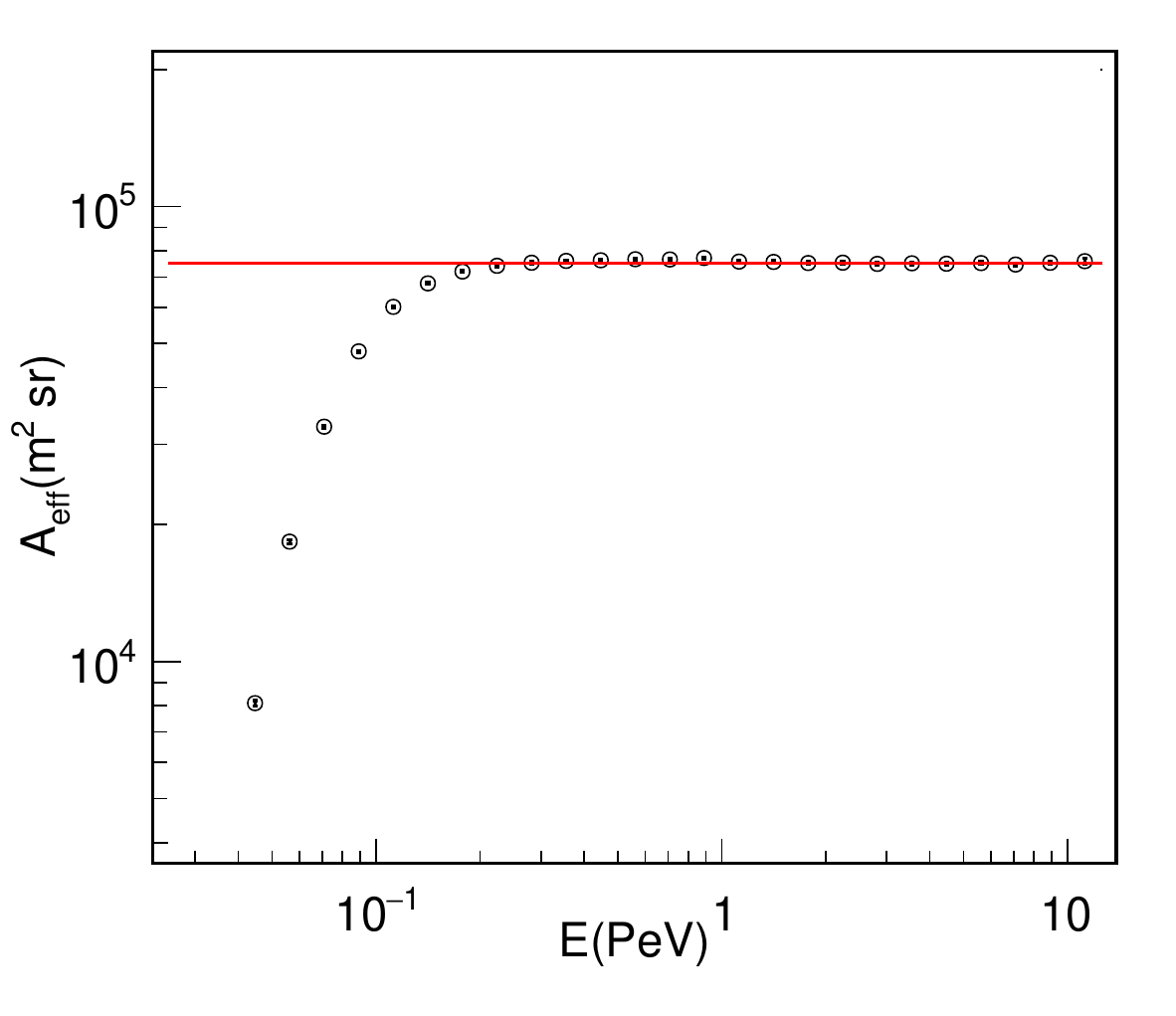}
    \caption{The effective aperture versus the energy of light component. The error bars indicate the statistical uncertainties. The red line represents an aperture of 75,000 m$^2$sr.}
    \label{fig:aperture}
\end{figure}

The effective aperture of light component is defined as the product of the geometric factor and selection efficiency. The effective aperture is calculated as follows:
\begin{equation}
    A_{\text{eff}}(E) = A_{\text{gen}} \times \frac{N_{\text{select}}(E)}{N_{\text{thrown}}(E)},
    \label{for:Aperture}
\end{equation}
where $A_{\text{gen}} = S\int^{\theta_1}_{\theta_2}\sin\theta\cos\theta d\theta\int^{2\pi}_{0}d\phi $ represents the geometrical factor of the MC generation aperture, $S$ denotes the total area at the ground level where the cores of the MC events are thrown, and the zenith angles of the thrown events are distributed from $\theta_{1}=35^{\circ}$ to $\theta_{2}=55^{\circ}$. 
$N_{\text{thrown}}(E)$ and $N_{\text{select}}(E)$ represent the numbers of generated events and the events that pass the event selections from the MC simulation, respectively. 
The effective aperture of hybrid observation, $A_{\text{eff}}(E) \sim$ 75,000 $\text{m}^{2}$sr, is shown in Fig.~\ref{fig:aperture}.

\section{Energy reconstruction}
\begin{figure}[h]
   \centering
     \includegraphics[width=0.5\linewidth]{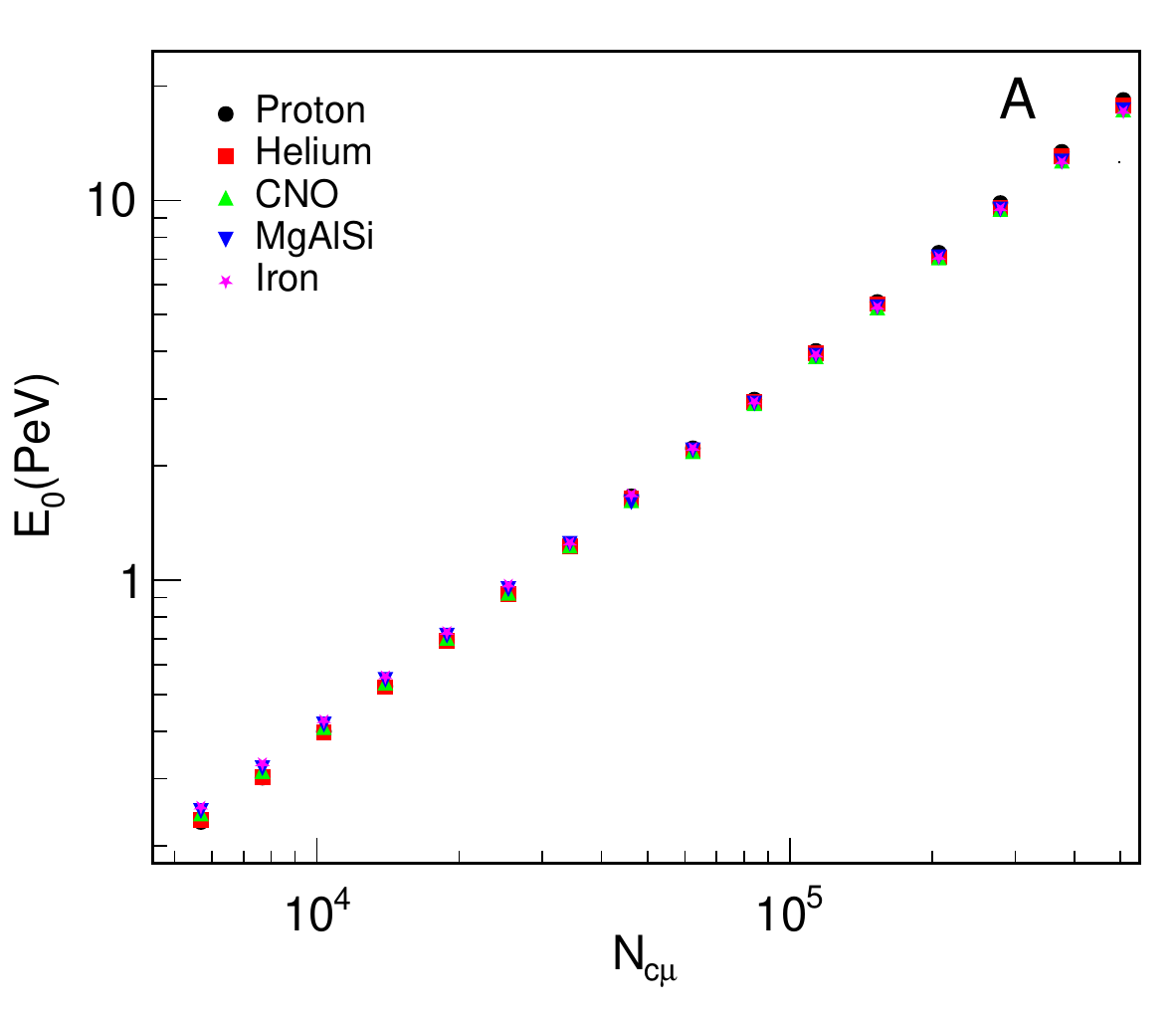}
     \centering\includegraphics[width=0.45\linewidth]{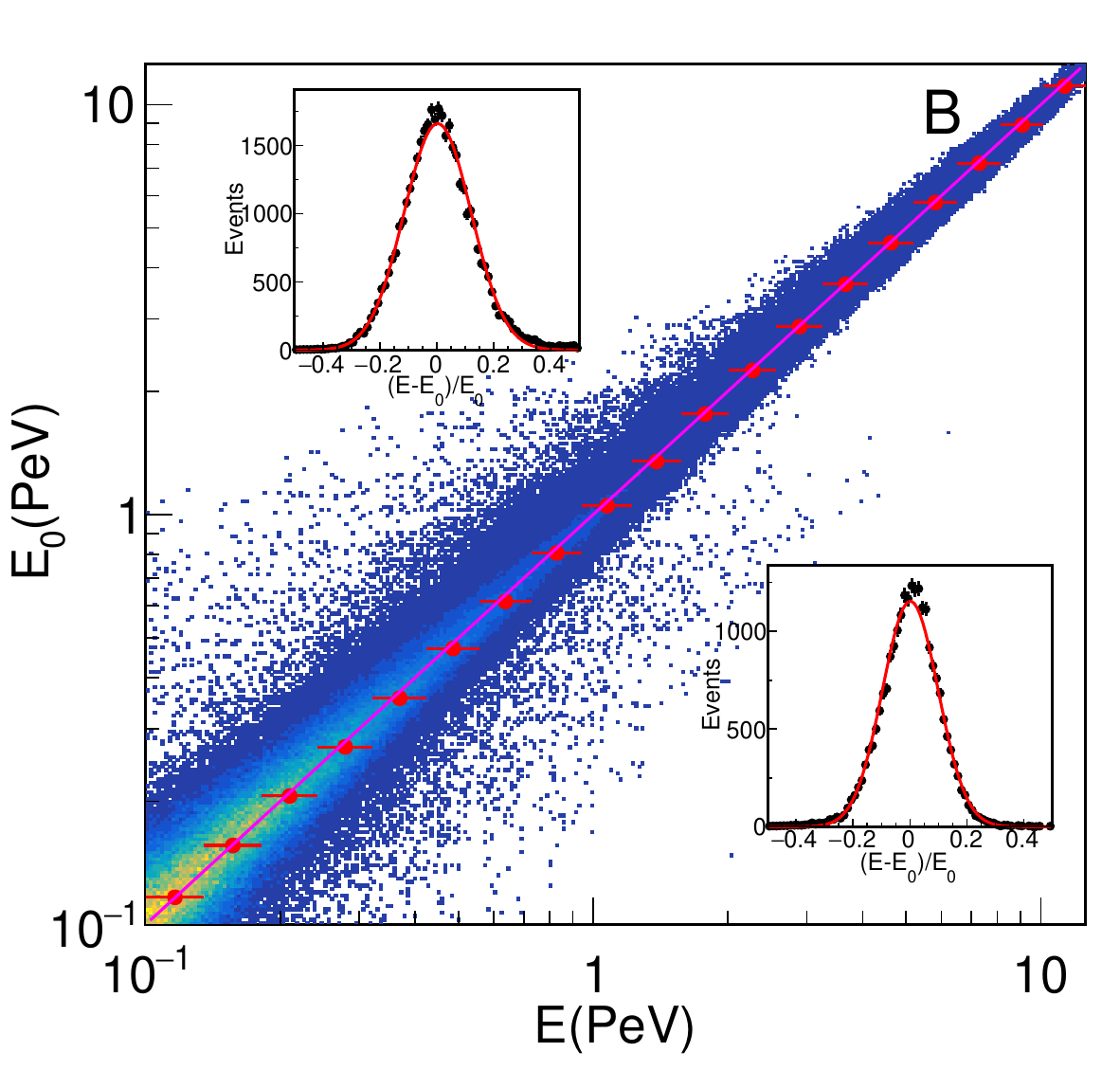}
    \caption{(A): The primary energy~($E_{0}$) versus the energy estimator~($N_{c\mu}$) for five mass components of the air shower. (B): Correlation between primary energy ($E_{0}$) and reconstructed energy ($E$) for light component from EPOS-LHC hadronic interaction model. The red points represent the mean values of the primary energy and reconstructed energy. The pink line represents the function $y=x$. Insets show relative energy deviation distributions at $\log_{10}(E/\text{PeV}) \in [-0.5, -0.4]$ (top-left) and $[0.0, 0.1]$ (bottom-right), with Gaussian fits (red lines).}
    \label{fig:RECE_}
\end{figure}

Extensive air showers initiated by nuclei such as protons and helium produce two main classes of secondary particles: an electromagnetic component dominated by $e^{\pm}$ and gamma rays, and a hadronic component, dominated by pions and kaons. Muons from $\pi^{\pm}$ or $k^{\pm}$ decay are measured by the KM2A muon detector array, providing the muon content representing the hadronic component. High-energy charged particles (primarily $e^{\pm}$) emit Cherenkov light when their speed exceeds the speed of light in air; the WFCTA telescopes measure these Cherenkov photons, representing the electromagnetic component. Combining Cherenkov photons and muon information significantly improves the energy reconstruction precision while reducing its dependence on the primary composition. This energy reconstruction method was first described in Ref.~\cite{Wang:2023qeg}.
The energy reconstruction method presented in the proton energy spectrum paper~\cite{LHAASO:2025byy} further demonstrates that the energy reconstruction estimator $N_{c\mu} = N_{ph}^{250} + 3.0 N_{\mu}$ shows a strong correlation with the primary energy and exhibits minimal dependence on composition, as shown in Fig.~\ref{fig:RECE_}. Here, $N_{ph}^{250}$ denotes the total number of Cherenkov photoelectrons in the image normalized to a reference distance of $R_{p} = 250$ m, and $N_{\mu}$ represents the number of muons within 40 to 200 m of the shower core. To ensure consistency of the energy scale for protons and helium nuclei in this study, the same energy reconstruction formula used in the proton study was applied in the present measurement:
\begin{equation}
    \log_{10}(E/\text{PeV})=p_{0}+p_{1}\log_{10}(N_{c\mu})+p_{2}\log_{10}^{2}(N_{c\mu}),
\end{equation}
where $p_{0}$, $p_{1}$ and $p_{2}$ are the fitting parameters, which are the same as those in the proton energy spectrum~\cite{LHAASO:2025byy}. 

\begin{figure}[h]
  \centering\includegraphics[width=0.45\linewidth]{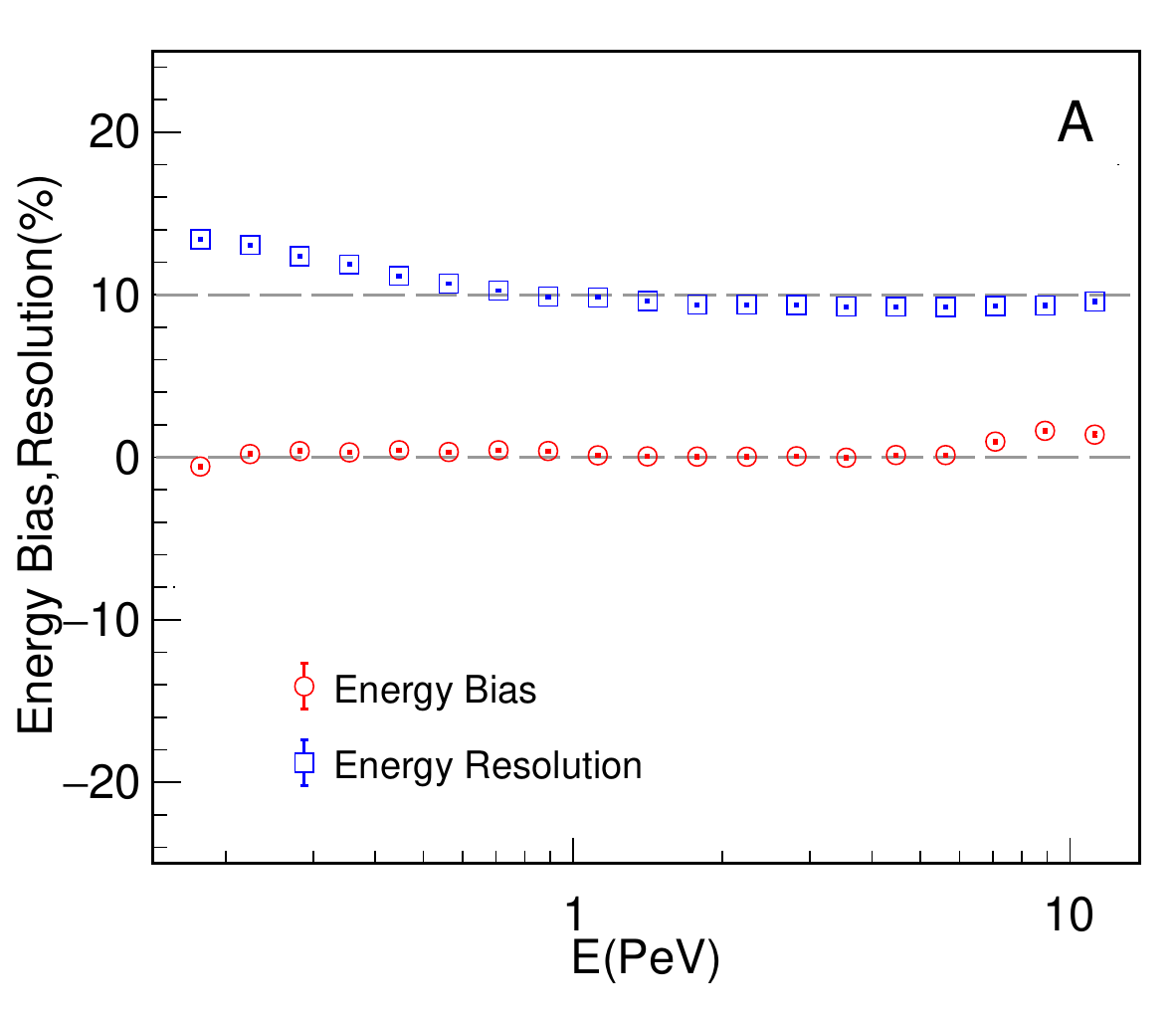}
  \centering\includegraphics[width=0.45\linewidth]{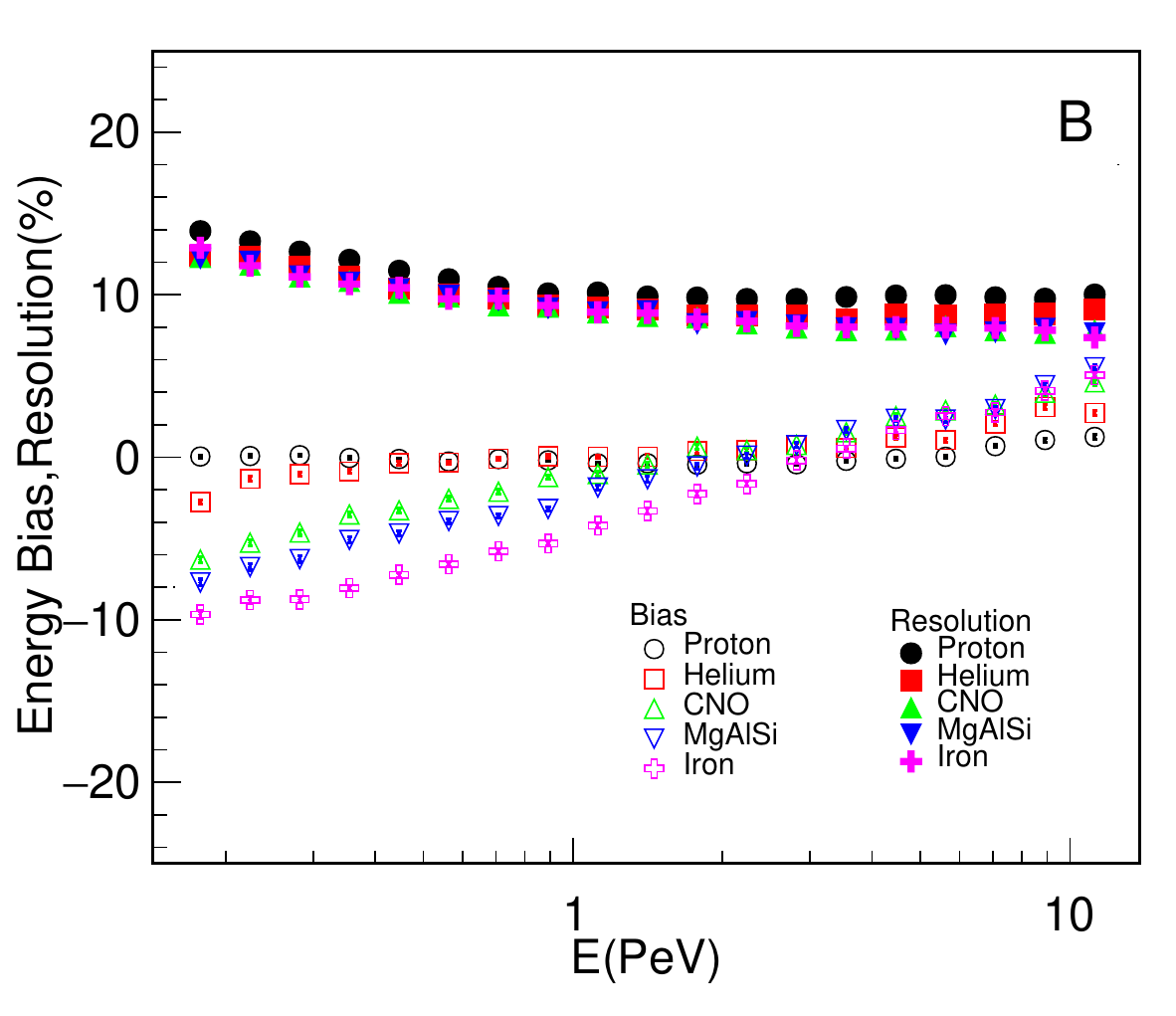}
    \caption{(A): Energy resolution (blue squares) and bias (red circles) versus energy for the light component of the air showers. (B): Energy resolution (solid symbols) and bias (open symbols) versus energy for the five components of the air showers.}
  \label{fig:RECE}
\end{figure}

Figure~\ref{fig:RECE} shows the energy reconstruction resolution and bias for the light component and individual cosmic ray species based on this reconstruction formula. The energy bias for the light component remains about 1.7\%, with biases of about 1\% for protons, about 2\% for helium nuclei, and about 8\% for CNO nuclei. Additionally, the difference in energy bias between protons and helium nuclei is about 2\%. 
The energy resolution for protons improves from approximately 15\% at around 150 TeV to approximately 10\% at around 900 TeV, and remains stable with increasing energy. This stability is beneficial for the accurate measurement of fine structures in the energy spectrum. The energy resolution for helium nuclei is slightly better by 1\% compared to protons and exhibits similar characteristics with increasing energy. The light component, a mixture of protons and helium nuclei, has energy resolution and bias values between those of protons and helium nuclei.

\section{Light component events selection and energy spectrum}
\subsection{Light component events selection}
The electromagnetic and muonic components in EAS are related to the primary energy and atomic number of the initiating particle. The superposition model~\cite{Horandel:2006jd} can be used to extend the simple approach from primary protons to particles with atomic number $A$. A nucleus with atomic number $A$ and energy $E_0$ is treated as $A$ independent nucleons, each with energy $E_0/A$. The number of muons is roughly proportional to $A^{0.1}$, and the maximum depth for iron-induced showers is approximately 150 g/cm$^2$ higher in the atmosphere compared to proton-induced showers~\cite{Horandel:2006jd}. For a given energy, the numbers of both electromagnetic particles ($N_e$) and muons ($N_{\mu}$) in a shower are sensitive to the type of primary particle. Heavier nuclei produce more muons and fewer electromagnetic particles.

\begin{figure}[h]
    \centering
    \includegraphics[width=0.5\linewidth]{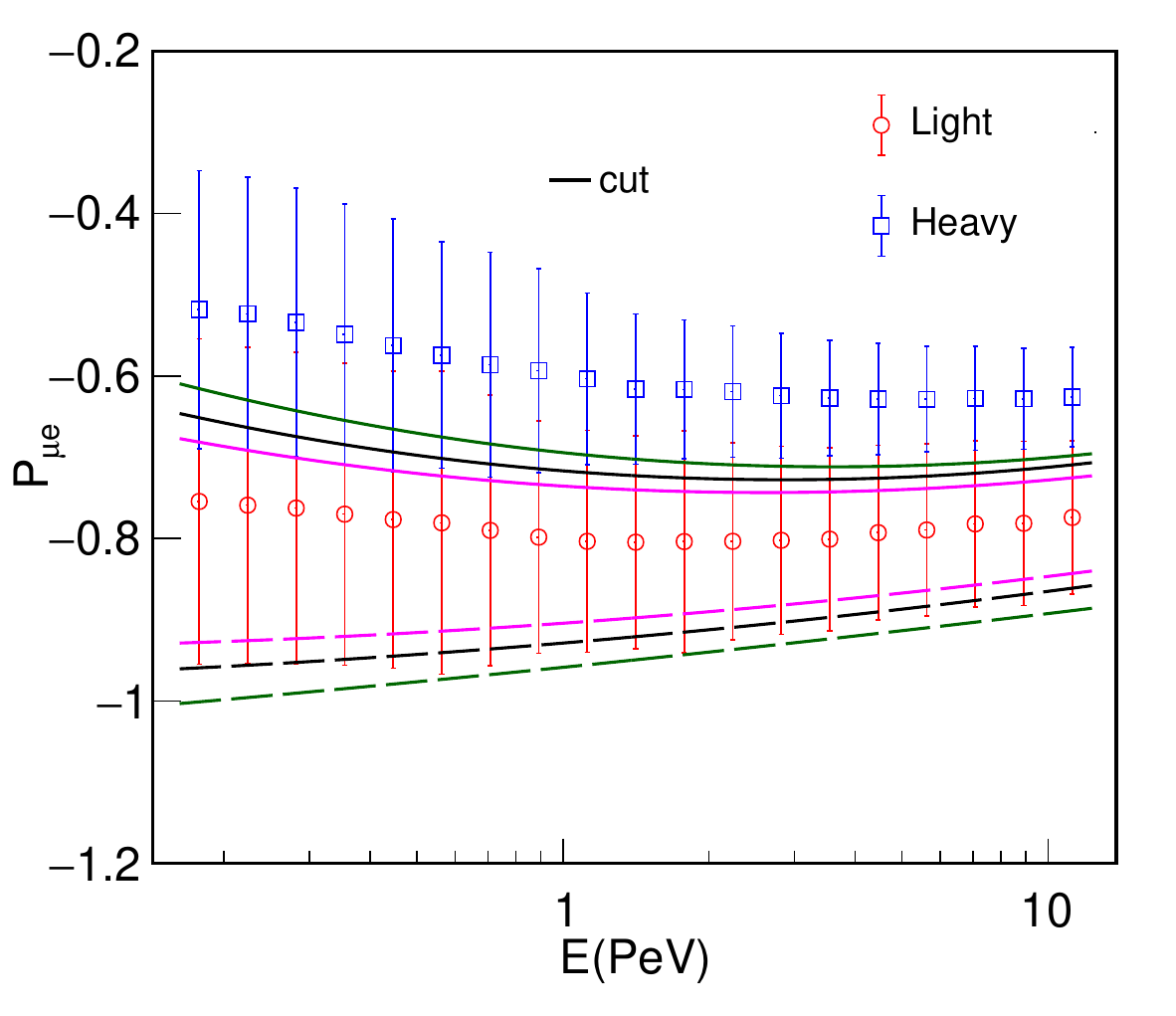}
    \caption{The distribution of average $P_{\mu e}$ of light component (red circles) and heavy component (blue squares) versus the energy of cosmic rays. The selected light-like events fall within the region bounded by the solid and dashed lines. The dotted and solid lines represent the selection cuts, with the pink, black, and green lines indicating selection efficiencies of 50\%, 60\%, and 70\% for the light component, respectively.}
    \label{fig:cutline}
\end{figure}

KM2A's combination of electromagnetic particle detectors (EDs)~\cite{LHAASO:2022lxa} and muon detectors (MDs)~\cite{LHAASO:2015pei} enables simultaneous measurement of these particle information, which is essential for composition discrimination. In this study, we define a composition-sensitive parameter, $P_{\mu e}$, as follows: 
\begin{equation}
    P_{\mu e}=\log_{10}(\frac{N_{\mu}}{N_{e}^{0.82}}),
    \label{for_Nue}
\end{equation} 
where the $N_{\mu}$ and $N_{e}$ are the number of muons and electromagnetic particles, respectively, within a ring from 40~m to 200~m with respect to the shower core. 
Figure~\ref{fig:cutline} illustrates the relation between $P_{\mu e}$ and energy for both light and heavy components of cosmic rays, revealing a weak correlation for the light component within this range. 

The energy spectrum of the light component is a mixture of protons and helium, with the ratio of protons to helium significantly influencing this spectrum. Thus, it is crucial to maintain the ratio of protons to helium~(H/He) unchanged. 
To maintain the H/He unchanged after the component selection, it is essential to keep the same selection efficiencies for the two species. Therefore, two cuts of the $P_{\mu e}$ are implemented in this work.
The first cut excludes events with higher $P_{\mu e}$ values to minimize contamination from heavy component events, as illustrated by the solid black line in Fig.~\ref{fig:cutline}. However, the ratio of protons to helium in the selected samples differs from that in the original cosmic rays due to the varying selection efficiencies for the two species (with the efficiency for helium at approximately 65\%, while that for protons exceeds 80\%). An additional cut, represented by the dotted line, is applied to reject events with lower $P_{\mu e}$ values, ensuring that the selection efficiencies for both protons and helium nuclei remain consistent, thereby keeping the H/He unchanged.

\begin{figure}[h]
    \centering
    \includegraphics[width=0.4\linewidth]{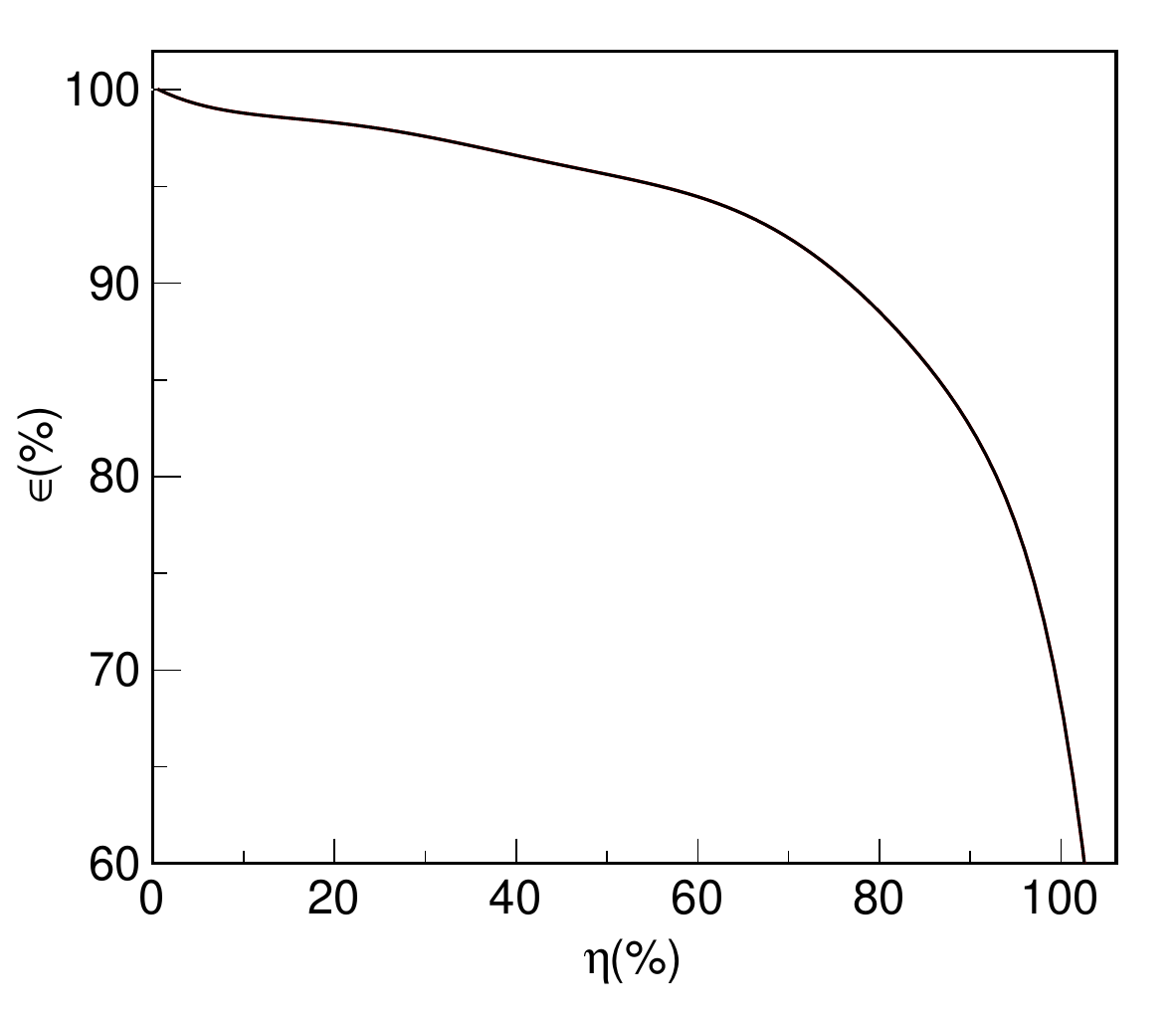}
    \caption{ 
    \textcolor{black}{Purity ($\epsilon$) as a function of selection efficiency ($\eta$) for the light component in the energy range of $\log_{10}(E/\text{PeV}) \in [0.2, 0.4]$ is assessed under different light event selection criteria.}
    }
    \label{fig:ROC}
\end{figure}

To clearly illustrate how the purity and selection efficiency of the light component vary with the threshold of the selection variable, this study employs the Receiver Operating Characteristic (ROC) curve. The ROC curve is one of the standard metrics in Toolkit for Multivariate Data Analysis (TMVA)~\cite{TMVA:2007ngy} for evaluating classifier performance, illustrating the relationship between the light component purity (vertical axis) and selection efficiency (horizontal axis) under different cut thresholds, as shown by the red curve in Fig.~\ref{fig:ROC}. Based on this curve, the final selection criteria were determined.
The selection results are summarized in panel B of Fig.~\ref{fig:divide}.
The purity ($\epsilon$) of the light component, defined as $\epsilon = N_L/N_S$, exceeds 80\% when the selection efficiency ($\eta = N_L/N_{L}^{0}$) is approximately 60\%. Here, $N_{S}$ represents the light-like component events with contamination from heavy component after composition selection, $N_{L}$ represents the light component events without heavy component contamination after composition selection, and $N_{L}^{0}$ represents the light component events before composition selection.

\subsection{Light component energy spectrum\label{sec:supp_light}}

\begin{figure}[h]
    \centering
    \includegraphics[width=0.6\linewidth]{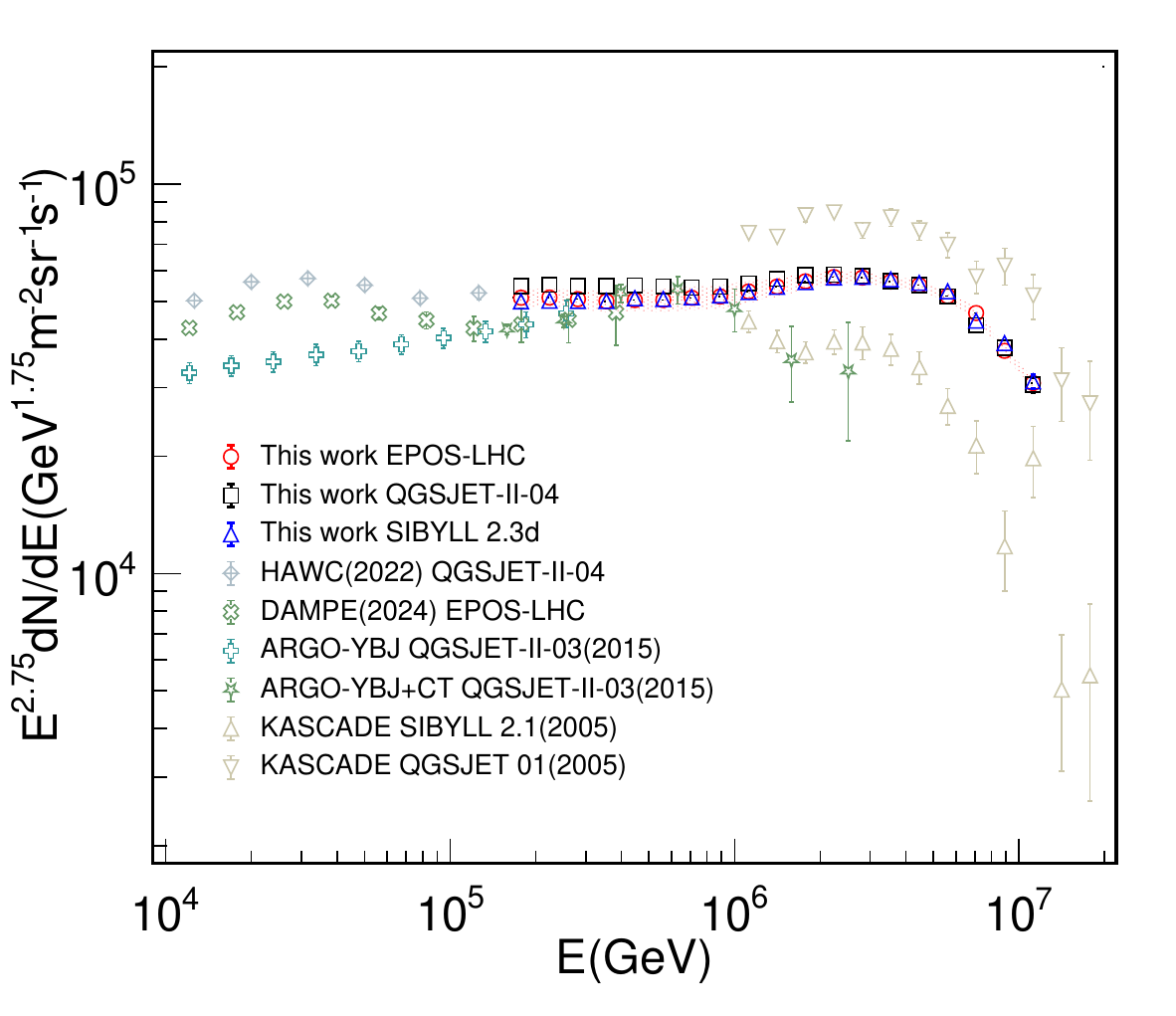}
    \caption{The differential energy spectra of light component obtained by the present work using three hadronic interaction models. The results are compared with DAMPE~\cite{DAMPE:2023pjt}, HAWC~\cite{HAWC:2022zma}, ARGO-YBJ~\cite{ARGO-YBJ:2015yal}, ARGO-YBJ+CT~\cite{ARGO-YBJ:2015isx} and KASCADE~\cite{KASCADE:2005ynk}.
    }
    \label{fig:hadronic}
\end{figure}

The energy spectrum of the light component is calculated as:
\begin{equation}
    \Phi_{\rm L}(E) = \frac{\Delta N_{S}(E)}{\Delta E \times A_{eff}(E) \times T} \times \frac{\epsilon}{\eta},
    \label{for_spe}
\end{equation}
where $\Delta N_{S}(E)$ represents the number of selected light-like events within the energy bin $\Delta E$.
$T \sim 900$ h and $A_{\text{eff}}(E) \sim$ 75,000 m$^{2}$sr denote the effective exposure time and effective aperture, respectively.
Additionally, $\epsilon$ and $\eta$ denote the purity and selection efficiency for the light component, respectively. 
Figure~\ref{fig:hadronic} shows the light component energy spectrum (scaled by $E^{2.75}$) in the range 0.16~–~13 PeV, as derived using different hadronic interaction models, and compares it with results from other experiments. The light component flux and associated errors are tabulated in Table~\ref{tab:light}.

\begin{figure}[h]
    \centering
    \includegraphics[width=0.6\linewidth]{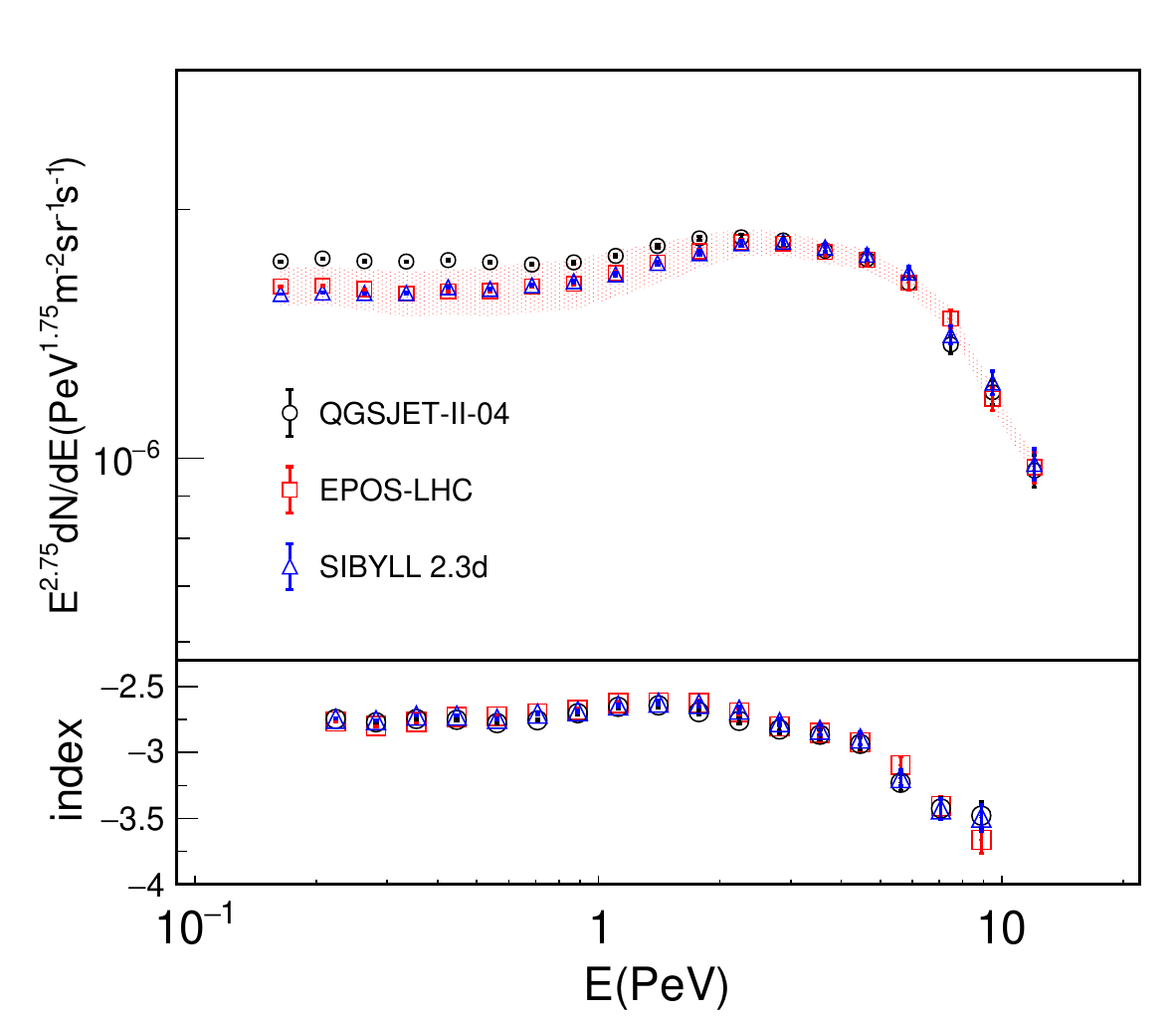}
    \caption{Top panel: Energy spectra of light component scaled by $E^{2.75}$ versus energy. Different markers indicate results from different hadronic interaction models, with error bars representing statistical uncertainties. The shaded region represents the total systematic uncertainty (excluding that from hadronic interaction models), which is plotted on the light component spectrum based on the EPOS-LHC model. Bottom panel: Spectral indices of the light component energy spectra derived from different hadronic interaction models. Each index was fitted using a single power-law function with three adjacent points. Error bars show fitting uncertainties.}
    \label{fig:light_had}
\end{figure}

\subsection{Systematic uncertainty analysis\label{sec:light_sys}}
Since the light component and proton spectra use the same energy reconstruction formula, their energy scale systematic uncertainties are the same. The total systematic uncertainty on the energy measurement is about 4\%. The main contributors are $\sim$1.5\% from SiPM camera calibration, $\sim$1\% from mirror reflectivity calibration, $\sim$1\% from $N_{\mu}$ calibration, $\sim$1\% from absolute humidity absorption correction, $\sim$2\% from aerosol absorption estimation, $\sim$0.5\% from air pressure correction, and $\sim$1.7\% from hadronic interaction models. Further details can be found in the Supplementary Materials of the proton energy spectrum~\cite{LHAASO:2025byy}.

For the light component energy spectrum flux, the dominant systematic uncertainties arise from composition models, component selection, atmospheric pressure, absolute humidity, night-sky background light, and hadronic interaction models. 
These system uncertainties are summarized in Table~\ref{table:systematicUncertainty} and are introduced and quantified below.

\noindent{\bf Systematic uncertainty on component selection:} 
To assess the impact of selection cuts on the light component energy spectrum,  we varied the cut threshold to achieve selection efficiencies of 50\% and 70\%(Fig.~\ref{fig:cutline}). Relative to the 60\% efficiency case, the resulting light component flux varied by less than 1\%.

\begin{figure}[h]
    \centering  \includegraphics[width=0.45\linewidth]{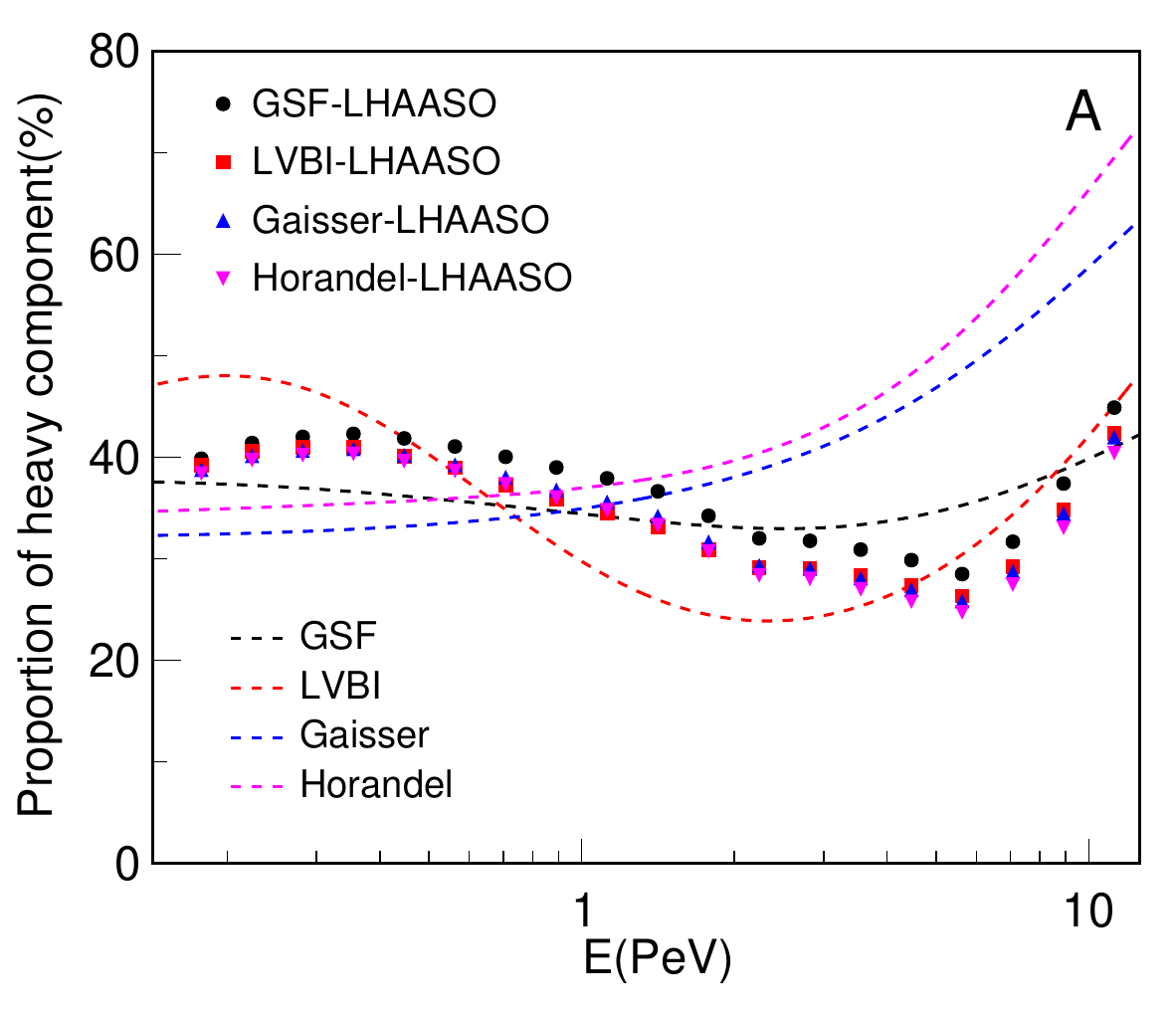}
    \includegraphics[width=0.45\linewidth]{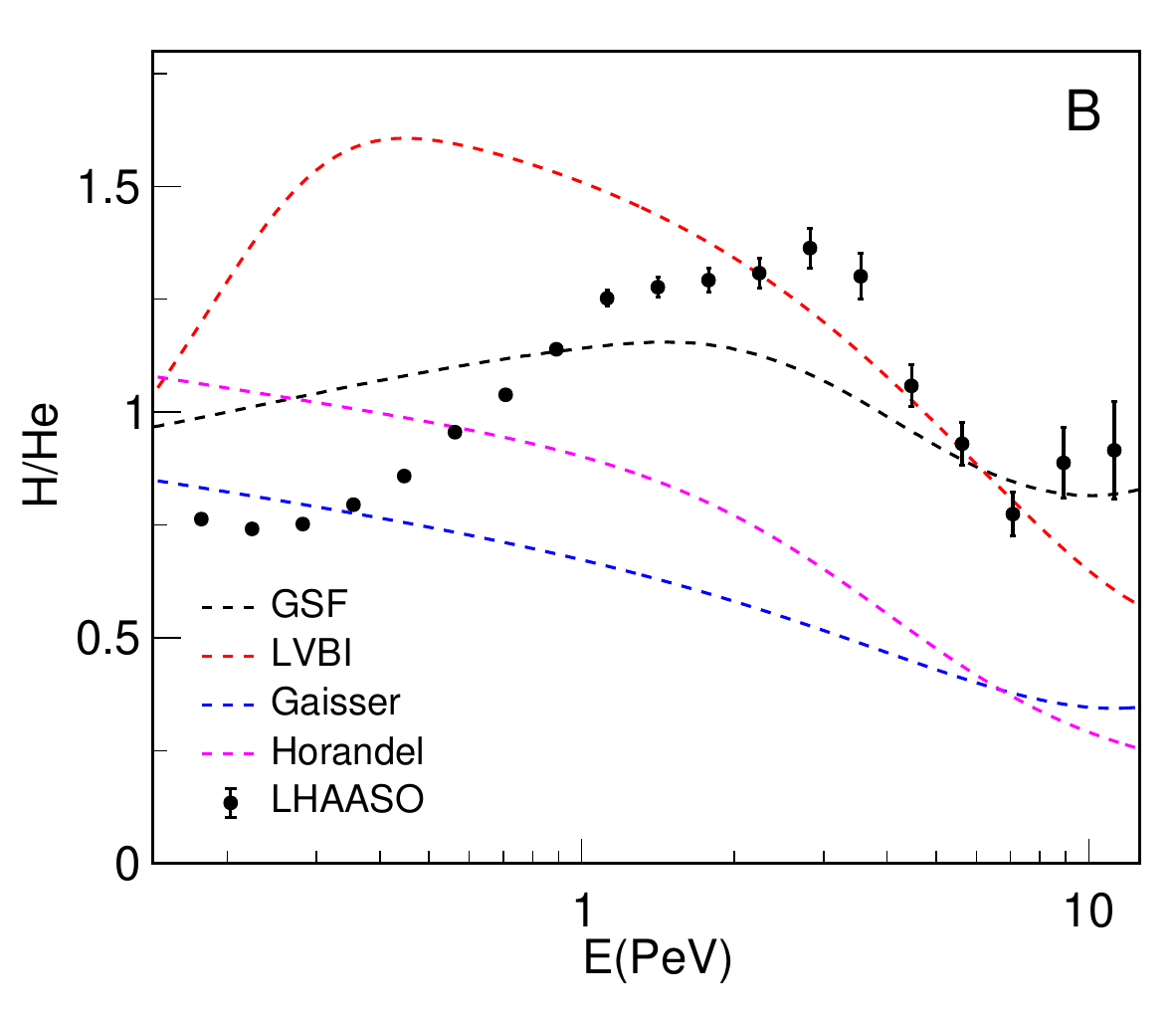}
    \caption{(A): The proportion of heavy component of cosmic rays in the 0.16~–~13~PeV range with the different composition models used in this work. The solid points are derived from the composition models modified by LHAASO primary measurements. (B): The proton to helium ratio (H/He) versus the energy across different cosmic ray composition models. The solid points are derived from LHAASO primary measurements.}
    \label{fig:fit_model}
\end{figure}

\noindent{\bf Systematic uncertainty on cosmic rays composition models:} 
Different composition models predict varying proportions for the light component (proton, helium) and heavy component (CNO, MgAlSi, iron) (see Fig.~\ref{fig:fit_model}). Under a fixed selection efficiency, the purity of the light component depends on the assumed composition model. Consequently, the light component spectrum is also composition model dependent. The relationship between the purity and the light component spectrum is introduced in Section ~\ref{sec:supp_light}.

We evaluated the systematic uncertainty introduced by the GSF~\cite{Dembinski:2017zsh}, Gaisser~\cite{Gaisser:2013bla}, Hörandel~\cite{Hoerandel:2002yg}, and LVBI~\cite{Lv:2024wrs} composition models for the light component spectrum and found a variation of about 10\%.
Incorporating LHAASO measurements of the proton and helium spectra, we updated these four models (detailed in Section~\ref{sec:new_model}). With these updates, the maximum discrepancy in the proportion of the heavy component decreased from $\sim$20\% to $\sim$5\%. Despite the updates, the relative proportions among the CNO, MgAlSi, and iron groups are unchanged in each model. The maximum discrepancy in the CNO proportion remains about $\sim$14\%, which is the dominant source of contamination in the light component. Overall, inter-model consistency is improved, and the systematic uncertainty induced by the composition model on the light component spectrum is reduced to less than 6\%.

\noindent{\bf Systematic uncertainty on atmosphere pressure:}
During the observation period, the atmospheric pressure at the LHAASO site varied by about 14 hPa, corresponding to a change in atmospheric depth of $\sim$14 g/cm$^{2}$. In the proton energy spectrum paper~\cite{LHAASO:2025byy}, the dependence of $N_{\mu}$ and $N_e$ on atmospheric pressure is described in detail, and the constant-intensity-cut (CIC) method is used to correct for pressure variations. Here, $N_{\mu}$ and $N_e$ are corrected to P = 586.6 hPa (consistent with the simulation setup). After correction, the residual impact of pressure variations on $N_{\mu}$ and $N_e$ is less than 1\%.

To assess residual pressure effects on the light component energy spectrum flux, we split the LHAASO data into two subsets: $>$595 hPa and $<$595 hPa. The difference in energy spectra between these two samples and the overall dataset is found to be less than 2\%.

\noindent{\bf Systematic uncertainty on night-sky background light:}
This analysis utilizes LHAASO data collected during periods when the moon was outside the telescope's direct field of view. We still need to consider the potential influence of scattered moonlight on the observational data. To evaluate the effect of scattered moonlight on our measurements, we categorized the data into two subsets: nights with moonlight and nights without moonlight. The difference in energy spectra between these two samples and the overall dataset
is found to be less than 2\%.

\noindent{\bf Systematic uncertainty on absolute humidity:}
In the proton energy spectrum paper~\cite{LHAASO:2025byy}, we discuss in detail the attenuation of Cherenkov photons during atmospheric propagation and identify water vapor as the dominant absorber. At the LHAASO site, the Aerosol Optical Depth (AOD) is measured with a Solar Radiometer, which quantifies aerosol-induced absorption and scattering in the daytime vertical atmospheric column. Using air temperature and relative humidity from the LHAASO site, we calculate the absolute humidity and establish its relationship with AOD. The observed attenuation of Cherenkov photons agrees with the calculated expectation to within 1\%. In the data analysis, we use absolute humidity to correct the Cherenkov photons to an AOD of zero (consistent with the simulation settings). The impact of residual aerosol components on Cherenkov-photon attenuation is also estimated in the proton energy spectrum paper. Further details can be found in the Supplementary Materials of the proton energy spectrum~\cite{LHAASO:2025byy}.

We categorize the data by absolute humidity into two datasets: a high-humidity set with absolute humidity greater than 1.53$\times$10$^{-6}$~g/cm$^{3}$ and a low-humidity set below this threshold. The difference in energy spectra between these two samples and the overall dataset is found to be less than 1\%.

\noindent{\bf Systematic uncertainty on hadronic interaction models:}
To assess the uncertainties associated with high-energy hadronic interaction models, we compared the light component energy spectra obtained with the EPOS-LHC, QGSJET-II-04, and SIBYLL 2.3d hadronic interaction models (Figs.~\ref{fig:hadronic} and ~\ref{fig:light_had}). When compared to the spectrum obtained using the EPOS-LHC model, the maximum discrepancy observed among the models is less than 9\%.

\begin{table*}[h]
    \renewcommand\arraystretch{1.5}
    \centering
    \caption{Table of LHAASO light energy spectrum. The first, second error on the flux represents the statistical uncertainty, systematic uncertainty, respectively.}
    \begin{tabular}{ccccc}
    \hline\hline
       \multirow{2}{*}{}Energy & Number of Events & flux$\pm$stat$\pm$ syst (QGSJET-II-04) & flux$\pm$stat$\pm$ syst(EPOS-LHC) & flux$\pm$stat$\pm$ syst(SIBYLL 2.3d) \\
       log$_{10}$(E/PeV) &  & PeV$^{-1}$ m$^{-2}$ s$^{-1}$ sr$^{-1}$ & PeV$^{-1}$ m$^{-2}$ s$^{-1}$ sr$^{-1}$ & PeV$^{-1}$ m$^{-2}$ s$^{-1}$ sr$^{-1}$\\
        \hline
        $-$0.8$\sim$$-$0.7 & 1298020 & 
        $(2.008\pm0.002\pm0.098)\times$10$^{-4}$ 
        &$(1.873\pm0.002\pm0.101)\times$10$^{-4}$
        &$(1.833\pm0.002\pm0.094)\times$10$^{-4}$ \\
        \hline
        $-$0.7$\sim$$-$0.6 & 892072 & $(1.074\pm0.001\pm0.050)\times$10$^{-4}$ &$(0.996\pm0.001\pm0.052)\times$10$^{-4}$ &$(0.977\pm0.001\pm0.050)\times$10$^{-4}$ \\
        \hline
        $-$0.6$\sim$$-$0.5 & 596301 & 
        $(5.661\pm0.007\pm0.287)\times$10$^{-5}$ 
        &$(5.236\pm0.007\pm0.305)\times$10$^{-5}$ 
        &$(5.179\pm0.007\pm0.301)\times$10$^{-5}$ \\
        \hline
        $-$0.5$\sim$$-$0.4 & 395921 & 
        $(3.003\pm0.005\pm0.160)\times$10$^{-5}$
        &$(2.744\pm0.004\pm0.171)\times$10$^{-5}$ 
        &$(2.750\pm0.004\pm0.167)\times$10$^{-5}$ \\
        \hline
        $-$0.4$\sim$$-$0.3 & 263669 & 
        $(1.600\pm0.003\pm0.085)\times$10$^{-5}$ 
        &$(1.467\pm0.003\pm0.092)\times$10$^{-5}$ 
        &$(1.484\pm0.003\pm0.091)\times$10$^{-5}$ \\
        \hline
        $-$0.3$\sim$$-$0.2 & 173713 & 
        $(8.445\pm0.020\pm0.470)\times$10$^{-6}$ 
        &$(7.792\pm0.019\pm0.533)\times$10$^{-6}$ 
        &$(7.839\pm0.019\pm0.499)\times$10$^{-6}$ \\
        \hline
        $-$0.2$\sim$$-$0.1 & 114818 & 
        $(4.455\pm0.013\pm0.252)\times$10$^{-6}$ 
        &$(4.191\pm0.012\pm0.295)\times$10$^{-6}$ 
        &$(4.203\pm0.012\pm0.275)\times$10$^{-6}$  \\
        \hline
        $-$0.1$\sim$0.0 & 76205 & 
        $(2.379\pm0.008\pm0.129)\times$10$^{-6}$ 
        &$(2.242\pm0.008\pm0.157)\times$10$^{-6}$
        &$(2.257\pm0.008\pm0.144)\times$10$^{-6}$ \\
        \hline
        0.0$\sim$0.1 & 51305 & 
        $(1.286\pm0.006\pm0.063)\times$10$^{-6}$ 
        &$(1.226\pm0.005\pm0.081)\times$10$^{-6}$ 
        &$(1.222\pm0.005\pm0.073)\times$10$^{-6}$ \\
        \hline
        0.1$\sim$0.2 & 34698 & 
        $(7.021\pm0.037\pm0.292)\times$10$^{-7}$
        &$(6.699\pm0.036\pm0.381)\times$10$^{-7}$ 
        &$(6.695\pm0.036\pm0.353)\times$10$^{-7}$ \\
        \hline
        0.2$\sim$0.3 & 23423 & 
        $(3.806\pm0.025\pm0.134)\times$10$^{-7}$ 
        &$(3.671\pm0.024\pm0.173)\times$10$^{-7}$ 
        &$(3.653\pm0.024\pm0.170)\times$10$^{-7}$ \\
        \hline
        0.3$\sim$0.4 & 15857 & 
        $(2.026\pm0.016\pm0.062)\times$10$^{-7}$ 
        &$(1.998\pm0.016\pm0.077)\times$10$^{-7}$ 
        &$(1.994\pm0.016\pm0.081)\times$10$^{-7}$ \\
        \hline
        0.4$\sim$0.5 & 10433 & 
        $(1.066\pm0.010\pm0.030)\times$10$^{-7}$ 
        &$(1.057\pm0.010\pm0.036)\times$10$^{-7}$ 
        &$(1.063\pm0.010\pm0.041)\times$10$^{-7}$ \\
        \hline
        0.5$\sim$0.6 & 6794 & 
        $(5.503\pm0.067\pm0.142)\times$10$^{-8}$ 
        &$(5.484\pm0.067\pm0.176)\times$10$^{-8}$ 
        &$(5.554\pm0.067\pm0.180)\times$10$^{-8}$ \\
        \hline
        0.6$\sim$0.7 & 4413 & 
        $(2.852\pm0.043\pm0.071)\times$10$^{-8}$ 
        &$(2.848\pm0.043\pm0.087)\times$10$^{-8}$ 
        &$(2.889\pm0.043\pm0.085)\times$10$^{-8}$ \\
        \hline
        0.7$\sim$0.8 & 2783 & 
        $(1.416\pm0.027\pm0.035)\times$10$^{-8}$ 
        &$(1.418\pm0.027\pm0.043)\times$10$^{-8}$ 
        &$(1.459\pm0.027\pm0.043)\times$10$^{-8}$
        \\
        \hline
        0.8$\sim$0.9 & 1701 & 
        $(6.338\pm0.159\pm0.154)\times$10$^{-9}$ 
        &$(6.811\pm0.165\pm0.219)\times$10$^{-9}$ 
        &$(6.508\pm0.161\pm0.203)\times$10$^{-9}$ \\
        \hline
        0.9$\sim$1.0 & 945 & 
        $(2.946\pm0.096\pm0.065)\times$10$^{-9}$ 
        &$(2.892\pm0.094\pm0.090)\times$10$^{-9}$ 
        &$(3.027\pm0.097\pm0.085)\times$10$^{-9}$ \\
        \hline
        1.0$\sim$1.1 & 544 & 
        $(1.258\pm0.055\pm0.039)\times$10$^{-9}$ 
        &$(1.269\pm0.054\pm0.036)\times$10$^{-9}$ 
        &$(1.281\pm0.055\pm0.042)\times$10$^{-9}$
        \\
         \hline\hline
    \end{tabular}
    \label{tab:light}
\end{table*}

\section{Helium energy spectrum}
\subsection{Methodology}
\begin{figure}[h]
    \centering
    \includegraphics[width=0.56\linewidth]{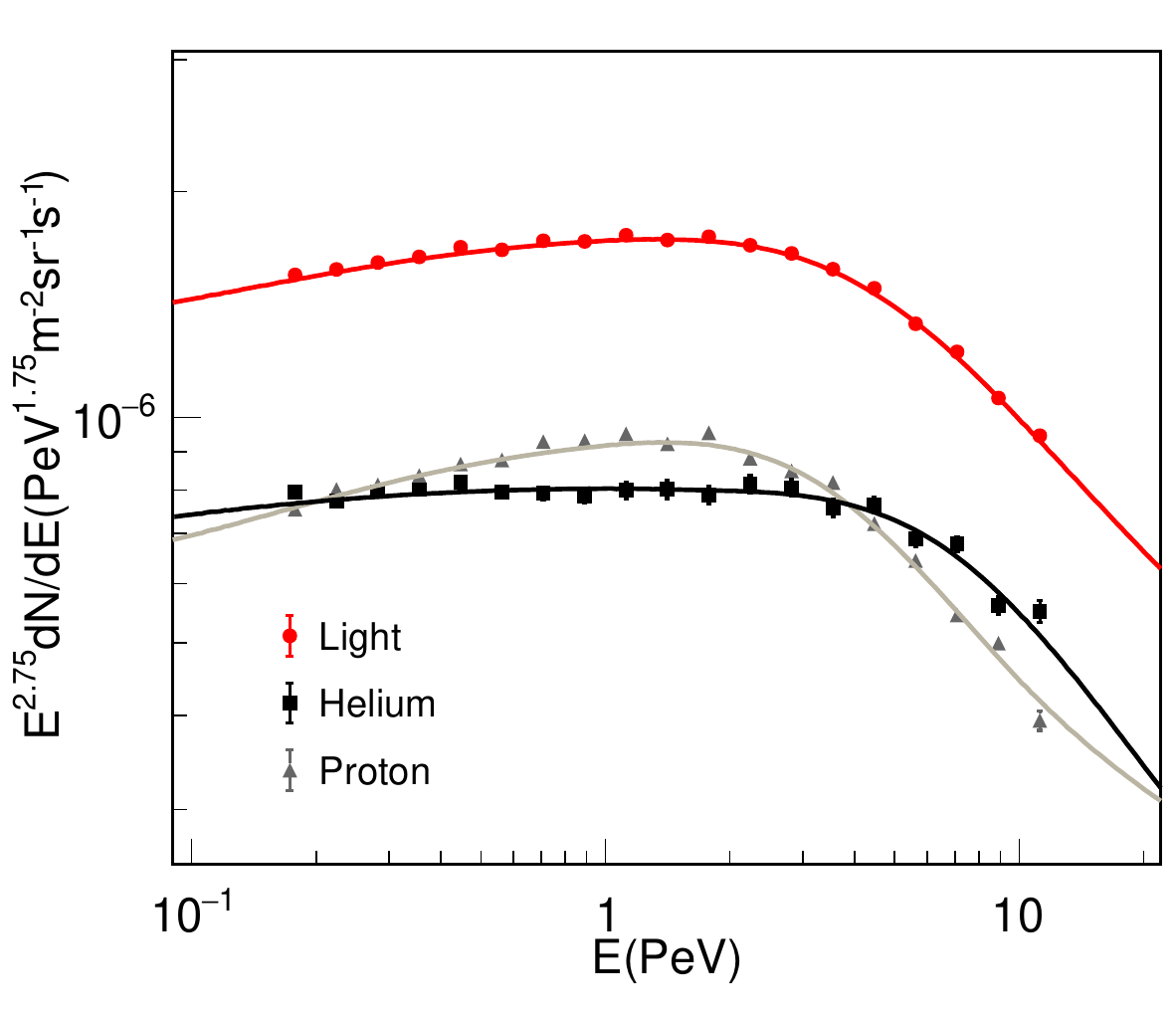}
    \caption {\textcolor{black}{Comparison between the assumed (solid curves) and reconstructed (dots) cosmic ray spectra for the light component (red), helium (black), and protons (gray). The reconstructed spectra, derived from simulation using the reconstructed energy, reproduce the assumed input spectra within 2\% for the light component and 5\% for helium. Error bars represent statistical uncertainties.}}
    \label{fig:MC_spectrum}
\end{figure}

The helium energy spectrum ($\Phi_{\rm He}$) is derived by subtracting the proton spectrum ($\Phi_{\rm H}$) from the measured light component spectrum ($\Phi_{\rm L}$), expressed as $\Phi_{\rm He} = \Phi_{\rm L} - \Phi_{\rm H}$. The following measures were implemented:
\begin{itemize}
\item 
The energy reconstruction formula for the light component is consistent with the methodology used in LHAASO's previously published analysis of the proton energy spectrum~\cite{LHAASO:2025byy}. This means that the same energy reconstruction formula is applied to both protons and helium nuclei. This approach ensures that the energy scale remains uniform across the analyses, with discrepancies of about 2\%. When reconstructing the energy flux of helium nuclei in bins of width 0.1 in $\log_{10}(E/\text{GeV})$, this minor discrepancy has an insignificant impact on the overall results.

\item The light component sample was selected using a double-cut method. This approach keeps the proton-to-helium ratio almost unchanged before and after selection, with variations of less than 5\% across the GSF~\cite{Dembinski:2017zsh}, Gaisser~\cite{Gaisser:2013bla}, Hörandel~\cite{Hoerandel:2002yg}, and LVBI~\cite{Lv:2024wrs} composition models.

\item When deriving the helium spectrum through $\Phi_{\rm He} = \Phi_{\rm L} - \Phi_{\rm H}$, both statistical and systematic uncertainties from the proton spectrum propagate into the helium spectrum. Minimizing these uncertainties in the proton spectrum is therefore critical. After obtaining the initial helium spectrum, we apply an iterative H/He ratio method (Section~\ref{sec:proton-iteration}) to measure the proton spectrum using a larger event sample. This process yields the proton spectrum with reduced statistical and systematic uncertainties, which subsequently lowers the uncertainties in the helium spectrum.

\end{itemize}
To validate the accuracy of the helium spectrum measurement method, we systematically verify it using simulated data, with the GSF model as an example. Figure~\ref{fig:MC_spectrum} presents the reconstructed energy spectrum from the simulation analysis (dots) alongside the model-predicted spectrum (solid line). The reconstructed spectra show good agreement with the model predictions. Specifically, the reconstructed spectra—derived from simulations using the reconstructed energy--reproduce the assumed input spectra with deviations within 2\% for the light component and 5\% for helium, which confirms the reliability of the proposed method.

\subsection{Helium energy spectrum\label{sec:Helium}}
\begin{figure}[h]
    \centering
    \includegraphics[width=0.56\linewidth]{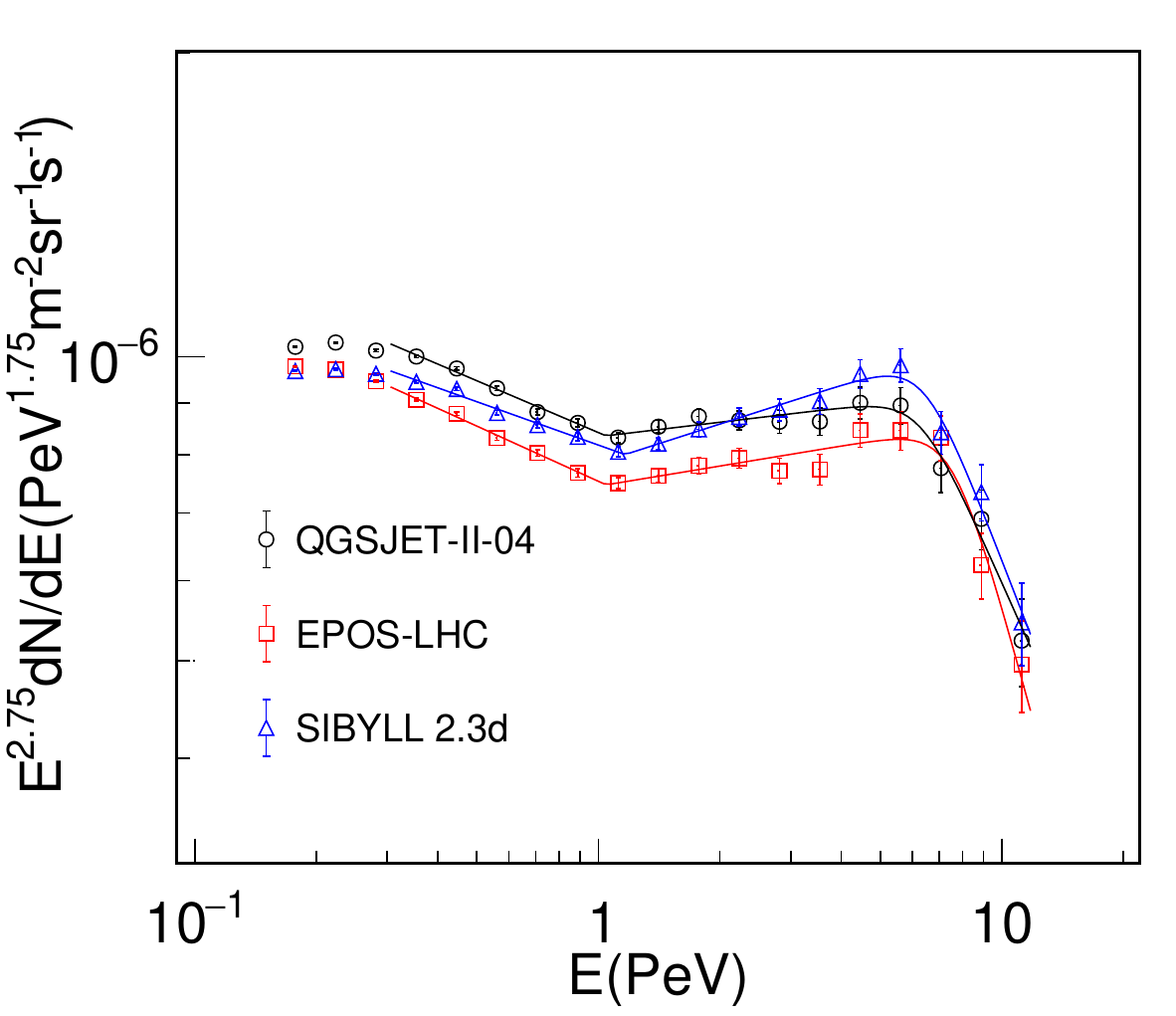}
    \caption{Fits of Eq.~\ref{for:fit3} to the helium spectral flux (0.3~–~13~PeV) were performed using different hadronic interaction models. Error bars represent statistical uncertainties, while solid lines depict the fitting functions. Colors correspond to the results from individual models.}
    \label{fig:He_fit}
\end{figure}

The helium spectra, obtained using the EPOS-LHC, QGSJET-II-04, and SIBYLL 2.3d hadronic interaction models, are presented in Fig.~\ref{fig:He_fit}.
The helium flux $\Phi_{\rm He}(E)$ and associated errors are tabulated in Table~\ref{tab:helium}. In Fig.~\ref{fig:He_fit}, a fit of LHAASO data with a `` double broken power law'' (DBPL) is shown in the energy range from 0.30~PeV to 13~PeV:
\begin{equation}
    \Phi(E) = \Phi_{0}\left(\frac{E}{0.10~\text{PeV}}\right)^{\gamma_1}\left(1+\left(\frac{E}{E_{h}}\right)^{1/w_{1}}\right)^{(\gamma_2-\gamma_1)w_{1}}\left(1+\left(\frac{E}{E_{k}}\right)^{1/w_{2}}\right)^{(\gamma_3-\gamma_2)w_{2}},
    \label{for:fit3}
\end{equation}
where $\Phi_{0}$ is the flux normalization, $E_{h}$ and $E_{k}$ are the energy positions where the spectral index changes, $\gamma_{1}$, $\gamma_{2}$ and $\gamma_{3}$ are the energy spectral indices, and $w_{1}$ and $w_{2}$ are the parameters describing the smoothness of the break.

The DBPL fit yields a chi-squared value of $\chi^{2}/ndf =$ 10.72/10, where $ndf$ denotes the number of degrees of freedom. The fit returns spectral indices $\gamma_1 = -2.93\pm0.01^{+0.06}_{-0.03}$, $\gamma_2 = -2.68\pm0.02^{+0.02}_{-0.06}$, and the hardening energy $E_{h} = 1.05\pm0.05^{+0.22}_{-0.13}$~PeV. 
The first smoothness parameter $w_{1}$ is fixed at 0.005 due to the apparent sharpness in the hardening structure.
The onset of a flux softening above a few PeV is also observed, with second break energy $E_{k}=7.6\pm0.7^{+0.3}_{-0.3}$~PeV and spectral index $\gamma_{3}=-4.2\pm0.5^{+0.5}_{-0.5}$. 
Given the relatively large uncertainties of the data in the highest energy bins, the second smoothness parameter $w_{2}$ cannot be effectively constrained and is kept fixed at a value of $w_{2} = 0.1$.
The fitting parameters and their associated uncertainties derived from different hadronic interaction models are summarized in Table~\ref{tab:error}. The first error reflects the fitting error introduced by statistical fluctuations, while the second error indicates the maximum uncertainty arising from systematic effects.

For the EPOS-LHC hadronic model, the change in index $\gamma_2 - \gamma_1$ is shown to be significantly different from zero, exceeding 13.9$\sigma$ when accounting for statistical errors. In the case of QGSJET-II-04, this value exceeds 12.3$\sigma$, while for SIBYLL 2.3d, it exceeds 15.0$\sigma$.

For the EPOS-LHC hadronic model, the change in index $\gamma_3 - \gamma_2$ is shown to be different from zero, exceeding 6.8$\sigma$ when accounting for statistical errors. In the case of QGSJET-II-04, this value exceeds 7.2$\sigma$, while for SIBYLL 2.3d, it exceeds 8.9$\sigma$. The fitting results of the three hadronic interaction models are presented in Fig.~\ref{fig:He_fit}.

The first three data points deviate from the single-index power-law, 
i.e. by 31.7$\sigma$, 8.7$\sigma$, and 0.7$\sigma$ in the EPOS-LHC hadronic model, respectively. This implies a potential structure at energies below 0.3 PeV.

\begin{table*}[t!]
    \renewcommand\arraystretch{1.5}
    \centering
    \caption{Parameters and their associated
uncertainties from fitting the helium spectrum using Eq.~\ref{for:fit3} with fixed $w_{1}$ and $w_{2}$, based on different high-energy hadronic interaction models.}
    \begin{tabular}{ccccccc}
    \hline\hline
       \multirow{2}{*}{}Model & $\Phi_{0}$@0.1PeV &  E$_{h}$ & $\gamma_{1}$ & $\gamma_{2}$ & E$_{k}$ & $\gamma_{3}$\\
        & PeV$^{-1}$ m$^{-2}$ s$^{-1}$ sr$^{-1}$ & PeV & &  & PeV\\
        \hline
        \multirow{2}{*}{}EPOS-LHC & (6.43 & 1.05 & $-$2.93 & $-$2.68  & 7.6 & $-$4.2 \\
        &$\pm$ 0.08 $^{+0.56}_{-0.14}$) $\times10^{-4}$ & $\pm$ 0.05 $^{+0.22}_{-0.13}$ & $\pm$ 0.01 $^{+0.06}_{-0.03}$ & $\pm$ 0.02 $^{+0.02}_{-0.06}$ & $\pm$ 0.7$^{+0.3}_{-0.3}$ & $\pm$ 0.5$^{+0.5}_{-0.5}$\\
        \hline
        \multirow{2}{*}{}QGSJET-II-04 & (7.00 & 1.05 & $-$2.92 & $-$2.70 & 6.3 & $-$3.7 \\
        & $\pm$ 0.08 $^{+0.66}_{-0.06}$) $\times10^{-4}$ & $\pm$ 0.06 $^{+0.09}_{-0.15}$ & $\pm$ 0.01 $^{+0.04}_{-0.01}$ & $\pm$ 0.02 $^{+0.02}_{-0.06}$ & $\pm$0.7$^{+0.5}_{-0.3}$ & $\pm$0.3$^{+0.2}_{-0.2}$\\
        \hline
        \multirow{2}{*}{}SIBYLL 2.3d & (6.38 & 1.16 & $-$2.89 & $-$2.62 & 6.4 & $-$3.8 \\
        & $\pm$ 0.06 $^{+0.61}_{-0.08}$) $\times10^{-4}$ & $\pm$ 0.07 $^{+0.17}_{-0.16}$ & $\pm$ 0.01 $^{+0.04}_{-0.01}$ & $\pm$ 0.02 $^{+0.02}_{-0.05}$ & $\pm$0.6$^{+0.5}_{-0.3}$ & $\pm$0.3$^{+0.1}_{-0.2}$\\
        \hline\hline
    \end{tabular}
    \label{tab:error}
\end{table*}

\begin{table*}[t!]
    \renewcommand\arraystretch{1.5}
    \centering
    \caption{Table of LHAASO helium energy spectrum. The first, second error on the flux represents the statistical uncertainty, systematic uncertainty, respectively.}
    \begin{tabular}{cccc}
    \hline\hline
       \multirow{2}{*}{}Energy & flux$\pm$stat$\pm$syst(QGSJET-II-04) & flux$\pm$stat$\pm$syst(EPOS-LHC) & flux$\pm$stat$\pm$syst(SIBYLL 2.3d)\\
       log$_{10}$(E/PeV)  & PeV$^{-1}$ m$^{-2}$ s$^{-1}$ sr$^{-1}$ & PeV$^{-1}$ m$^{-2}$ s$^{-1}$ sr$^{-1}$ & PeV$^{-1}$ m$^{-2}$ s$^{-1}$ sr$^{-1}$ \\
        \hline
        $-$0.8$\sim$$-$0.7 & 
        $(1.186\pm0.002\pm0.105)\times$10$^{-4}$ & 
        $(1.134\pm0.002\pm0.100)\times$10$^{-4}$  & 
        $(1.122\pm0.002\pm0.099)\times$10$^{-4}$ \\
        \hline
        $-$0.7$\sim$$-$0.6 & 
        $(6.358\pm0.013\pm0.554)\times$10$^{-5}$ & 
        $(5.975\pm0.012\pm0.521)\times$10$^{-5}$ & 
        $(5.983\pm0.012\pm0.520)\times$10$^{-5}$ \\
        \hline
        $-$0.6$\sim$$-$0.5 & 
        $(3.314\pm0.008\pm0.309)\times$10$^{-5}$ & 
        $(3.091\pm0.008\pm0.289)\times$10$^{-5}$ & 
        $(3.143\pm0.008\pm0.293)\times$10$^{-5}$ \\
        \hline
        $-$0.5$\sim$$-$0.4 & 
        $(1.736\pm0.006\pm0.179)\times$10$^{-5}$ & 
        $(1.573\pm0.005\pm0.163)\times$10$^{-5}$ & 
        $(1.638\pm0.005\pm0.169)\times$10$^{-5}$
        \\
        \hline
        $-$0.4$\sim$$-$0.3 & 
        $(8.967\pm0.036\pm0.980)\times$10$^{-6}$ & 
        $(8.090\pm0.034\pm0.887)\times$10$^{-6}$ & 
        $(8.558\pm0.034\pm0.934)\times$10$^{-6}$
        \\
        \hline
        $-$0.3$\sim$$-$0.2 & 
        $(4.551\pm0.024\pm0.611)\times$10$^{-6}$ & 
        $(4.063\pm0.023\pm0.548)\times$10$^{-6}$ & 
        $(4.302\pm0.023\pm0.578)\times$10$^{-6}$
        \\
        \hline
        $-$0.2$\sim$$-$0.1 & 
        $(2.288\pm0.016\pm0.338)\times$10$^{-6}$ & 
        $(2.084\pm0.015\pm0.309)\times$10$^{-6}$ &
        $(2.221\pm0.015\pm0.328)\times$10$^{-6}$
        \\
        \hline
        $-$0.1$\sim$0.0 & 
        $(1.185\pm0.010\pm0.185)\times$10$^{-6}$ & 
        $(1.057\pm0.010\pm0.166)\times$10$^{-6}$ &   
        $(1.148\pm0.010\pm0.180)\times$10$^{-6}$
        \\
        \hline
        0.0$\sim$0.1 & 
        $(6.084\pm0.070\pm0.948)\times$10$^{-7}$ & 
        $(5.483\pm0.068\pm0.859)\times$10$^{-7}$ & 
        $(5.890\pm0.067\pm0.918)\times$10$^{-7}$ \\
        \hline
        0.1$\sim$0.2 & 
        $(3.311\pm0.047\pm0.449)\times$10$^{-7}$ & 
        $(2.960\pm0.045\pm0.405)\times$10$^{-7}$ & 
        $(3.185\pm0.045\pm0.432)\times$10$^{-7}$
        \\
        \hline
        0.2$\sim$0.3 & 
        $(1.799\pm0.031\pm0.202)\times$10$^{-7}$ & 
        $(1.608\pm0.030\pm0.183)\times$10$^{-7}$ & 
        $(1.747\pm0.030\pm0.196)\times$10$^{-7}$ \\
        \hline
        0.3$\sim$0.4 & 
        $(9.468\pm0.201\pm0.859)\times$10$^{-8}$ & 
        $(8.686\pm0.200\pm0.805)\times$10$^{-8}$ & 
        $(9.542\pm0.196\pm0.865)\times$10$^{-8}$ \\
        \hline
        0.4$\sim$0.5 & 
        $(5.010\pm0.130\pm0.409)\times$10$^{-8}$ & 
        $(4.480\pm0.131\pm0.379)\times$10$^{-8}$ & 
        $(5.151\pm0.128\pm0.419)\times$10$^{-8}$ \\
        \hline
        0.5$\sim$0.6 & 
        $(2.661\pm0.083\pm0.198)\times$10$^{-8}$ & 
        $(2.386\pm0.084\pm0.185)\times$10$^{-8}$ & 
        $(2.787\pm0.082\pm0.206)\times$10$^{-8}$ \\
        \hline
        0.6$\sim$0.7 & 
        $(1.475\pm0.053\pm0.095)\times$10$^{-8}$ & 
        $(1.385\pm0.053\pm0.092)\times$10$^{-8}$ & 
        $(1.575\pm0.052\pm0.100)\times$10$^{-8}$ \\
        \hline
        0.7$\sim$0.8 & 
        $(7.781\pm0.326\pm0.465)\times$10$^{-9}$ & 
        $(7.353\pm0.325\pm0.452)\times$10$^{-9}$ & 
        $(8.529\pm0.323\pm0.501)\times$10$^{-9}$ \\
        \hline
        0.8$\sim$0.9 & 
        $(3.580\pm0.192\pm0.217)\times$10$^{-9}$ & 
        $(3.838\pm0.197\pm0.233)\times$10$^{-9}$ & 
        $(3.886\pm0.190\pm0.232)\times$10$^{-9}$ \\
        \hline
        0.9$\sim$1.0 & 
        $(1.693\pm0.116\pm0.105)\times$10$^{-9}$  & 
        $(1.524\pm0.114\pm0.098)\times$10$^{-9}$  & 
        $(1.800\pm0.115\pm0.111)\times$10$^{-9}$ \\
        \hline
        1.0$\sim$1.1 & 
        $(6.809\pm0.680\pm0.457)\times$10$^{-10}$  & 
        $(6.449\pm0.672\pm0.443)\times$10$^{-10}$  & 
        $(7.097\pm0.669\pm0.475)\times$10$^{-10}$ \\
         \hline\hline
    \end{tabular}
    \label{tab:helium}
\end{table*}

\subsection{Systematic uncertainties of helium spectrum\label{sec:helium_sys}}

For the helium energy spectrum, as with the light component, the energy scale systematic uncertainties are the same as those for the proton energy spectrum~\cite{LHAASO:2025byy}. The total systematic uncertainty in the energy measurement is about 4\%. 

The primary sources of systematic uncertainties in the heulim energy spectrm flux include uncertainties from the observational environment (atmospheric pressure, absolute humidity, night-sky background light) and hadronic interaction models, as well as from component selection and composition models for the proton and light components. It should be noted that the helium flux is approximately half that of the light component, and thus propagating the absolute systematic uncertainty of the light component to the helium spectrum results in a significantly larger relative uncertainty. These systematic uncertainties are summarized in Table~\ref{table:systematicUncertainty} and are introduced below.

\noindent{\bf Observational environment:} Following the same procedure used for the light component, we split the data into two subsets representing different observational environments, obtain the helium spectrum for each subset, and compare them with the full-sample result. The discrepancies are considered as the systematic uncertainty due to observing conditions.

\noindent{\bf Composition models:} We compared the helium spectra obtained with composition models of GSF-LHAASO, Gaisser-LHAASO, Hörandel-LHAASO, and LVBI-LHAASO. The flux differences among these models are approximately 10\% at 300 TeV and about 5\% at 6 PeV.

\noindent{\bf Component selection:} Comparing the selection efficiency of the light component between 50\%, 60\%, and 70\%, the helium flux change obtained is less than 2\%. This means that the systematic uncertainty due to component selection is less than 2\%.

\begin{figure}[h]
    \centering
    \includegraphics[width=0.6\linewidth]{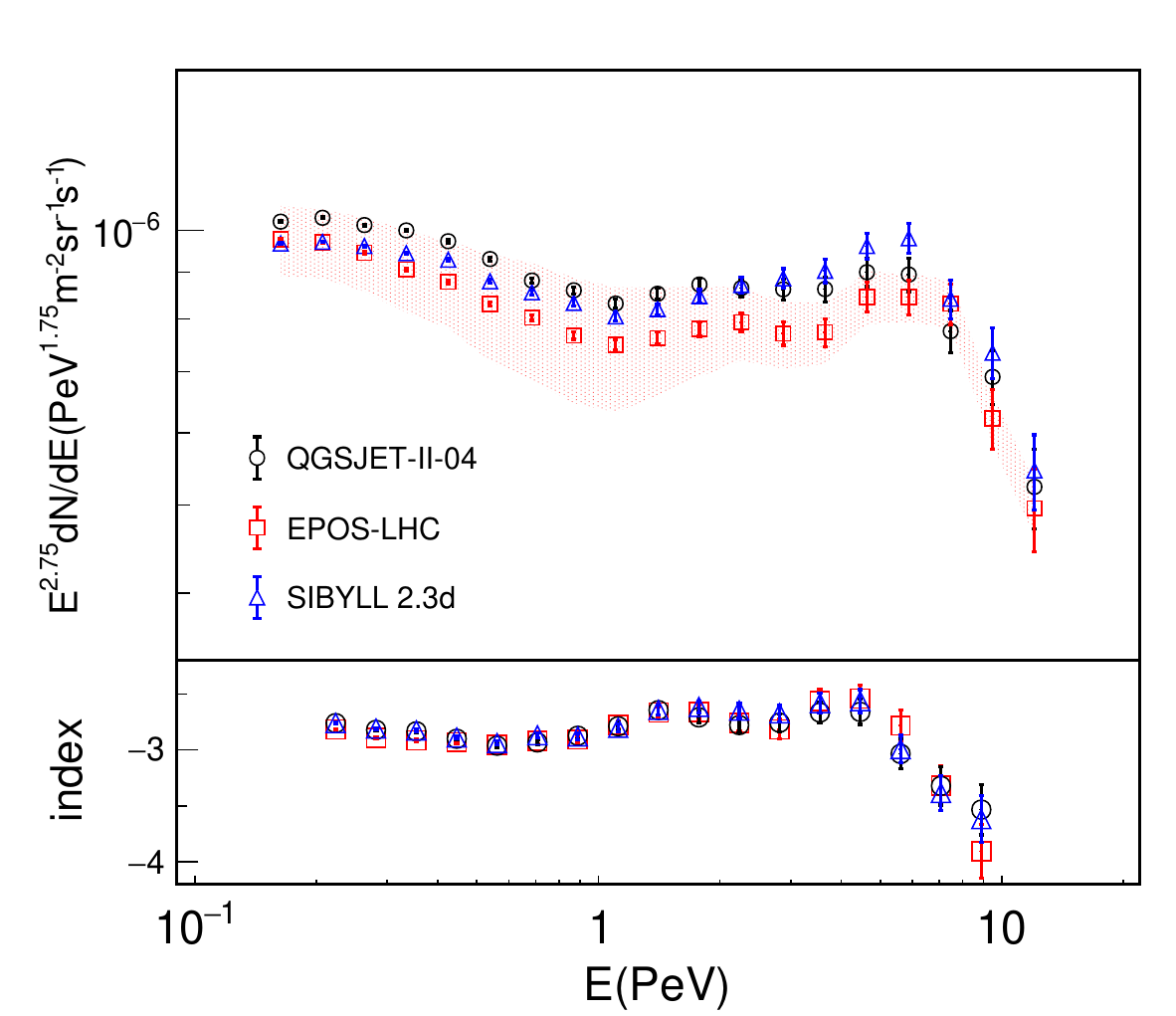}
    \caption{Top panel: Energy spectra of helium scaled by $E^{2.75}$ versus energy. Different markers indicate results from different hadronic interaction models, with error bars representing statistical uncertainties ($\sqrt{\sigma_{\rm light}^2 + \sigma_{\rm proton}^2}$). The shaded region represents the total systematic uncertainty (excluding that from hadronic interaction models), which is plotted on the helium spectrum based on the EPOS-LHC model. Bottom panel: Spectral indices of the helium energy spectra derived from different hadronic interaction models. Each index was fitted using a single power-law function with three adjacent points. Error bars show fitting uncertainties.}
    \label{fig:helium_had}
\end{figure}

\noindent{\bf Hadronic interaction models:} The light component energy spectrum and proton energy spectrum are measured with different hadronic interaction models: EPOS-LHC, QGSJET-II-04, and SIBYLL 2.3d.~The helium energy spectrum is then obtained using these same models. Figure~\ref{fig:helium_had} shows the helium spectra obtained with EPOS-LHC, QGSJET-II-04, and SIBYLL 2.3d. Relative to EPOS-LHC, the mean flux differences are about 13\% for QGSJET-II-04 and about 10\% for SIBYLL 2.3d on average.

\begin{table*}[htbp]
  \centering  
  \caption{Systematic uncertainty sources for the light energy spectrum and helium energy spectrum.}
  \label{tab:sys_he}
  \begin{threeparttable}
    \begin{tabular}{c c c}
      \hline
      \hline
      Source & ~~~~~~~~~~~~~~~~~~~~~~Uncertainty of Light~~~~~~~~~~&~~~~~~~Uncertainty of Helium\\
      \hline 
      \hline    \\ 
       Composition models & ~~~~~~~~~~~~~~~~~~~~~~~~~$\sim6\%@300$TeV,$\sim2\%@6$PeV~~~~~~~~~&~~~~$\sim10\%@300$TeV, $\sim5\%@6$PeV \\  \\   
      Component selection & ~~~~~~~~~~~~~~~~~~~~~~$<1\%$~~~&~~~$<2\%$ \\ \\   
      Atmosphere pressure& ~~~~~~~~~~~~~~~~~~~~~$<2\%$~~~&~~~$<3\%$ \\  \\    
      Absolute humidity& ~~~~~~~~~~~~~~~~~~~~~$<1\%$~~~&~~~$<2\%$ \\  \\   
      Background light& ~~~~~~~~~~~~~~~~~~~~~~$<2\%$~~~&~~~$<3\%$ \\  \\  
      Hadronic interaction models& ~~~~~~~~~~~~~~~~~~~~~~$\sim9\%@300$TeV, $\sim3\%@6$PeV~~~&~~~~~~~$\sim10\%@300$TeV, $\sim15\%@6$PeV \\
      \hline
      \hline
    \end{tabular}
  \end{threeparttable}
  \label{table:systematicUncertainty}  
\end{table*}

\begin{figure}[h]
    \centering
   \includegraphics[width=0.6\linewidth]{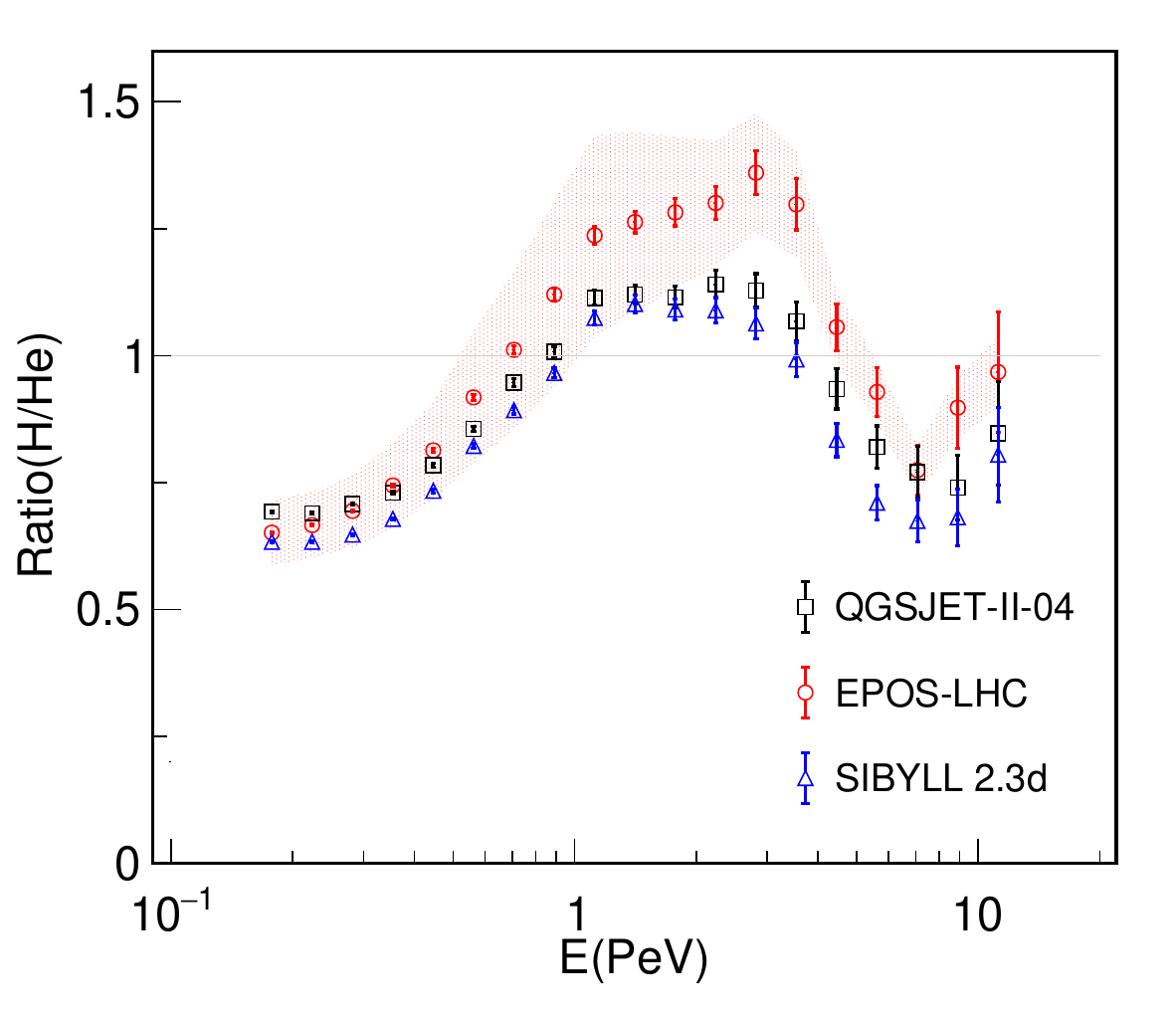}
    \caption{The proton to helium flux ratio (H/He) as a function of energy is shown for various hadronic interaction models. The shaded region indicates the systematic uncertainty derived from the EPOS–LHC hadronic interaction model.}
    \label{fig:hhe_had}
\end{figure}

Figure~\ref{fig:hhe_had} shows the H/He ratio under different hadronic interaction models.  The shaded area indicates the total systematic uncertainty (excluding contributions from hadronic interaction models): about 11\% around 300 TeV and about 6\% around 6 PeV. 

\begin{figure}[h]
    \centering
    \includegraphics[width=1.0\linewidth]{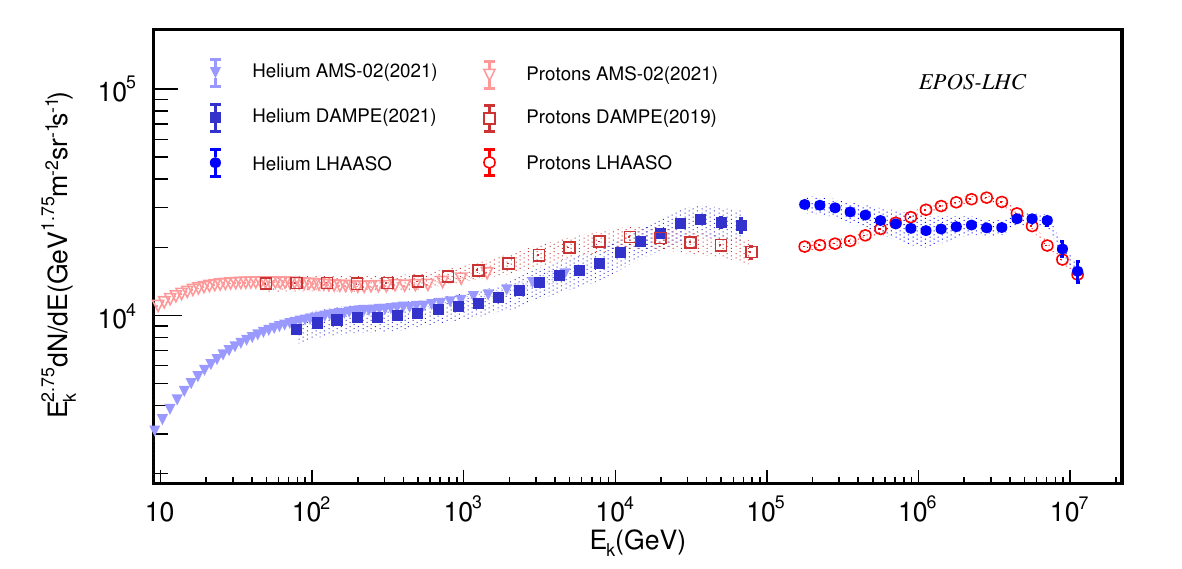}
     \caption{ Energy spectra of helium (blue solid geometries) and protons (red hollow geometries), scaled by $E^{2.75}$, as a function of kinetic energy. Error bars denote statistical uncertainties, shaded bands denote the systematic uncertainties. For comparison, the helium and proton spectra includes data from AMS-02~\cite{AMS:2021nhj}, DAMPE~\cite{Alemanno:2021gpb,DAMPE:2019gys}. }
    \label{fig:other_spec_sys}
\end{figure}

\begin{figure}[h]
    \centering
    \includegraphics[width=1.0\linewidth]{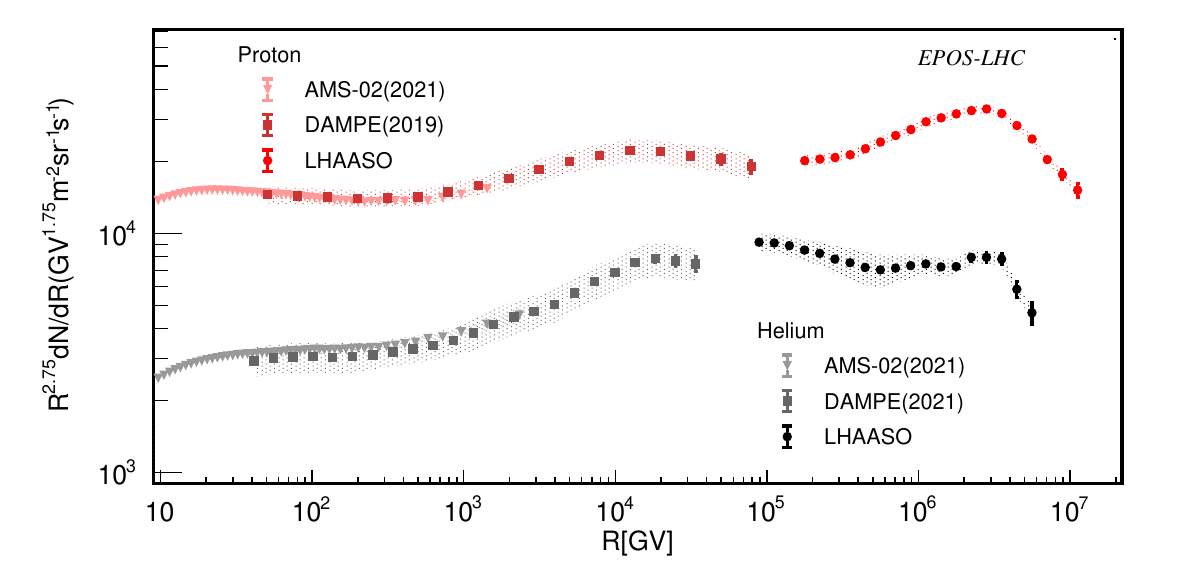}
     \caption{Rigidity spectra of helium nuclei (black solid geometries) and protons (red solid geometries) scaled by $R^{2.75}$. Error bars indicate statistical uncertainties. Shaded bands denote the systematic uncertainties. For comparison, proton and helium spectra from AMS-02~\cite{AMS:2021nhj} and DAMPE~\cite{Alemanno:2021gpb, DAMPE:2019gys} are shown.}
    \label{fig:other_spec_R_sys}
\end{figure}

\subsection{Obtaining larger sample proton energy spectra using H/He iteration \label{sec:proton-iteration}}
To obtain the proton energy spectrum with high purity proton samples ($\sim$89\% purity), the LHAASO proton energy spectrum paper adopted strict composition selection criteria, retaining approximately 25\% of the proton samples~\cite{LHAASO:2025byy}. Under these strict selection criteria, helium became the dominant contaminant in the proton samples, whereas the contamination from heavier components such as CNO and heavier nuclei can be ignored~\cite{LHAASO:2025byy}. 

\begin{figure}[h]
    \centering
    \includegraphics[width=0.6\linewidth]{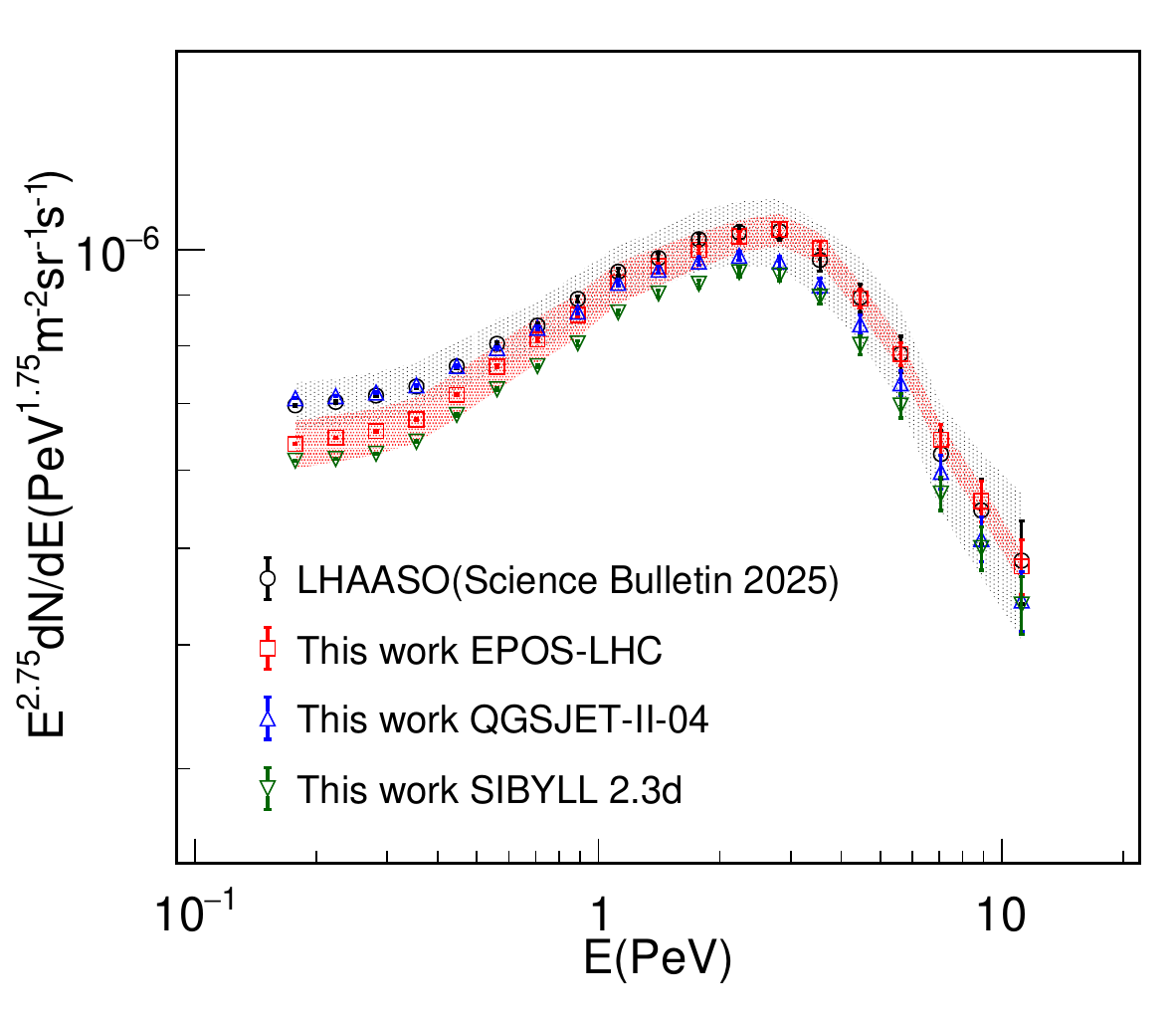}
    \caption{The energy spectra of the protons multiplied by $E^{2.75}$ as a function of energy, black symbols show the results obtained with the EPOS-LHC hadronic interaction model from the proton spectrum paper~\cite{LHAASO:2025byy}, red, green and blue geometries represent the results based on different hadronic interaction models after the iteration in this work. The shaded regions denote the total systematic uncertainty (excluding that from hadronic interaction models).}
    \label{fig:proton}
\end{figure}

In this work, we measured the helium spectrum and subsequently obtained the H/He ratio. We then employed iterative methods to relax the selection criteria for proton samples, retaining 50\% of the proton samples. In this case, the contamination from heavier components is less than 3\% below 300 TeV and around 1\% above 300 TeV. The contamination of protons by helium will increase, but a precise proton spectrum can still be obtained after applying a correction based on the H/He ratio.
The iteration was terminated once the proton spectrum converged with that of the preceding iteration; a total of four iterations were performed. The updated proton spectrum is shown in Fig.~\ref{fig:proton}. Compared to our previously published results~\cite{LHAASO:2025byy}, the proton event statistics have doubled.
By retaining more proton samples, the systematic uncertainty associated with hadronic interaction models has been reduced from 17\% to 12\%. Additionally, using an updated composition model provided by LHAASO measurements, the systematic uncertainty due to composition-model variations decreases from 7\% to 2\% around 3 PeV. Table~\ref{tab:proton} presents the updated proton flux along with the corresponding statistical and systematic uncertainties derived using three hadronic interaction models.

\begin{table*}[ht]
    \renewcommand\arraystretch{1.5}
    \centering
    \caption{Table of LHAASO proton energy spectrum. The first, second error on the flux represents the statistical uncertainty, systematic uncertainty, respectively.}
    \begin{tabular}{cccc}
    \hline\hline
       \multirow{2}{*}{}Energy & flux$\pm$stat$\pm$syst(QGSJET-II-04) & flux$\pm$stat$\pm$syst(EPOS-LHC) & flux$\pm$stat$\pm$syst(SIBYLL 2.3d)\\
       log$_{10}$(E/PeV)  & PeV$^{-1}$ m$^{-2}$ s$^{-1}$ sr$^{-1}$ & PeV$^{-1}$ m$^{-2}$ s$^{-1}$ sr$^{-1}$ & PeV$^{-1}$ m$^{-2}$ s$^{-1}$ sr$^{-1}$ \\
        \hline
        $-$0.8$\sim$$-$0.7 & 
        $(0.822\pm0.001\pm0.046)\times$10$^{-4}$ & 
        $(0.739\pm0.001\pm0.041)\times$10$^{-4}$  & 
        $(0.711\pm0.001\pm0.039)\times$10$^{-4}$ \\
        \hline
        $-$0.7$\sim$$-$0.6 & 
        $(4.384\pm0.007\pm0.239)\times$10$^{-5}$ & 
        $(3.982\pm0.006\pm0.217)\times$10$^{-5}$ & 
        $(3.790\pm0.006\pm0.207)\times$10$^{-5}$ \\
        \hline
        $-$0.6$\sim$$-$0.5 & 
        $(2.347\pm0.004\pm0.125)\times$10$^{-5}$ & 
        $(2.146\pm0.004\pm0.115)\times$10$^{-5}$ & 
        $(2.036\pm0.004\pm0.109)\times$10$^{-5}$ \\
        \hline
        $-$0.5$\sim$$-$0.4 & 
        $(1.267\pm0.003\pm0.067)\times$10$^{-5}$ & 
        $(1.171\pm0.003\pm0.062)\times$10$^{-5}$ & 
        $(1.112\pm0.003\pm0.059)\times$10$^{-5}$
        \\
        \hline
        $-$0.4$\sim$$-$0.3 & 
        $(7.035\pm0.020\pm0.375)\times$10$^{-6}$ & 
        $(6.581\pm0.018\pm0.351)\times$10$^{-6}$ & 
        $(6.280\pm0.018\pm0.335)\times$10$^{-6}$
        \\
        \hline
        $-$0.3$\sim$$-$0.2 & 
        $(3.894\pm0.013\pm0.208)\times$10$^{-6}$ & 
        $(3.729\pm0.013\pm0.199)\times$10$^{-6}$ & 
        $(3.537\pm0.012\pm0.189)\times$10$^{-6}$
        \\
        \hline
        $-$0.2$\sim$$-$0.1 & 
        $(2.167\pm0.009\pm0.115)\times$10$^{-6}$ & 
        $(2.108\pm0.009\pm0.112)\times$10$^{-6}$ &
        $(1.982\pm0.009\pm0.105)\times$10$^{-6}$
        \\
        \hline
        $-$0.1$\sim$0.0 & 
        $(1.194\pm0.006\pm0.061)\times$10$^{-6}$ & 
        $(1.185\pm0.006\pm0.060)\times$10$^{-6}$ &   
        $(1.109\pm0.006\pm0.056)\times$10$^{-6}$
        \\
        \hline
        0.0$\sim$0.1 & 
        $(6.776\pm0.042\pm0.327)\times$10$^{-7}$ & 
        $(6.782\pm0.041\pm0.327)\times$10$^{-7}$ & 
        $(6.328\pm0.040\pm0.305)\times$10$^{-7}$ \\
        \hline
        0.1$\sim$0.2 & 
        $(3.709\pm0.028\pm0.167)\times$10$^{-7}$ & 
        $(3.739\pm0.027\pm0.168)\times$10$^{-7}$ & 
        $(3.510\pm0.027\pm0.158)\times$10$^{-7}$
        \\
        \hline
        0.2$\sim$0.3 & 
        $(2.006\pm0.018\pm0.083)\times$10$^{-7}$ & 
        $(2.062\pm0.018\pm0.085)\times$10$^{-7}$ & 
        $(1.906\pm0.018\pm0.079)\times$10$^{-7}$ \\
        \hline
        0.3$\sim$0.4 & 
        $(10.796\pm0.121\pm0.415)\times$10$^{-8}$ & 
        $(11.297\pm0.121\pm0.435)\times$10$^{-8}$ & 
        $(10.393\pm0.116\pm0.400)\times$10$^{-8}$ \\
        \hline
        0.4$\sim$0.5 & 
        $(5.653\pm0.078\pm0.209)\times$10$^{-8}$ & 
        $(6.094\pm0.080\pm0.225)\times$10$^{-8}$ & 
        $(5.478\pm0.076\pm0.202)\times$10$^{-8}$ \\
        \hline
        0.5$\sim$0.6 & 
        $(2.842\pm0.049\pm0.104)\times$10$^{-8}$ & 
        $(3.097\pm0.051\pm0.113)\times$10$^{-8}$ & 
        $(2.766\pm0.048\pm0.101)\times$10$^{-8}$ \\
        \hline
        0.6$\sim$0.7 & 
        $(1.378\pm0.031\pm0.051)\times$10$^{-8}$ & 
        $(1.463\pm0.031\pm0.054)\times$10$^{-8}$ & 
        $(1.314\pm0.029\pm0.048)\times$10$^{-8}$ \\
        \hline
        0.7$\sim$0.8 & 
        $(6.381\pm0.184\pm0.237)\times$10$^{-9}$ & 
        $(6.827\pm0.183\pm0.254)\times$10$^{-9}$ & 
        $(6.060\pm0.172\pm0.225)\times$10$^{-9}$ \\
        \hline
        0.8$\sim$0.9 & 
        $(2.758\pm0.108\pm0.103)\times$10$^{-9}$ & 
        $(2.973\pm0.107\pm0.111)\times$10$^{-9}$ & 
        $(2.622\pm0.100\pm0.098)\times$10$^{-9}$ \\
        \hline
        0.9$\sim$1.0 & 
        $(1.253\pm0.064\pm0.048)\times$10$^{-9}$  & 
        $(1.368\pm0.065\pm0.052)\times$10$^{-9}$  & 
        $(1.227\pm0.062\pm0.047)\times$10$^{-9}$ \\
        \hline
        1.0$\sim$1.1 & 
        $(5.766\pm0.394\pm0.221)\times$10$^{-10}$  & 
        $(6.241\pm0.395\pm0.239)\times$10$^{-10}$  & 
        $(5.714\pm0.380\pm0.219)\times$10$^{-10}$ \\
         \hline\hline
    \end{tabular}
    \label{tab:proton}
\end{table*}

\section{Update of composition models\label{sec:new_model}}
Cosmic ray composition models, such as GSF~\cite{Dembinski:2017zsh}, Gaisser~\cite{Gaisser:2013bla}, Hörandel~\cite{Hoerandel:2002yg}, and LVBI~\cite{Lv:2024wrs}, provide accurate descriptions of the measured energy spectra for components below 100 TeV. However, due to differences in extrapolation approaches, these models exhibit significant variations in their predicted composition ratios. As shown in Fig.~\ref{fig:fit_model}, the proportion of light component differs by up to approximately $\sim$20\% among the models. Recently, LHAASO has released proton energy spectrum measurement~\cite{LHAASO:2025byy}, and light component and helium energy spectra measurements in this work above 100 TeV. Using these latest measurements, a new composition model can be developed. The steps involved are as follows:

\begin{itemize}
\item Preliminary helium energy spectrum: the preliminary helium energy spectrum is measured based on the GSF composition model.

\item Replace the proton and helium spectra in the composition model (such as the GSF composition model) with the proton and helium spectra measured by LHAASO.

\item 
Determine the proportion of heavy component (CNO, MgAlSi, and iron) in all particles by fitting the $P_{\mu e}$ distribution: keep the energy spectrum characteristics of heavy component the same as their energy spectrum characteristics in the original composition model, including their relative proportions. Replace the proton and helium nuclear spectra in the original composition model with the measurement results from LHAASO. Adjust the proportion of heavy component to achieve the best match between the $P_{\mu e}$ distribution from the simulation and the $P_{\mu e}$ distribution from data. When the simulated $P_{\mu e}$ distribution best matches the experimental data's $P_{\mu e}$ distribution, determine the final proportion of heavy component. In this way, new composition models are obtained and named GSF-LHAASO, Gaisser-LHAASO, Hörandel-LHAASO, and LVBI-LHAASO.

\item With the updated composition model (e.g., GSF-LHAASO), the light component spectrum measurements are then updated.
\end{itemize}

\begin{figure}[ht]
    \centering
    \includegraphics[width=0.49\linewidth]{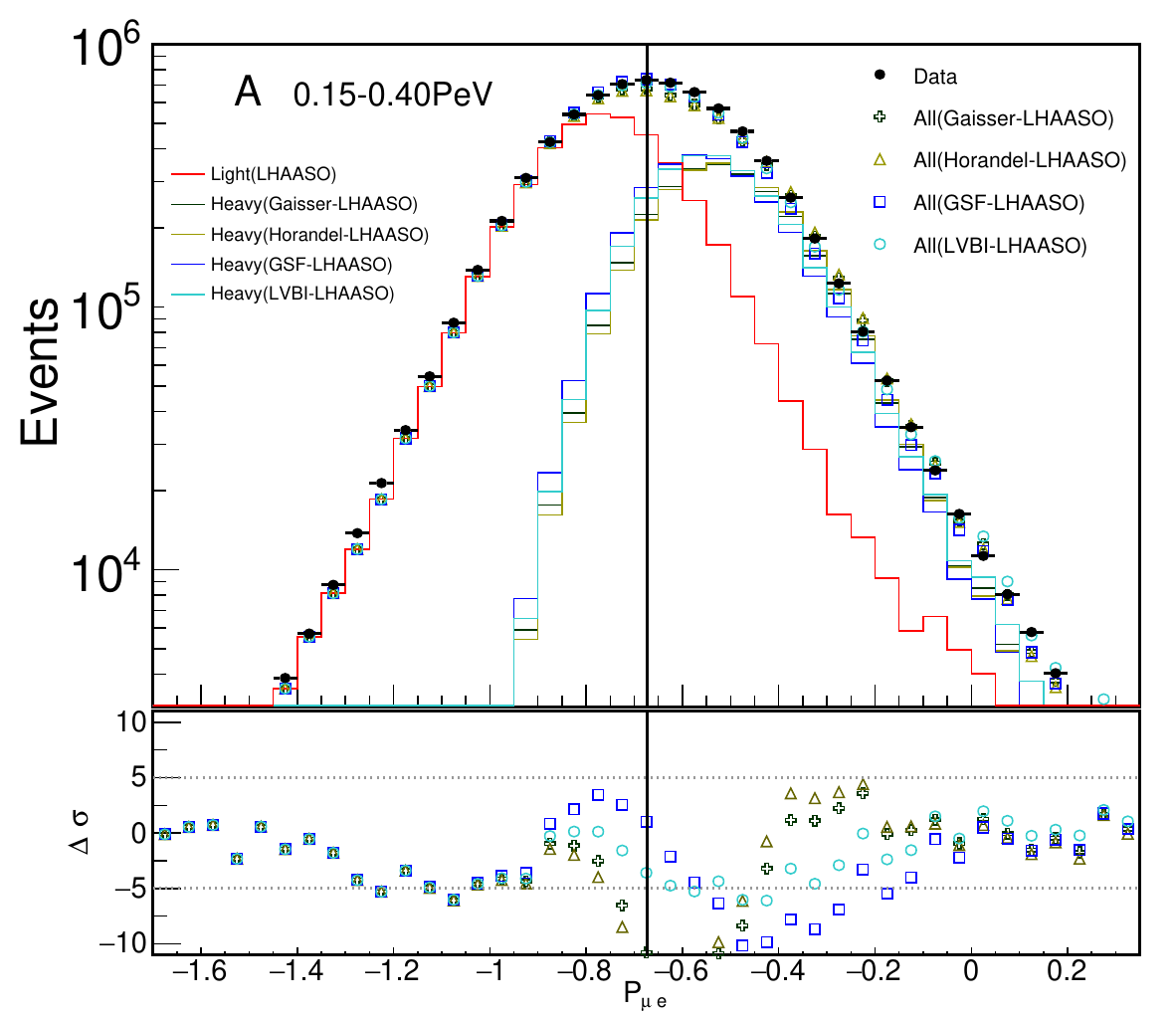}
    \hfill
    \includegraphics[width=0.49\linewidth]{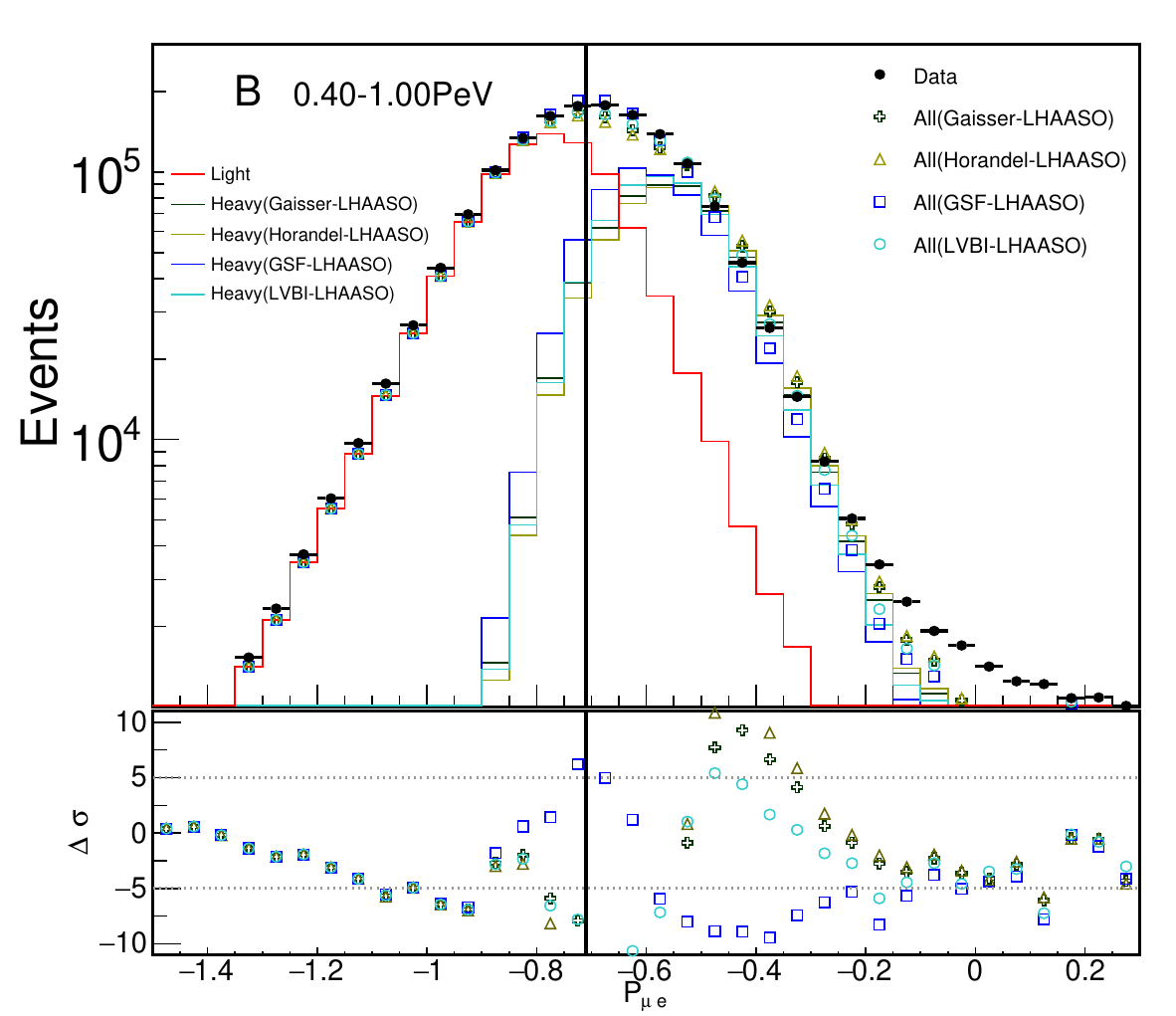}
    \hfill
    \includegraphics[width=0.49\linewidth]{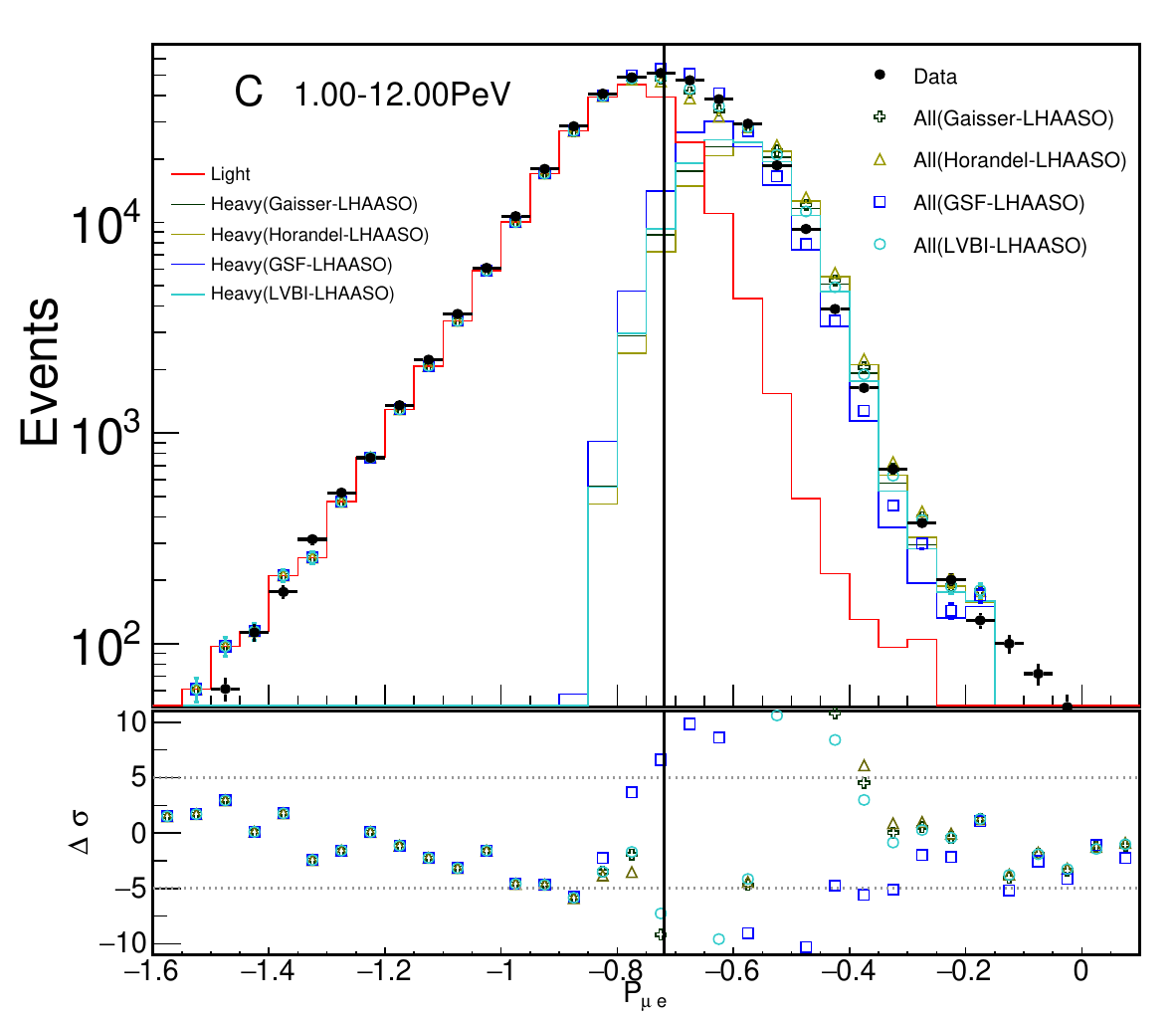}
    \caption{Upper: Distributions of $P_{\mu e}$ for events in the energy intervals 0.16~–~0.40 PeV (A), 0.40~–~1.00 PeV (B), and 1.00~–~13.00 PeV (C). LHAASO experimental data are indicated by solid black points. Open symbols correspond to predictions from different composition models. The red line shows the $P_{\mu e}$ distribution derived from the light component (proton and helium) obtained in this work. The dark green, light green, dark blue and light blue lines represent $P_{\mu e}$ distributions based on heavy component from different composition models. Simulations use the EPOS-LHC hadronic interaction model.
    Bottom: Open symbols show deviations between experimental data and simulations for each composition model. Gray lines mark the $\pm 5\sigma$ range. The solid black line indicates the maximum $P_{\mu e}$ threshold used for light component selection.}
     \label{fig:epos}
\end{figure}

\begin{figure}[h]
    \centering
    \includegraphics[width=0.49\linewidth]{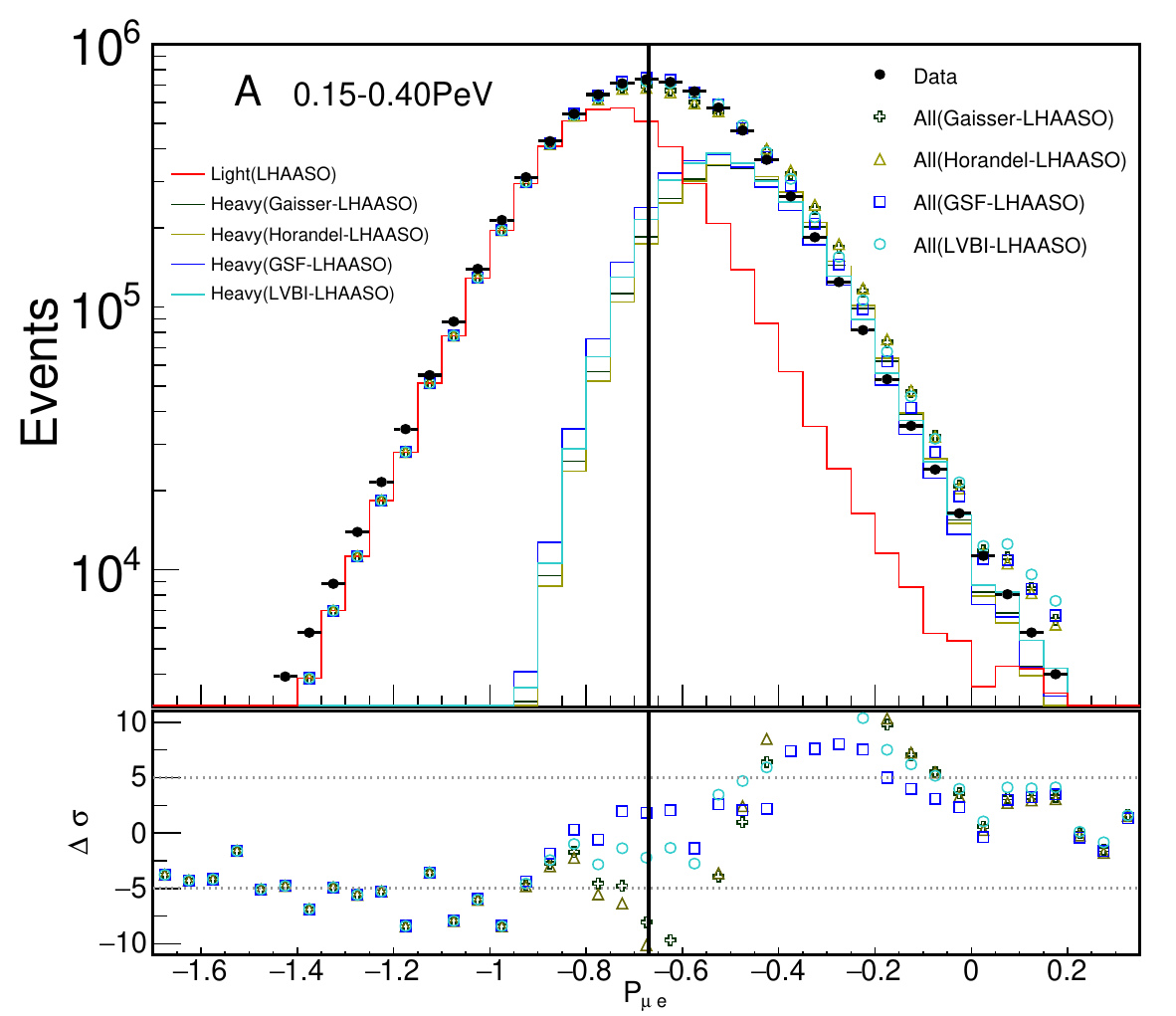}
    \hfill
    \includegraphics[width=0.49\linewidth]{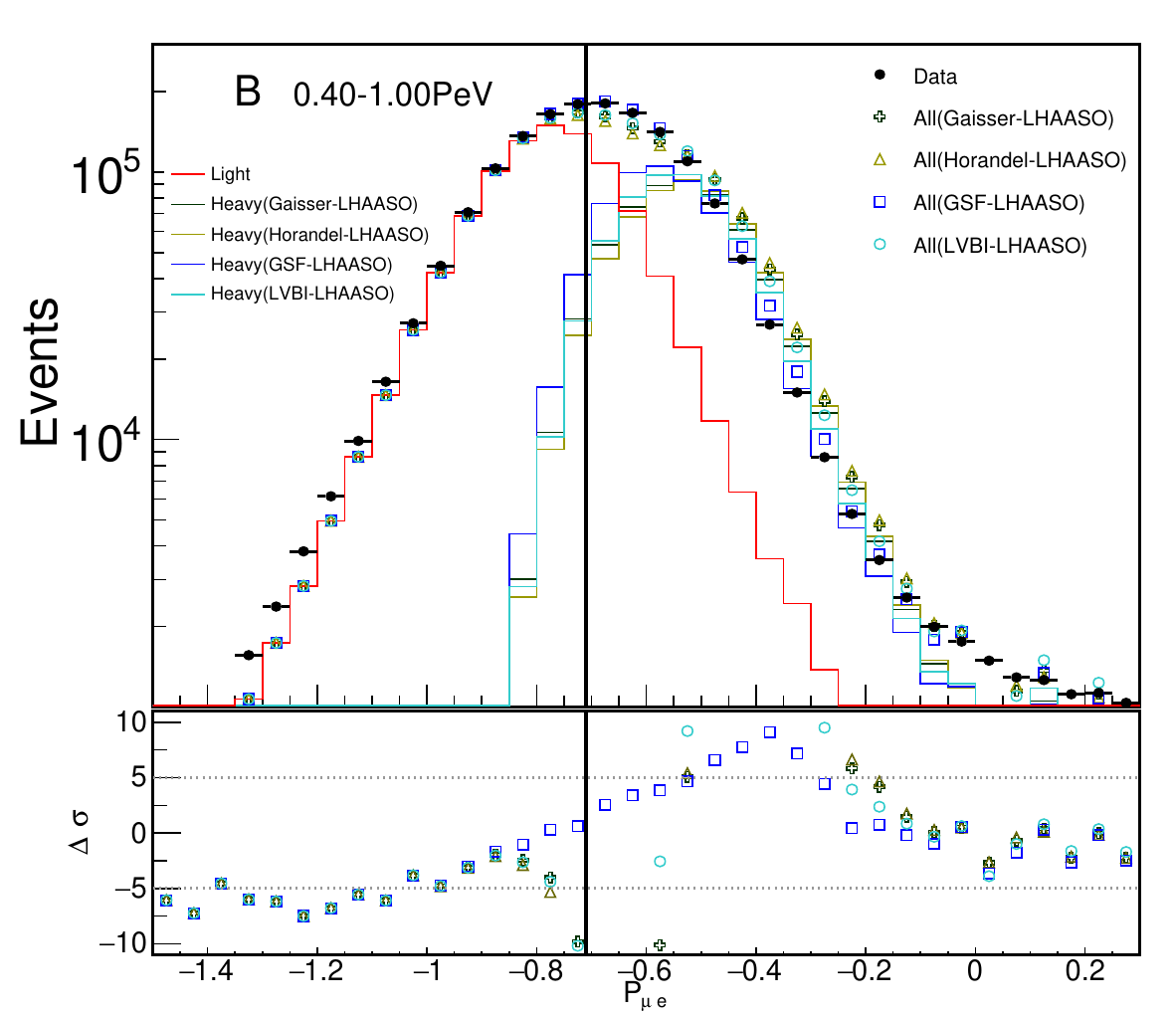}
    \hfill
    \includegraphics[width=0.49\linewidth]{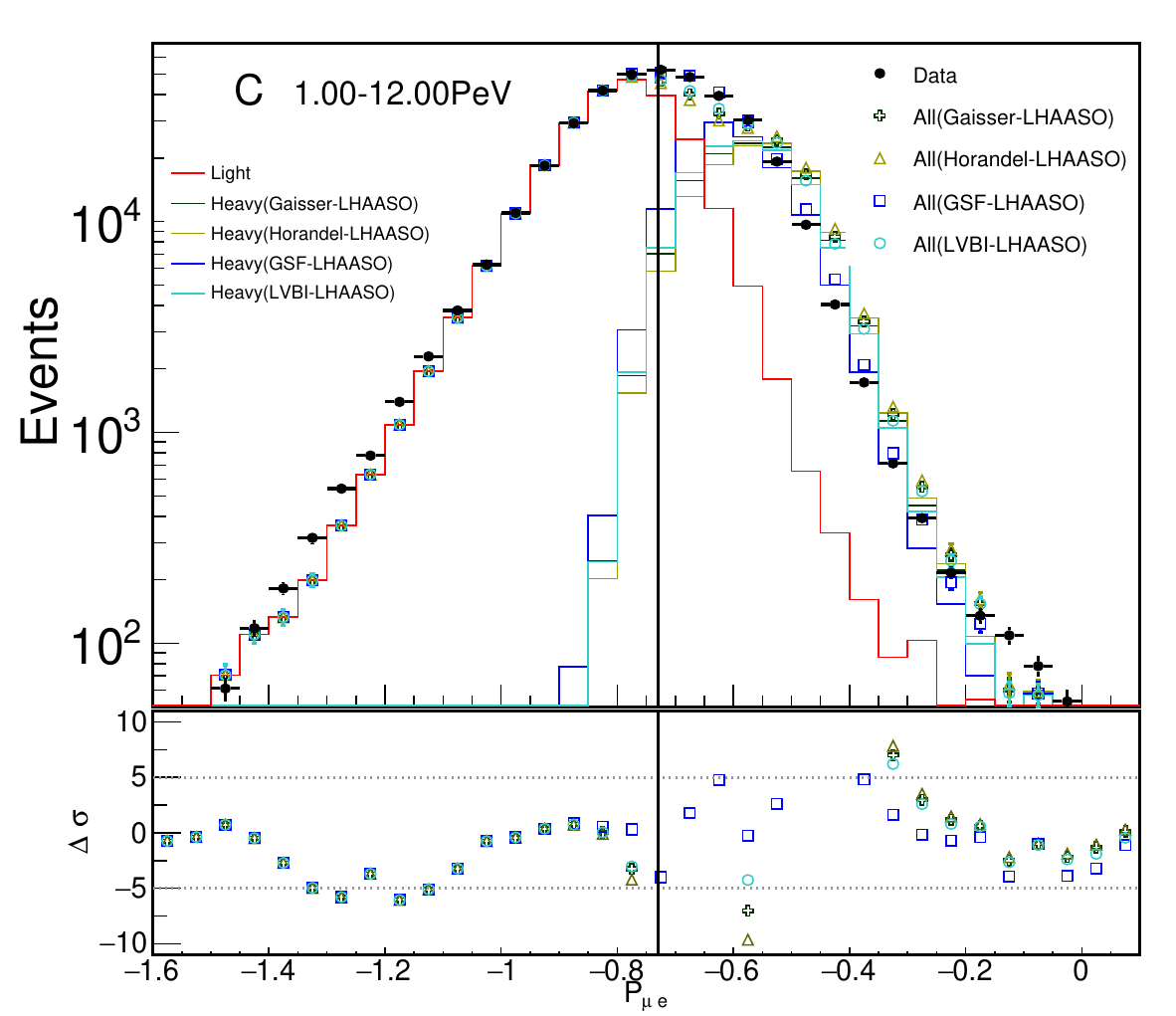}
    \caption{Upper: Distributions of $P_{\mu e}$ for events in the energy intervals 0.16~–~0.40 PeV (A), 0.40~–~1.00 PeV (B), and 1.00~–~13.00 PeV (C). LHAASO experimental data are indicated by solid black points. Open symbols correspond to predictions from different composition models. The red line shows the $P_{\mu e}$ distribution derived from the light component (proton and helium) obtained in this work. The dark green, light green, dark blue and light blue lines represent $P_{\mu e}$ distributions based on heavy component from different composition models. Simulations use the QGSJET-II-04 hadronic interaction model.
    Bottom: Open symbols show deviations between experimental data and simulations for each composition model. Gray lines mark the $\pm 5\sigma$ range. The solid black line indicates the maximum $P_{\mu e}$ threshold used for light component selection.}
    \label{fig:qgsj}
\end{figure}

\begin{figure}[h]
    \centering
    \includegraphics[width=0.49\linewidth]{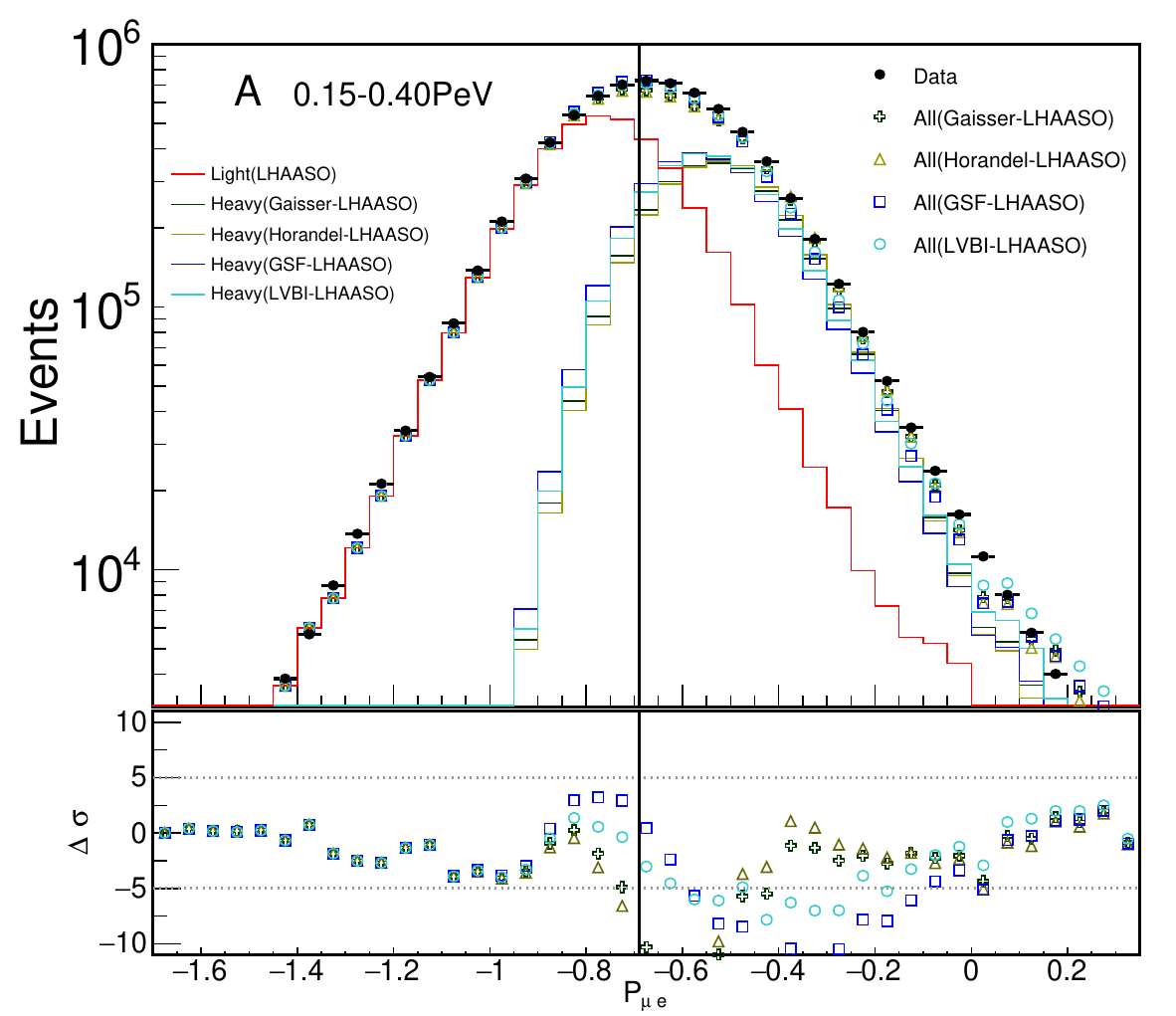}
    \hfill
    \includegraphics[width=0.49\linewidth]{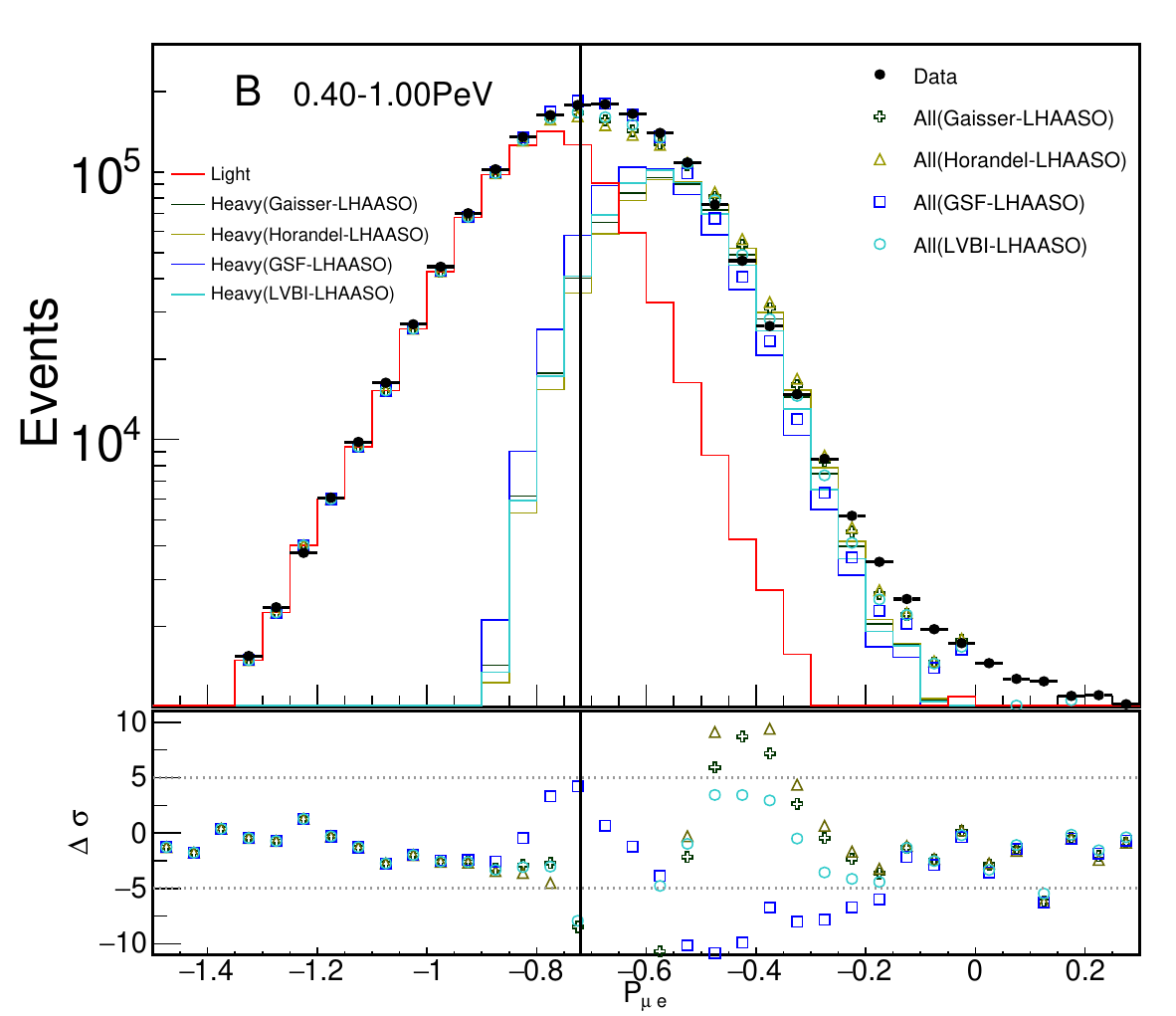}
    \hfill
    \includegraphics[width=0.49\linewidth]{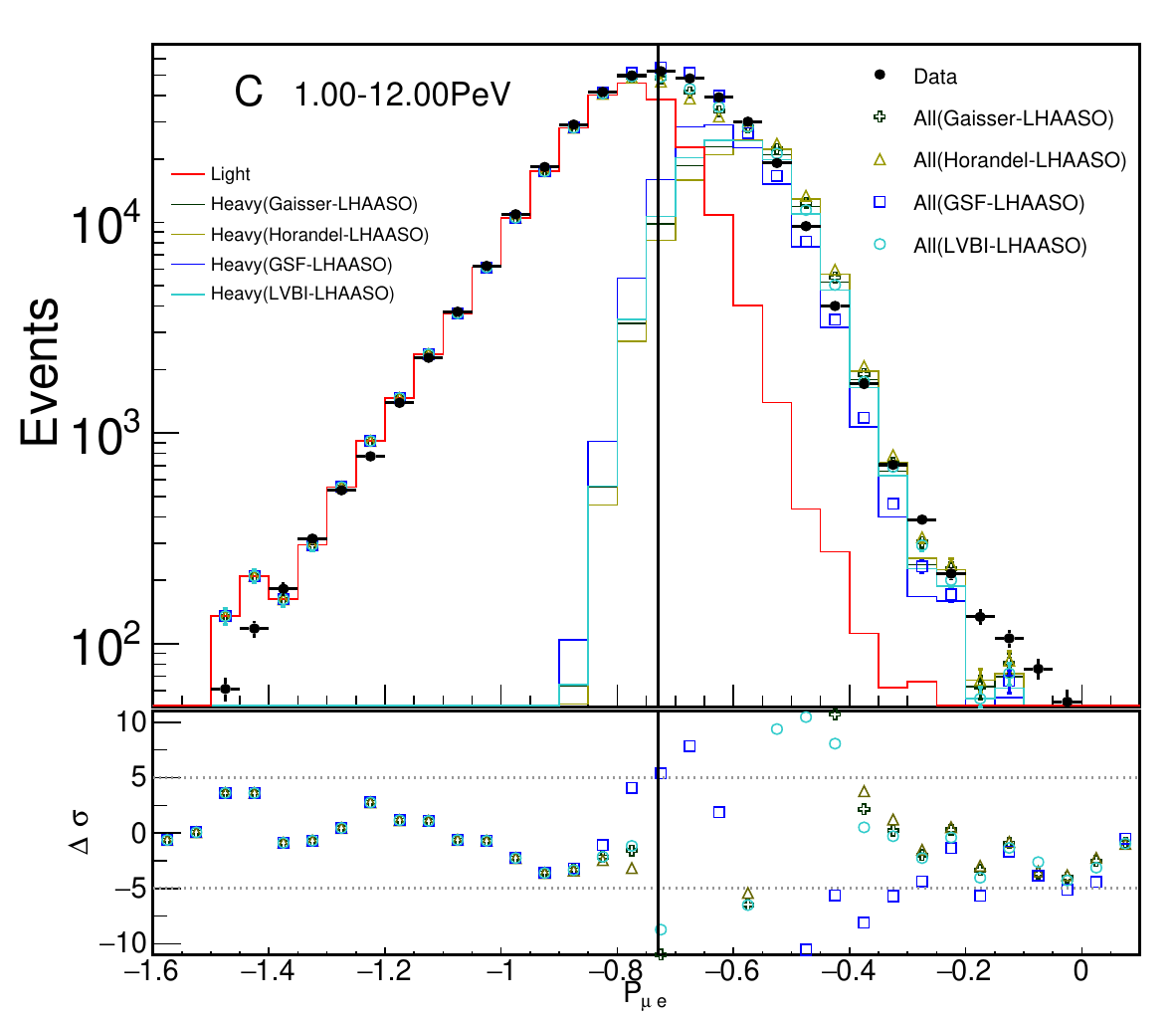}
    \caption{Upper: Distributions of $P_{\mu e}$ for events in the energy intervals 0.16~–~0.40 PeV (A), 0.40~–~1.00 PeV (B), and 1.00~–~13.00 PeV (C). LHAASO experimental data are indicated by solid black points. Open symbols correspond to predictions from different composition models. The red line shows the $P_{\mu e}$ distribution derived from the light component (proton and helium) obtained in this work. The dark green, light green, dark blue and light blue lines represent $P_{\mu e}$ distributions based on heavy component from different composition models. Simulations use the SIBYLL 2.3d hadronic interaction model.
    Bottom: Open symbols show deviations between experimental data and simulations for each composition model. Gray lines mark the $\pm 5\sigma$ range. The solid black line indicates the maximum $P_{\mu e}$ threshold used for light component selection.}
    \label{fig:siby}
\end{figure}

As shown in Fig.~\ref{fig:epos}, Fig.~\ref{fig:qgsj} and Fig.~\ref{fig:siby}, the interpretation of the $P_{\mu e}$ distribution depends on the composition model used to estimate the contamination of heavy elements and also exhibits some dependence on the hadronic interaction model. For $P_{\mu e}\leq-$0.9, the main components are protons and helium, and the proportion of heavy component is very small and can be ignored. The energy spectrum of protons and helium has been accurately measured in this work; therefore, the $P_{\mu e}$ distribution in this range can be used to study the differences between MC and data under different hadronic interaction models. From the comparison between MC and data, EPOS-LHC and SIBYLL 2.3d show slightly better performance compared to QGSJET-II-04, without significant systematic bias. However, their deviation from experiment data remains within $\pm$5$\sigma$, making it difficult to draw a clear conclusion. 
Although the composition models are updated in this work, we only simply replace the proton energy spectrum and the helium energy spectrum according to the LHAASO measurements, and adjust the proportion of heavy component in the all-particle spectrum. The relative proportions and spectral characteristics between CNO, MgAlSi, and iron remain consistent with the original model. Therefore, for larger $P_{\mu e}$ values, as the proportion of heavy component such as CNO, MgAlSi, and iron gradually increases, $P_{\mu e}$ exhibits a clear dependence on the composition models.
The $P_{\mu e}$ distribution of experiment data lies between the $P_{\mu e}$ distributions predicted by different composition models and different hadronic interaction models. With known compositions, especially single-element samples, it will be possible to study how to improve the hadronic interaction model, thereby enhancing the measurement accuracy of energy spectra and compositions. This will be carried out in future research.

\end{document}